\documentclass[11pt,a4paper,twoside]{article}
\usepackage{graphicx,xcolor,textpos}

\usepackage{helvet}
\usepackage[english]{babel}
\usepackage[normalem]{ulem}
\usepackage{amsmath}
\usepackage{amsthm}
\usepackage{thmtools,thm-restate}
\usepackage{bbm}
\usepackage{amssymb}
\usepackage{hyperref}
\usepackage{relsize}
\usepackage[margin=0.7in]{geometry}
\usepackage{physics}
\usepackage{enumitem}
\usepackage{mathtools}
\usepackage{changepage}
\usepackage{caption}
\usepackage{subcaption}
\usepackage{verbatim}
\usepackage{url}
\usepackage{standalone}
\usepackage{tikz}
\usepackage{ifthen}
\usetikzlibrary{calc,patterns,angles,quotes}

\makeatletter
\newcommand\Label[1]{&\refstepcounter{equation}(\theequation)\ltx@label{#1}&}
\makeatother

\newcommand{\stkout}[1]{\ifmmode\text{\sout{\ensuremath{#1}}}\else\sout{#1}\fi}
\newcommand{\IAC}{\textrm{IAC}}
\newcommand{\cone}{\textrm{Cone}}
\renewcommand{\d}{\text{d}}

\newcommand{\defeq}{\vcentcolon=}
\newcommand{\Hsplit}{{H_{\text{split}}}}
\newcommand{\HsplitTilde}{{\tilde{H}_{\text{split}}}}
\newcommand{\phiTilde}{\tilde{\phi}}

\let\originalleft\left
\let\originalright\right
\renewcommand{\left}{\mathopen{}\mathclose\bgroup\originalleft}
\renewcommand{\right}{\aftergroup\egroup\originalright}

\newcommand{\UU}{\mathcal U}

\newcommand{\BB}{\mathcal B}
\newcommand{\PP}{\mathcal P}
\newcommand{\HH}{\mathcal H}
\newcommand{\ZZ}{\mathbb Z}
\newcommand{\CC}{\mathbb C}
\newcommand{\TT}{\mathbb T}
\renewcommand{\AA}{\mathcal A}

\newcommand{\RR}{\mathbb R}
\newcommand{\NN}{\mathbb{N}}
\newcommand{\id}{\mathbbm{1}}

\newcommand{\Ad}[1]{\textrm{Ad}\left(#1\right)}
\newcommand{\Aut}[1]{\textrm{Aut}\left(#1\right)}
\newcommand{\QAut}[1]{\textrm{QAut}_1\left(#1\right)}

\newcommand{\ue}{\underset{\text{u.e.}}{\sim}}

\theoremstyle{definition}
\newtheorem{theorem}{Theorem}[section]
\newtheorem{definition}[theorem]{Definition}
\newtheorem{lemma}[theorem]{Lemma}
\newtheorem{remark}[theorem]{Remark}

\newtheorem{conjecture}[theorem]{Conjecture}

\def\version{0}

\newcommand{\versionDifference}[2]{\ifthenelse{\version=0}{#1}{#2}}

\title{SPT indices emerging from translation invariance in two-dimensional quantum spin systems}
\author{Tijl Jappens\footnote{Institute for Theoretical Physics, Katholieke Universiteit Leuven.}}
\date{\today}

\numberwithin{equation}{section}
\begin{document}
	\maketitle 
	\begin{abstract}
		We consider SPT-phases with on-site $G$ (where $G$ is any finite group) symmetry for two-dimensional quantum spin systems. We then impose translation invariance in one direction and observe that on top of the $H^3(G,\TT)$-valued index constructed in \cite{ogata2021h3gmathbb}, an additional $H^2(G,\TT)$-valued index emerges. We also show that if we impose translation invariance in two directions, on top of the expected $H^3(G,\TT)\oplus H^2(G,\TT)\oplus H^2(G,\TT)$ valued index, an additional $H^1(G,\TT)$-valued index emerges.
	\end{abstract}
	\section{Introduction}
	In 2010, Chen, Gu and Wen \cite{chen_gu_wen_2011} introduced the notion of symmetry-protected topological phases of matter (SPT). They first considered one-dimensional matrix product states and later on, Chen, Liu and Wen \cite{Chen_2011}, extended this to two-dimensional projective entangled pair models. Afterwards, in \cite{Chen_2013} this concept was extended to other dimensions. This last reference used a setup involving so-called lattice non-linear sigma models and provided an $H^{d+1}(G,U(1))$ valued index for any $d$-dimensional lattice non-linear sigma model. The standard setup for defining SPT phases is in quantum spin systems (defined in section \ref{sec:QuasiLocalC*Algebra}). To define SPT phases in this setting, one needs to fix a finite group $G$ with an on-site group action $g\mapsto U_g$ (with $g\in G$), defining a global group action $\beta_g$ (also defined in section \ref{sec:QuasiLocalC*Algebra}). A state $\omega$, is then called $G$-invariant, if $\omega\circ\beta_g=\omega$. Afterwards, one needs to restrict the space of states. We\footnote{There are other definitions possible like for instance the concept of invertible G-invariant states used in \cite{kapustin2021classification} or the unique gapped ground state of a G invariant interaction presented in the introduction of \cite{ogata2021h3gmathbb}.} ask that our states are short-range entangled (SRE), see definition \ref{def:sre}. A state $\omega$ is short-range entangled if there is a disentangler $\gamma$. This is a locally generated automorphism, produced by a one-parameter family of interactions $\Phi\in\BB_{F_\phi}([0,1])$ ($\gamma=\gamma_{0;1}^\Phi$ as defined in section \ref{sec:Interactions}) satisfying that $\omega\circ\gamma$ is a product state. Combining these two definitions, we say that $\omega$ is an SPT state if it is short-range entangled and $G$-invariant.
	\\\\
	One then defines an equivalence relation on these SPT states by saying that $\omega_1$ is equivalent to $\omega_2$ with respect to the on-site group action $U$ if there exists a $G$-invariant one-parameter family of interactions $\Phi\in\BB_{F_\phi}([0,1])$ such that $\omega_1\circ\gamma^\Phi_{0;1}=\omega_2$ (again, see section \ref{sec:Interactions} for the definitions). One can then extend this equivalence to the stronger notion of being stably equivalent. This goes as follows: one defines an operation called stacking that takes two SPT states $\omega_1$ and $\omega_2$ and outputs a third one $\omega_1\otimes_{\text{stack}}\omega_2$ (and similarly for the group action). This $\otimes_\text{stack}$ is defined in section \ref{sec:QuasiLocalC*Algebra}. In short, it is the tensor product at the level of the on-site Hilbert space. One then defines another equivalence relation and says that two SPT states with on-site group actions $(\omega_1,U_1)$ and $(\omega_2,U_2)$ are stably equivalent if and only if there exist trivial\footnote{A trivial state is a $G$-invariant product state that transforms trivially under the on-site group action (see remark \ref{rem:NontrivialProductState}).} SPTs with their group actions, $(\phi_1,\tilde{U}_1)$ and $(\phi_2,\tilde{U}_2)$ and a ($G$ independent) unitary $V$ mapping between the respective on-site Hilbert spaces\footnote{The reader can think of this $V$ as being a unitary transformation of the on-site group action.}, such that
	\begin{enumerate}
		\item $V^\dagger U_{1,g}\otimes \tilde{U}_{1,g}V=U_{2,g}\otimes \tilde{U}_{2,g}$ for all $g\in G$.
		\item $\omega_1\otimes_{\text{stack}}\phi_1\circ i_V$ (where $i_V$ is defined in section \ref{sec:QuasiLocalC*Algebra}) is equivalent to $\omega_2\otimes_{\text{stack}}\phi_2$ with respect to the above on-site group action.
	\end{enumerate}
	In one spatial dimension so-called matrix product states (\cite{Chen_2011},\cite{pollman2012symmetry},\cite{schuch2011MatrixProduct}) can be used to construct non-trivial SPT states. Later on and more generally it was shown in \cite{ogata2019classification} that any $G$-invariant state satisfying the split property (this includes any SRE state) carries an $H^2(G,\TT)$-valued\footnote{By $H^n(G,\TT)$ we mean the n-th (Borel) group cohomology of $G$ with coefficients in the torus $\TT$ (seen as a (group) module under addition with trivial group action).} index and that this index is constant on the (stable) equivalence classes. Later on in \cite{kapustin2021classification} it was then shown that this classification problem is complete (the index, seen as a map from the set of stable equivalence classes to $H^2(G,\TT)$ is injective.). In two spatial dimensions, more recently in \cite{ogata2021h3gmathbb}, it was proven there is an $H^3(G,\TT)$-valued index that is constant on the (stable) equivalence classes\footnote{In \cite{ogata2021h3gmathbb}, there is no mention of the stacking operation but using remark \ref{rem:StackingH3ValuedIndex} and remark \ref{rem:OnSiteUnitaryTransformationH3ValuedIndex}, one can extend it to the stable equivalence class.}. In this paper, we will denote this $H^3(G,\TT)$-valued index of a state $\omega$ by $\textrm{Index}_{\text{2D}}(\omega)$.
	\\\\
	We've used stacking to define the stable equivalence relation. Stacking however plays a more general role. It is widely believed that it allows us to define an (abelian) group structure on the set of stable equivalence classes\footnote{The stacking with a trivial state will then be related to the identity of this group}. Although it is nowhere stated explicitly, the author thinks that the literature (like, for example, \cite{kapustin2021classification}) suggests the following conjecture:
	\begin{conjecture}\label{conj:StableEquivGroupStructure}
		Take some fixed group and a fixed lattice. Let $S$ be the set of stable SPT classes defined through the above equivalence class. Let $*$ be the map
		\begin{equation}
			*:S^2\rightarrow S:(\expval{\omega_1,U_1}_{\sim},\expval{\omega_2,U_2}_{\sim})\mapsto \expval{\omega_1\otimes_\text{stack}\omega_2,U_1\otimes U_2}_{\sim}.
		\end{equation}
		Then $*$ is well-defined (independent of the choice of representative) and $S$ equipped with the $*$ operation forms an abelian group.
	\end{conjecture}
	Before proceeding, we make three remarks on this conjecture:
	\begin{enumerate}
		\item The claim that $*$ is well-defined and abelian is trivial. Namely, by letting $i_V$ be the automorphism that exchanges the two algebras that are being stacked, one can exchange the order of the stacking freely. By using
		\begin{equation}
			\omega_1\circ\gamma^{\Phi_1}_{0;1}\otimes_\text{stack}\omega_2\circ\gamma^{\Phi_2}_{0;1}=(\omega_1\otimes_{\text{stack}}\omega_2)\circ\gamma^{\Phi_1\otimes\id+\id\otimes\Phi_2}_{0;1}\sim\omega_1\otimes_{\text{stack}}\omega_2,
		\end{equation}
		one can show that this is a well-defined map.
		\item A representative of the inverse is sometimes called a $G$-inverse.
		\item From a category point of view, if one assumes that this conjecture is true (for some subclass of groups that is a category), it is not hard to see that it gives rise to a contravariant functor from this category of groups to the category of abelian groups (say $S[G]$). More specifically, to each group morphism $f:G_1\rightarrow G_2$, one can find a group morphism $\tilde{f}:S[G_2]\rightarrow S[G_1]:\expval{\omega,U}_\sim\mapsto \expval{\omega,U\circ f}_\sim$ and this $\tilde{f}$ is indeed invariant on the choice of representative (if two states can be connected while preserving the symmetry action $\beta$, they can certainly be connected while preserving the symmetry $\beta\circ f$).
	\end{enumerate}
	In what follows, we will call $S$ with the above group structure the stable SPT classification. In this paper, we consider the 2D case with the (stable) equivalence relation as presented here but with one notable difference. We will only consider states, interactions, and on-site symmetries that have a translation symmetry in one direction. There is a conjecture about the SPT classification of such states, see section 3.1.5.1 of \cite{xiong2019classification} for a more general and detailed exposition (which was partially based on \cite{Chen_2013}).
	\begin{conjecture}\label{conj}
		The set of translation invariant SPTs under stable equivalence\footnote{In the stable equivalence relation for translation invariant states we ask that the on-site symmetry is translation invariant and that any path connecting two states and each trivial state is not only G-invariant but also translation invariant.} also satisfies conjecture \ref{conj:StableEquivGroupStructure}. Let $g$ be the inclusion map from the space of 2d translation invariant SPT states to the more general set of 2d SPT states. Let $f$ be the map that takes a 1d SPT state and outputs a translation invariant 2d SPT state by taking the tensor product in the direction of the translation symmetry. The sequence
		\begin{equation}
			0\rightarrow\{\text{1d SPT states}\}\stackrel{f}{\rightarrow}\{\text{2d translation invariant SPT states}\}\stackrel{g}{\rightarrow}\{\text{2d SPT states}\}\rightarrow 0
		\end{equation}
		induces a sequence (of group morphisms) on the stable equivalence classes. By this, we mean that the class of $f(\phi)$ only depends on the class of $\phi$ and similarly for $g$. Moreover, this induced sequence is exact and split.
	\end{conjecture}
	This implies that
	\begin{equation}
		\{\text{2d translation invariant SPT classification}\}\simeq \{\text{1d SPT classification}\}\oplus \{\text{2d SPT classification}\}.
	\end{equation}
	Since a 1d SPT state carries an $H^2(G,\TT)$ valued index and a 2d SPT state carries an $H^3(G,\TT)$ valued index, this means that 2d SPT states with a translation symmetry should carry an $H^2(G,\TT)\oplus H^3(G,\TT)$ valued index. This is consistent with the result of section XII of \cite{Chen_2013}.
	\\\\
	The first result of this paper is the construction of an $H^2(G,\TT)$ valued index (which we will denote by $\textrm{Index}$) on the space of translation invariant SPT states that is consistent with conjectures \ref{conj:StableEquivGroupStructure} and \ref{conj}. By being consistent with the conjecture it is meant (among other things) that, if $\textrm{Index}_1$ is the one-dimensional SPT index from \cite{ogata2019classification}, then $\textrm{Index}(f(\omega))=\textrm{Index}_1(\omega)$ for any 1d SPT $\omega$. This construction highly relies on the objects that are present in the construction of the 2d SPT index in \cite{ogata2021h3gmathbb}.
	\\\\
	We also look at the case where there is a second translation symmetry (so the system is translation invariant in both $x$ and $y$ directions). For such systems, there is the following conjecture (see again \cite{xiong2019classification} and \cite{Chen_2013}):
	\begin{conjecture}\label{conj2}
		The space of SPTs, translation invariant in two directions under stable equivalence also satisfies conjecture \ref{conj:StableEquivGroupStructure}. Let $g$ be the inclusion map from the space of 2d SPT states that are translation invariant in both $x$ and $y$ directions to the more general set of 2d SPT states with translation invariance in the $x$ direction only. Let $f$ be the map that takes a 1d translation invariant SPT state and outputs a 2d SPT state that is translation invariant in both $x$ and $y$ directions by taking the tensor product in the $y$ direction. The sequence
		\begin{equation}
			0\rightarrow\left\{\begin{matrix}\text{1d translation invariant}\\ \text{SPT states}\end{matrix}\right\}\stackrel{f}{\rightarrow}\left\{\begin{matrix}\text{2d SPT states with two}\\ \text{translation symmetries}\end{matrix}\right\}\stackrel{g}{\rightarrow}\left\{\begin{matrix}\text{2d SPT states translation}\\ \text{invariant in $x$ direction}\end{matrix}\right\}\rightarrow 0
		\end{equation}
		induces a sequence (of group morphisms) on the stable equivalence classes. By this, we mean that the class of $f(\phi)$ only depends on the class of $\phi$ and similarly for $g$. Moreover, this induced sequence is exact and split.
	\end{conjecture}
	Translation invariant SPT states in one spatial dimension are known to carry an $H^2(G,\TT)\oplus H^1(G,\TT)$ valued index (see section \ref{sec:OneDimensionalIndices} for a construction of these indices consistent with the rest of the setup). From this and the last two conjectures, we expect there to be an
	\begin{equation}\label{eq:2TranslationsIntroduction}
		H^3(G,\TT)\oplus H^2(G,\TT)\oplus H^2(G,\TT)\oplus H^1(G,\TT)
	\end{equation}
	valued index. The first part of the index is just the above mentioned $\textrm{Index}_{2d}$. The following two parts can be related to the case of a single translation symmetry as follows. Let $\mu$ be the automorphism that rotates the lattice by 90° and let $\textrm{Index}(\omega)$ be the index as constructed before (this is the second part of \ref{eq:2TranslationsIntroduction}). The third part of the index is now given by $\textrm{Index}(\omega\circ\mu)$ (the state rotated by 90°). The last part of the index requires a different construction altogether. It can be thought of as being the charge in the Brillouin zone.
	\\\\
	A final remark:
	\begin{remark}
		As noticed in \cite{Chen_2013}. The groups in which these indices take values can be concisely written out. Using that
		\begin{align}
			H^0(\ZZ,\TT)&\cong \TT& H^1(\ZZ,\TT)&\cong \TT& H^2(\ZZ,\TT)&\cong 0&H^3(\ZZ,\TT)&\cong 0
		\end{align}
		and inserting this into the K\"unneth formula for $\ZZ\times G$ gives
		\begin{align}
			H^2(\ZZ \times G,\TT)&\cong \bigoplus_{i=0}^2 H^i(\ZZ,\TT)\otimes H^{2-i}(G,\TT)\cong H^2(G,\TT)\oplus H^1(G,\TT)\\
			H^3(\ZZ \times G,\TT)&\cong \bigoplus_{i=0}^3 H^i(\ZZ,\TT)\otimes H^{3-i}(G,\TT)\cong H^3(G,\TT)\oplus H^2(G,\TT).
		\end{align}
		This result can in turn be used to work out the K\"unneth formula for $\ZZ^2\times G$
		\begin{equation}
			H^3(\ZZ^2\times G,\TT)\cong \bigoplus_{i=0}^3 H^i(\ZZ,\TT)\otimes H^{3-i}(\ZZ\times G,\TT)\cong H^3(G,\TT)\oplus H^2(G,\TT)\oplus H^2(G,\TT)\oplus H^1(G,\TT).
		\end{equation}
	\end{remark}
	The layout of this paper is as follows: First in section \ref{sec:Setup} we explain the setup, define the concept of locally generated automorphisms, and give the algebraic definition of group cohomology. In section \ref{sec:Results_1} we state the two results for SRE states. In section \ref{sec:Results_2}, we state the first result using the weaker assumption \ref{assumption} and the second result using the weaker assumption \ref{assumption:2Translations}. It is using these weaker assumptions that we then prove the statements. The rest of the paper provides a proof of these statements. The results in this paper rely heavily on the methods that were developed in \cite{ogata2021h3gmathbb}.
	\subsection*{Acknowledgements}
	T.J. was supported in part by the FWO under grant G098919N.\\\\
	We are grateful to the referees of Communications in Mathematical Physics for their constructive input during the review process. For example, we are very grateful for the proposed changes of lemma \ref{lem:SplittingOfUnitary}, the mistake they found in section \ref{sec:examples} (the previous assumption was too weak), and the mistakes they found in section \ref{sec:AllIndicesInvariantUnderLGA}.
	\subsection*{Data availability statement}
	No new data were created or analyzed in this study. Data sharing is not applicable to this article.
	\versionDifference{\subsection*{ArXiv version}This manuscript is a (more extensive) ArXiv version of a paper that will be published in communications in mathematical physics. To access the journal version of the paper, see \url{https://rdcu.be/dwtjE}.}{\subsection*{ArXiv version}This paper is a more compact version of the ArXiv version \cite{jappens2023spt}. We will sometimes refer to this version.}
	\section{Setup and definitions}\label{sec:Setup}
	In this paper, we work in the two-dimensional lattice $\ZZ^2$. We will first need some specific subsets of $\ZZ^2$, so let
	\begin{align}
		L&\defeq \left\{(x,y)\in\ZZ^2\left|x<0\right.\right\},&R&\defeq \left\{(x,y)\in\ZZ^2\left|x\geq 0\right.\right\}\\
		U&\defeq \left\{(x,y)\in\ZZ^2\left|y\geq 0\right.\right\},&D&\defeq \left\{(x,y)\in\ZZ^2\left|y<0\right.\right\}
	\end{align}
	be the left, right, upper, and lower half-planes respectively, and let
	\begin{equation}
		C_\theta\defeq \left\{(x,y)\in\ZZ^2\left|\tan(\theta)\leq\abs{\frac{y}{x}}\right.\right\}
	\end{equation}
	be the horizontal cone (the green area in figure \ref{fig:SetupWithQAutomorphism}). We will use $\tau$ to denote the bijection that moves every element of $\ZZ^2$ one site upward. Similarly, we let $\nu$ denote the bijection that translates every element of $\ZZ^2$ by one site to the right. In what follows we will sometimes need to widen our cone or other subsets of $\ZZ^2$ vertically by one site. For this purpose we define $W$ such that for any $\Gamma\subset\ZZ^2$,
	\begin{equation}
		W(\Gamma)\defeq \Gamma\cup\tau(\Gamma)\cup\tau^{-1}(\Gamma).
	\end{equation}
	For example, the red area in figure \ref{fig:SetupWithQAutomorphism} is $W(C_\theta)^c$. In the later part, we will also need to rotate our lattice by $90^\circ$ (clockwise) so call $\mu$ the bijection on $\ZZ^2$ that does precisely this.\footnote{We will not discuss rotation invariant states in this paper. The $\mu$ is only defined to make certain constructions work.} Finally, we will need to define the horizontal lines and the finite horizontal line
	\begin{align}
		L_j&\defeq \{(i,j)|i\in\ZZ\}&L_j^n&\defeq \{(i,j)|i\in\ZZ\cap[-n,n]\}.
	\end{align}
	\subsection{Quasi local \texorpdfstring{$C^*$}{}-algebra}\label{sec:QuasiLocalC*Algebra}
	The setup in this paper will be very similar to the setup in \cite{ogata2021h3gmathbb} and is just the standard setup for defining quantum spin systems. For the rest of this paper, take $d\in\NN_+$ arbitrary. This number will be called the on-site dimension\footnote{In principle in the case where there is only one translation we can let $d$ be $x$-dependent but for simplicity let us take $d$ constant over the lattice.}. For each $z\in\ZZ^2$, let $\AA_{\{z\}}$ be an isomorphic copy of $\BB(\CC^d)$ (the bounded operators on $\CC^d$). In what follows, let $\mathfrak{G}_{\ZZ^2}$ be the set of finite subsets of $\ZZ^2$. For any $\Lambda\in\mathfrak{G}_{\ZZ^2}$, we set $\AA_\Lambda=\bigotimes_{z\in\Lambda}\AA_{\{z\}}$. We define the local algebra as $\AA_{\text{loc}}=\bigcup_{\Lambda\in \mathfrak{G}_{\ZZ^2}}\AA_{\Lambda}$ and define the quasi local $C^*$ algebra as the norm closure of the local algebra ($\AA=\overline{\AA_{\text{loc}}}$). Similarly, for any (possibly infinite) subset $\Gamma\subset\ZZ^2$ we set $\AA_{\text{loc},\Gamma}=\bigcup_{\Lambda\in \mathfrak{G}_{\Gamma}}\AA_{\Lambda}$ and $\AA_\Gamma=\overline{\AA_{\text{loc},\Gamma}}$.\\\\
	Throughout this paper, we will sometimes require a stronger form of locality than merely being in the norm completion of the local algebra. Specifically when $\Gamma=L_j$ is a horizontal line. We will say that an operator $A\in\AA_{L_j}$ is summable if and only if there exists a sequence $A_n\in\AA_{L_{j}^n}$ on the finite horizontal lines such that
	\begin{equation}
		\sum_{i=0}^{\infty}\norm{A-A_n}<\infty.
	\end{equation}
	We will need to define some automorphisms on this algebra using the bijections defined previously. To this end, define the isomorphisms
	\begin{align}
		\tau&:\AA_\Gamma\rightarrow\AA_{\tau(\Gamma)}&\nu&:\AA_\Gamma\rightarrow\AA_{\nu(\Gamma)}&\mu&:\AA_\Gamma\rightarrow\AA_{\mu(\Gamma)}
	\end{align}
	as the translation upwards, the translation to the right, and the right-handed rotation respectively. By construction, these isomorphisms are automorphisms if $\Gamma$ is taken to be $\ZZ^2$ (because $d$ is a constant throughout the lattice). Clearly if $\Gamma=\ZZ$ (by which we mean again the $x$-axis $(\ZZ,0)$) only one of these translation automorphisms descends to an automorphism on $\AA_{\ZZ}$. This automorphism is then also simply denoted as $\nu$.\\\\
	We will need to define one additional class of automorphisms. For any set of unitaries $V_i\in\UU(\CC^d)$ (labelled by $i\in\ZZ^2$), we define the (unique) automorphism $i_V^\Gamma\in\Aut{\AA_\Gamma}$ (for all $\Gamma\subset \ZZ^2$) such that $i_V^\Gamma (A)=\Ad{\otimes_{i\in I} V_i}(A)$ for all $I\subset\Gamma$ finite and $A\in\AA_I$. This class of automorphisms will be used to define a group action on the algebra. More specifically, let $G$ be a finite group and let $U_i\in\hom(G,\UU(\AA_i))$ (for $i\in\ZZ^2$) be the on-site group action. Let $\beta^\Gamma\in\hom(G,\Aut{\AA_\Gamma})$ (for any $\Gamma\subset\ZZ^2$) be such that
	\begin{equation}
		\beta^\Gamma_g(A)=i_{U(g)}^\Gamma=\Ad{\otimes_{i\in I} U_i(g)}(A)
	\end{equation}
	for any $g\in G$, $I\subset\Gamma$ finite and $A\in\AA_I$. Clearly, any unitary transformation of our representation $U(\cdot)\rightarrow V^\dagger U(\cdot) V$ induces a transformation of the group action $\beta_g\rightarrow i_{V}^{-1}\circ\beta_g\circ i_V$.\\\\
	Throughout our paper when we are discussing translation invariance in the vertical direction we will ask that the group action is translation invariant in the vertical direction ($\tau\circ\beta_g^\Gamma=\beta_g^{\tau(\Gamma)}\circ\tau$). If we also have translation invariance in the horizontal direction, we will also ask that the group action be translation invariant in that direction as well ($\nu\circ\beta_g^\Gamma=\beta_g^{\nu(\Gamma)}\circ\nu$). As it turns out, in showing that the $H^1$-valued index is consistent with all the conjectures of the introduction, we will even require that the on-site group action $U_i(g)$ is the same at each site. In what follows we will always assume this. This means that when we denote an on-site symmetry action we merely have to give one representation $U\in\hom(G,\UU(\CC^d))$, and when we denote a unitary transformation of the on-site Hilbert space we only have to give a single unitary $V\in \UU(\CC^d)$, we will use this notation frequently.\\\\
	By construction, there is a tensor product operation on this $C^*$ algebra in the sense that for any $\Gamma_1,\Gamma_2\subset\ZZ^2$ satisfying that $\Gamma_1\cap\Gamma_2=\emptyset$ we can define a bilinear, surjective map
	\begin{equation}
		\otimes:\AA_{\Gamma_1}\times\AA_{\Gamma_2}\rightarrow \AA_{\Gamma_1\cup\Gamma_2}.
	\end{equation}
	There is however a second tensor product operation on this $C^*$ algebra that we will use. We will call it the stacking operation. It is such that for any $\Gamma\subset\ZZ^2$ we define the bilinear, surjective map
	\begin{equation}
		\otimes_{\text{stack}}:\AA_{\Gamma}\times\AA_{\Gamma}\rightarrow \AA^2_{\Gamma}
	\end{equation}
	where $\AA^2$ is the quasi-local $C^*$ algebra on $\ZZ^2$ constructed from an on-site algebra $\BB(\CC^d)\otimes\BB(\CC^d)\simeq \BB(\CC^{d\times d})$\footnote{This can also be used to stack two $C^*$-algebras with different on-site Hilbert spaces, this was used in the introduction. In the rest of the paper we will for simplicity of notation (but without loss of generality) only stack two identical $C^*$-algebras on top of each other.}. We will also need to define the stacking of the group action by which we mean
	\begin{equation}
		(\beta_g\otimes_{\text{stack}}\beta_g)^\Lambda=\bigotimes_{z\in\Lambda}\Ad{U(g)\otimes U(g)_z}
	\end{equation}
	for any $\Lambda\in\mathfrak{G}_{\ZZ^2}$ (in fact stacking can be defined for any automorphism, not just the group action).
	\subsection{States on \texorpdfstring{$\AA$}{}}\label{sec:States}
	We will use $\PP(\AA)$ to denote the set of pure states on $\AA$ (see \cite{bratteli1979operator}). In this paper, we will often refer to the GNS triple. For its definition, construction, and properties we refer to \cite{bratteli1979operator}. We will sometimes write the tensor product of states $\omega\otimes\phi$ to be such that $(\omega\otimes\phi)(a\otimes b)=\omega(a)\phi(b)$ and similarly for the stacking tensor product. We will call a pure state $\phi$ on $\AA_\Gamma$ a product state if for any $\Lambda\subset\Gamma$, the restriction $\phi|_{\AA_\Lambda}$ is still pure. We will call a state $\omega$, $G$-invariant if $\omega\circ\beta_g=\omega$ (where $\beta_g=i_{U(g)}$ with $U$ an on site group action). Moreover, we say that a $G$-invariant product state $\phi$ is trivial with respect to $U$ (we will sometimes denote this as $(\phi,U)$ is trivial) if $\phi(U_i(g))=1$ for all $i\in\ZZ^2$.\\\\
	There is a natural way to construct a pure $G$-invariant state on the quasi local $C^*$-algebra over $\ZZ^2$, $\AA_{\ZZ^2}$ using a pure $G$-invariant state on the quasi local $C^*$-algebra over $\ZZ$, $\AA_{L_0}$\footnote{When applied to a trivial state on $\AA_{L_0}$ this will also give a trivial state on $\AA_{\ZZ^2}$.}.
	\begin{definition}\label{def:InfiniteTensorProduct}
		Let $\phi\in\PP(\AA_{L_0})$ be a state that is $G$-invariant under a group action $\beta_g^{L_0}$. Define $\phi_i\in\PP(\AA_{L_i})$ as $\phi_i:=\phi_0\circ\tau^{-i}$. We define the infinite tensor product state $\omega_\phi$ as $\omega_\phi:=\bigotimes_{j\in\ZZ}\phi_j$. This is a $G$-invariant pure state over $\ZZ^2$ that is invariant under the automorphism $\tau$.
	\end{definition}
	The proof that this is a well-defined state is done in definition \ref{def:InfiniteTensorProductState}.
	\subsection{Interactions and locally generated automorphisms}\label{sec:Interactions}
	An interaction $\Phi$ is a map
	\begin{equation}
		\Phi: \mathfrak{G}_{\ZZ^2}\rightarrow \AA_{\text{loc}}: I \mapsto \Phi(I)
	\end{equation}
	where $\Phi(I)\in\AA_I$ is hermitian ($\Phi(I)=\Phi(I)^*$). We will sometimes use the restriction of an interaction, $\Phi_\Gamma$ for $\Gamma\subset\ZZ^2$. This means setting all $\Phi(I)=0$ when $I$ is not in $\Gamma$. Following the work done in \cite{doi:10.1063/1.5095769}, will use $\norm{\Phi}_F$ to indicate the $F$-norm of $\Phi$ where $F$ is an $F$-function. We will use $\BB_{F}([0,1])$ to denote the one-parameter families of interactions with norm continuous therms and uniformly bounded $F$-norms. Just as was done in \cite{doi:10.1063/1.5095769}, for any $\Phi\in \BB_{F}([0,1])$, we use $\gamma^{\Phi}_{s;t}$ to denote the locally generated automorphism (LGA) generated by $\Phi$ from $s$ to $t$. Following \cite{ogata2021h3gmathbb}, we will fix a specific family of monotonically decreasing positive functions by saying that for any $0<\phi<1$ we define
	\begin{equation}\label{eq:OurFFunction}
		F_\phi:\RR^+\rightarrow\RR^+:r\mapsto \frac{\exp(-r^\phi)}{(1+r)^4}.
	\end{equation}
	This $F$-function is particularly useful to us because, as was shown in \cite{ogata2021h3gmathbb}, if $\Phi\in\BB_{F_\phi}([0,1])$, we can decompose the automorphism into different parts. More precisely, we can show that $\gamma^{\Phi}_{s;t}\in\QAut{\AA}$ where $\QAut{\AA}$ is defined in \ref{def:QAut}\footnote{In appendix \versionDifference{\ref{sec:properties-of-locally-generated-automorphisms-1d}}{B of \cite{jappens2023spt}} we show a stronger version of this statement.}.
	\\\\
	Sometimes we will say that an interaction is $G$-invariant. By this we simply mean that $\beta_g(\Phi(I))=\Phi(I)$ (for all $I\in\mathfrak{G}_{\ZZ^2}$). Similarly, if we say an interaction is translation invariant we mean that $\tau(\Phi(I))=\Phi(\tau(I))$ (for all $I\in\mathfrak{G}_{\ZZ^2}$). It should be clear that if an interaction is $G$-invariant (translation invariant), then the LGA it generates commutes with the group action (the translation automorphism) as well.
	\subsection{The stable equivalence class}\label{sec:StableEquivalenceClasses}
	We will now define a subset of these states that we will call short-range entangled (SRE):
	\begin{definition}\label{def:sre}
		Let $\omega\in\PP(\AA)$. We say that $\omega$ is an SRE state if and only if there exists a $\phi\in]0,1[$ such that there exists an interaction $\Phi\in\BB_{F_\phi}([0,1])$ such that $\omega\circ\gamma^\Phi_{0;1}$ is a product state. We then call $\gamma^\Phi_{0;1}$ a disentangler for $\omega$.
	\end{definition}
	On the set of $G$-invariant SRE states we will now define three stable equivalence classes depending on the additional imposed symmetry:
	\begin{definition}
		Take $(\omega_1,U_1)$ and $(\omega_2,U_2)$ elements of
		\begin{equation}
			S_0=\{(\omega,U)\in\PP(\AA)\times\hom(G,U(\CC^d))|\omega\text{ is SRE},\:\omega\circ i_{U(g)}=\omega\:\forall g\in G\}
		\end{equation}
		for two algebras $\AA_1$ and $\AA_2$\footnote{With two different algebras we simply mean two different on-site dimensions because we already fixed each on-site dimension to be identical.}. We say that $\expval{\omega_1,U_1}_\sim=\expval{\omega_2,U_2}_\sim$ if and only if there exists a $\phi\in]0,1[$ such that there exists a $G$-invariant interaction $\Phi\in\BB_{F_\phi}([0,1])$, trivial elements of $S_0$ (see subsection \ref{sec:States}), $(\phi_1,\tilde U_1)$ and $(\phi_2,\tilde U_2)$ for algebras $\tilde\AA_1$ and $\tilde\AA_2$ respectively and an on-site unitary such that
		\begin{align}
			\omega_1\otimes_{\text{stack}}\phi_1\circ i_V&=\omega_2\otimes_{\text{stack}}\phi_2\circ\gamma^\Phi_{0;1}&V^\dagger U_{1}(g)\otimes \tilde{U}_{1}(g)V&=U_{2}(g)\otimes \tilde{U}_{2}(g).
		\end{align}
	\end{definition}
	We also define similar equivalence classes $\expval{\cdot}_\sim^1,\expval{\cdot}_\sim^2$ and $\expval{\cdot}_\sim^3$ on
	\begin{align}
		\nonumber
		S_1&=\{(\omega,U)\in\PP(\AA)\times\hom(G,U(\CC^d))|\omega\text{ is SRE},\:\omega\circ i_{U(g)}=\omega\:\forall g\in G,\:\omega\circ\tau=\omega\}\\
		S_2&=\{(\omega,U)\in\PP(\AA)\times\hom(G,U(\CC^d))|\omega\text{ is SRE},\:\omega\circ i_{U(g)}=\omega\:\forall g\in G,\:\omega\circ\nu=\omega\}\\
		\nonumber
		S_3&=\{(\omega,U)\in\PP(\AA)\times\hom(G,U(\CC^d))|\omega\text{ is SRE},\:\omega\circ i_{U(g)}=\omega\:\forall g\in G,\:\omega\circ\tau=\omega,\:\omega\circ\nu=\omega\}
	\end{align}
	respectively. Here we now also demand that the interactions and the trivial states are invariant under the additional translation symmetries.
	\section{Results}
	To formulate the results, we will often use $H^n(G,\TT)$ to denote the n-th (Borel) group cohomology of $G$ with coefficients in the torus $\TT$ (seen as a (group) module under addition with trivial group action). For the definition of these group cohomology modules, we refer to \cite{benson1991representations}.
	\\\\
	This section will contain three different parts. First, we discuss the main result starting from some rather abstract assumptions. In the second part, we present the claim that these assumptions are implied by some more natural assumptions relating to short-range entanglement. In the last part of the section we compare our assumptions to those made in \cite{ogata2021h3gmathbb}.
	\subsection{Statement of the result in terms of Q-automorphisms}\label{sec:Results_2}
	\begin{figure}
		\centering
		\def\s{0.5}
		\resizebox{0.27\textwidth}{!}{%
			\begin{tikzpicture}
	\fill[fill=blue!30!white] (-4*\s,-4*\s) rectangle (4*\s,4*\s);
	\draw[draw=black,line width=0.3mm] (0,4*\s) -- (0,-4*\s);
	
	\node at (-2*\s,0) {$\alpha_L$};
	\node at (2*\s,0) {$\alpha_R$};
\end{tikzpicture}}
		$\qquad$
		\def\s{0.4}
		\resizebox{0.32\textwidth}{!}{%
			\begin{tikzpicture}
	\draw[draw=white,line width=0mm] (1+\s,0) coordinate (a) -- (\s,0) coordinate (b) -- (1+\s,1) coordinate (c);
	
	\fill[fill=green!30!white] (-1*\s,0) -- (-6*\s,-5*\s) -- (-6*\s,5*\s);
	\fill[fill=green!30!white] (1*\s,0) -- (6*\s,-5*\s) -- (6*\s,5*\s);
	\fill[fill=red!30!white] (0,1*\s) -- (4*\s,5*\s) -- (-4*\s,5*\s);
	\fill[fill=red!30!white] (0,-1*\s) -- (4*\s,-5*\s) -- (-4*\s,-5*\s);
		
	\draw[draw=black,line width=0.3mm]
	plot[smooth,samples=2,domain=0:4*\s] (\x,\x+1*\s);
	\draw[draw=black,line width=0.3mm]
	plot[smooth,samples=2,domain=-4*\s:0] (\x,-\x+1*\s);
	\draw[draw=black,line width=0.3mm]
	plot[smooth,samples=2,domain=0:4*\s] (\x,-\x-1*\s);
	\draw[draw=black,line width=0.3mm]
	plot[smooth,samples=2,domain=-4*\s:0] (\x,\x-1*\s);
	\draw[draw=black,line width=0.3mm]
	plot[smooth,samples=2,domain=0:5*\s] (\x+1*\s,\x);
	\draw[draw=black,line width=0.3mm]
	plot[smooth,samples=2,domain=0:5*\s] (\x+1*\s,-\x);
	\draw[draw=black,line width=0.3mm]
	plot[smooth,samples=2,domain=-5*\s:0] (\x-1*\s,-\x);
	\draw[draw=black,line width=0.3mm]
	plot[smooth,samples=2,domain=-5*\s:0] (\x-1*\s,\x);
		
	\draw[draw=black,line width=0.3mm] (1*\s,0) -- (3*\s,0);
	\pic [draw, ->, "$\theta$", angle eccentricity=1.5, angle radius=\s*1.5cm,line width=0.3mm] {angle=a--b--c};
		
	\node at (-4*\s,0) {$\eta^L_g$};
	\node at (4.5*\s,0) {$\eta^R_g$};
	\node at (0,3*\s) {$\Theta$};
\end{tikzpicture}}
		\caption{These figures indicate the support area of the different automorphisms in a particular decomposition of the $\textrm{QAut}$. The angle $\theta$ needs to be smaller than or equal to what was indicated here so that the $\Theta$ and the $\eta_g$ (possibly after widening by one site vertically and one site horizontally) commute.}
		\label{fig:SetupWithQAutomorphism}
	\end{figure}
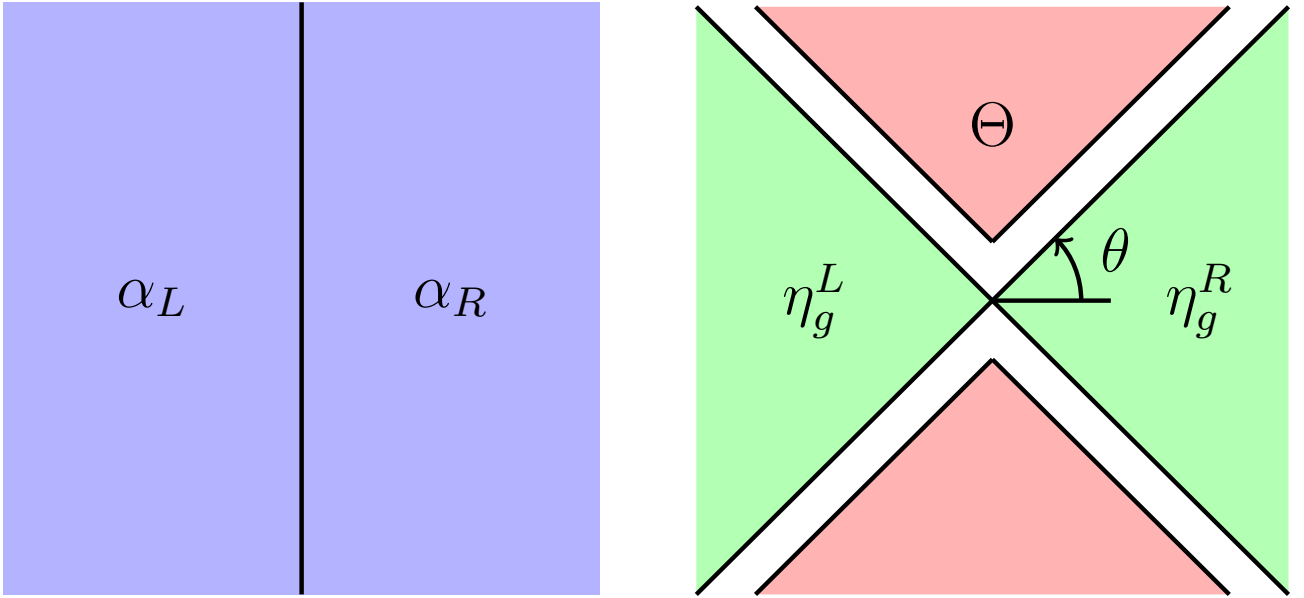
	Similarly what was done in \cite{ogata2021h3gmathbb}, we will define an index on the space of states that can be disentangled by a Q-automorphism. This class of automorphisms is defined as:
	\begin{definition}\label{def:QAut}
		Take $\alpha\in\Aut{\AA}$. We say that $\alpha\in\QAut{\AA}$ if and only if $\forall\theta\in]0,\pi/2[$ there exists an $\alpha_L\in\Aut{\AA_L},\alpha_R\in\Aut{\AA_R},V_1\in\UU(\AA)$ and a $\Theta\in\Aut{\AA_{W(C_\theta)^c}}$\footnote{Note that our definition of $\QAut{\AA}$ is slightly different from the definition of $\textrm{QAut}(\AA)$ from \cite{ogata2021h3gmathbb} because of this vertical widening $W$.} such that
		\begin{equation}
			\alpha=\Ad{V_1}\circ\alpha_L\otimes\alpha_R\circ\Theta.
		\end{equation}
	\end{definition}
	We will often write $\alpha_0$ to mean $\alpha_L\otimes\alpha_R$. In this paper, we will consider states that satisfy the following property (see figure \ref{fig:SetupWithQAutomorphism} for the support of the automorphisms):
	\begin{restatable}{assumption}{assumptionOne}\label{assumption}
		Let $\omega\in\PP(\AA)$ be
		\begin{enumerate}
			\item such that there exists an automorphism $\alpha\in\QAut{\AA}$ and a product state $\omega_0\in\PP(\AA)$ satisfying
			\begin{align}
				\omega=\omega_0\circ\alpha.
			\end{align}
			\item such that there exists a $\theta\in]0,\pi/2[$ for which there exists a map, $\tilde\beta:G\rightarrow\Aut{\AA}:g\mapsto\tilde\beta_g$, satisfying
			\begin{align}\label{eq:DecompositionOfBetaTilde}
				\omega\circ\tilde\beta_g&=\omega&\tilde\beta_g=\Ad{V_{g,2}}\circ\eta^L_g\otimes\eta^R_g\circ\beta^U_g
			\end{align}
			for some $V_{g,2}\in\UU(\AA),\eta^L_g\in\Aut{\AA_{\nu^{-1}(C_\theta\cap L)}}$ and $\eta^R_g\in\Aut{\AA_{\nu(C_\theta\cap R)}}$.
			\item translation invariant ($\omega\circ\tau=\omega$).
		\end{enumerate}
	\end{restatable}
	We will also require a similar subclass of automorphisms and states on the quasi-local $C^*$-algebra over $\ZZ$, $\AA_{L_0}$. We now define the assumption
	\begin{restatable}{assumption}{assumptionOneDimensionalOne}\label{assumption1d}
		Take $\phi\in\PP(\AA_{L_0})$ such that
		\begin{enumerate}
			\item $\phi\circ\beta_g=\phi$.
			\item there exist automorphisms $\tilde\alpha_L\in\Aut{\AA_{L_0\cap L}}$, $\tilde\alpha_R\in\Aut{\AA_{L_0\cap R}}$ and a summable operator $b\in\UU(\AA_{L_0})$ such that
			\begin{equation}
				\phi_0=\phi\circ\Ad{b}\circ\tilde\alpha_{L}\otimes\tilde\alpha_{R}
			\end{equation}
			is a product state.
		\end{enumerate}
	\end{restatable}
	We now state the result that using assumption \ref{assumption} we can define an $H^2(G,\TT)$-valued index.
	\begin{theorem}\label{thrm:ExistenceFirstIndex}
		For any $\AA$ satisfying the construction from section \ref{sec:QuasiLocalC*Algebra} and any choice of on-site group action $U\in\hom(G,U(\CC^d))$, there exists a map
		\begin{equation}
			\textrm{Index}^{\AA,U}:\{\omega\in\PP(\AA)|\omega\text{ satisfies assumption \ref{assumption}}\}\rightarrow H^2(G,\TT)
		\end{equation}
		that is well-defined (doesn't depend on any of the choices in assumption \ref{assumption}). This map will be consistent with conjecture \ref{conj} by which we mean that:
		\begin{enumerate}
			\item For any $G$ invariant and translation invariant family of interactions $\Phi\in\BB_{F_{\phi}}([0,1])$ (for some $0<\phi<1$) we have that $\textrm{Index}^{\AA,U}(\omega)=\textrm{Index}^{\AA,U}(\omega\circ\gamma^{\Phi}_{0;1})$.
			\item For $\phi\in\PP(\AA_{L_0})$, any one dimensional $G$-invariant state satisfying assumption \ref{assumption1d}, the state $\omega_\phi$ as defined in definition \ref{def:InfiniteTensorProduct} satisfies $\textrm{Index}^{\AA_{L_0},U}_{1d}(\phi)=\textrm{Index}^{\AA,U}(\omega)$.
			\item This index is a group homomorphism under the stacking operator. By this, we mean that
			\begin{equation}
				\textrm{Index}^{\AA^2,U\otimes U}(\omega_1\otimes_{\text{stack}}\omega_2)=\textrm{Index}^{\AA,U}(\omega_1)+\textrm{Index}^{\AA,U}(\omega_2).
			\end{equation}
			\item The index is independent on the choice of basis for the on-site Hilbert space by which we mean that for any on-site unitary $V$ we get that $\textrm{Index}^{\AA,\Ad{V^\dagger}(U)}(\omega\circ i_V)=\textrm{Index}^{\AA,U}(\omega)$.
		\end{enumerate}
	\end{theorem}
	\begin{proof}
		This statement is proven in section \ref{sec:ProofOfMainTheorems}.
	\end{proof}
	The assumption from which we can define the $H^1(G,\TT)$-valued index is the same one with an additional translation invariance:
	\begin{restatable}{assumption}{assumptionTwo}\label{assumption:2Translations}
		Let $\omega\in\PP(\AA)$ satisfy assumption \ref{assumption} with the additional property that it is also translation invariant in the other direction ($\omega\circ\nu=\omega$).
	\end{restatable}
	Similar to before, we also require an assumption for states on the quasi-local $C^*$-algebra over $\ZZ$.
	\begin{restatable}{assumption}{assumptionOneDimensionalTwo}\label{assumption1dWithTranslation}
		Let $\phi\in\PP(\AA_{L_0})$ satisfy assumption \ref{assumption1d} with the additional property that $\phi$ is translation invariant ($\phi\circ\nu=\phi$).
	\end{restatable}
	We are now ready to state the second main result:
	\begin{theorem}\label{thrm:ExistenceSecondIndex}
		For any $\AA$ satisfying the construction from section \ref{sec:QuasiLocalC*Algebra} and any choice of on-site group action $U\in\hom(G,U(\CC^d))$, there exists a map
		\begin{equation}
			\textrm{Index}^{\AA,U}_{2\text{ trans}}:\{\omega\in\PP(\AA)| \omega\text{ satisfies assumption \ref{assumption:2Translations}}\}\rightarrow H^1(G,\TT)
		\end{equation}
		that is well-defined (doesn't depend on any of the choices made in assumption \ref{assumption:2Translations}). This map will be consistent with conjecture \ref{conj2} by which we mean the following:
		\begin{enumerate}
			\item For any $G$ invariant family of interactions that is translation invariant in both directions $\Phi\in\BB_{F_\phi}([0,1])$ we have that $\textrm{Index}^{\AA,U}_{2\text{ trans}}(\omega)=\textrm{Index}^{\AA,U}_{2\text{ trans}}(\omega\circ\gamma^\Phi_{0;1})$.
			\item For $\phi\in\PP(\AA_{L_0})$, any one dimensional $G$-invariant and translation invariant state satisfying assumption \ref{assumption1d}, the state $\omega_\phi$ as defined in \ref{def:InfiniteTensorProduct}, satisfies $\textrm{Index}^{\AA_{L_0},U}_{1d\text{ trans}}(\phi)=\text{Index}^{\AA,U}_{2\text{ trans}}(\omega_\phi)$.
			\item This index is a group homomorphism under the stacking operator by which we mean that
			\begin{equation}
				\textrm{Index}_{\text{2 trans}}^{\AA^2,U\otimes U}(\omega_1\otimes_{\text{stack}}\omega_2)=\textrm{Index}_{\text{2 trans}}^{\AA,U}(\omega_1)+\textrm{Index}_{\text{2 trans}}^{\AA,U}(\omega_2).
			\end{equation}
			\item The index is independent of the choice of basis for the on-site Hilbert space. By this we mean that, for any on-site unitary $V$, we get that $\textrm{Index}^{\AA,\Ad{V^\dagger}(U)}_{\text{2 trans}}(\omega\circ i_V) =\textrm{Index}^{\AA,U}_{\text{2 trans}}(\omega)$.
		\end{enumerate}
	\end{theorem}
	\begin{proof}
		This statement is also proven in subsection \ref{sec:ProofOfMainTheorems}.
	\end{proof}
	\subsection{SRE implies previous assumptions}\label{sec:Results_1}
	The assumptions in the previous subsection are rather technical. However, it turns out that any SRE state, with some additional symmetries, naturally satisfies these assumptions. The precise statements and proofs are formulated in section \ref{sec:ProofOfMainResult1} but we will describe them in words here.
	\\\\
	As is shown in appendix \versionDifference{\ref{sec:properties-of-locally-generated-automorphisms-2d}}{C of \cite{jappens2023spt}}, which is based on the work in \cite{ogata2021h3gmathbb}, for any one parameter family of interactions $\Phi\in\BB_{F_\phi}([0,1])$, we have that $\gamma^\Phi_{0;1}\in\QAut{\AA}$. This implies the following. Take any SRE state $\omega$ (see definition \ref{def:sre}), with entangler $\gamma^\Phi_{0;1}$, and product state $\omega_0$ such that $\omega=\omega_0\circ \gamma^\Phi_{0;1}$. We can simply define the $\alpha$ from assumption \ref{assumption} to be equal to this $\gamma^\Phi_{0;1}$. Similarly, in \cite{ogata2021h3gmathbb} it has been shown that if this state has an additional on-site symmetry, we can use the $\Phi$ and $\omega_0$ to construct some $\tilde\beta_g$ that can be decomposed like in equation \eqref{eq:DecompositionOfBetaTilde} for any angle.
	\\\\
	It should also be said that we use the same definition of SRE (with the same $F$-function) for states over the quasi-local $C^*$-algebra over $\ZZ$, $\AA_{L_0}$. This is because we will use the 1D SRE states to create translation-invariant 2D SRE states. Since the $F$-norm of a sum of disjoint interactions has an $F$-norm given by the highest of the $F$-norms of its sub-interactions, this means that to ensure that the resulting state has a well-defined index, we will need a similar $F$-norm and not something weaker.
	\\\\
	Additionally, it turns out that the indices that are defined in the last subsection are independent of the choice of representative of the equivalence classes introduced in \ref{sec:StableEquivalenceClasses}.
	\\\\
	All in all, we will show the following theorem:
	\begin{theorem}\label{lem:IndicesConstantOnStableEquivalenceClasses}
		Take $(\omega,U)\in S_0$ (see subsection \ref{sec:StableEquivalenceClasses}) for the algebra $\AA$. The following statements now hold:
		\begin{enumerate}
			\item $\textrm{Index}^{\AA,U}_{\text{no trans}}(\omega)$ is well defined and independent on the choice of representative of $\expval{\omega}_\sim$.
			\item If additionally, $(\omega,U)\in S_1$, $\textrm{Index}^{\AA,U}(\omega)$ is also well defined and independent on the choice of representative of $\expval{\omega}_\sim^1$.
			\item If instead, $(\omega,U)\in S_2$, $\textrm{Index}^{\AA,U}(\omega\circ\mu)$ is well defined and independent on the choice of representative of $\expval{\omega}_\sim^2$.
			\item If additionally, $(\omega,U)\in S_3$, $\textrm{Index}^{\AA,U}_{\textrm{2 trans}}(\omega)=\textrm{Index}^{\AA,U}_{\textrm{2 trans}}(\omega\circ\mu)$ is well defined and independent on the choice of representative of $\expval{\omega}_\sim^3$.
		\end{enumerate}
	\end{theorem}
	\begin{proof}
		This is shown in section \ref{sec:ProofOf:lem:IndicesConstantOnStableEquivalenceClasses}.
	\end{proof}
	\subsection{Comparison with \texorpdfstring{\cite{ogata2021h3gmathbb}}{}}\label{sec:SomeRemarks}
	The first thing we would like to remark on is that in \cite{ogata2021h3gmathbb}, a different equivalence class on states is used. First of all the set of states considered is a priori different. Namely, for any state $\psi\in \PP(\AA)$, we define two different conditions
	\begin{enumerate}
		\item There exists a $\Phi\in\PP_{\text{SL}}$\footnote{This is defined in \cite{ogata2021h3gmathbb} but it essentially means bounded interaction that is connected to a trivial (without overlapping terms) interaction.} such that $\psi$ is the unique gapped groundstate of $\Phi$.
		\item $\psi$ is an SRE state.
	\end{enumerate}
	However, in theorem 5.1 of \cite{ogata2021h3gmathbb} it is proven that the first item implies the second item. If one then looks at the definition of the index there, one could as well have started from the second condition. This is precisely what we did in this paper. As a side note, one should also be able to prove that the second item implies the first item by using theorem D.5 of \cite{ogata2021h3gmathbb} with $\mathcal{K}_t$ the disentangler and $\Phi$ a gapped interaction that has the product state as its ground state. However, we won't work this out explicitly in this paper.\\\\
	When we add the group, the story changes considerably. Let us define the two different equivalence classes. To this end, let $\psi_1$ and $\psi_2$ be two (pure) $G$-invariant states that are unique gapped groundstates of $\Phi_0$ and $\Phi_1\in\PP_{\text{SL},\beta}$ respectively. The following two conditions define different equivalence classes:
	\begin{enumerate}
		\item There exists a $G$-invariant, bounded path of gapped interactions $\lambda\mapsto\tilde\Phi(\lambda)$ (see \cite{ogata2021h3gmathbb} for precise definitions) such that $\psi_1$ and $\psi_2$ are unique gapped groundstates of $\tilde{\Phi}(0)$ and $\tilde{\Phi}(1)$ respectively.
		\item There exists a $G$-invariant LGA, $\gamma^\Psi_{0;1}$ such that $\psi_1\circ\gamma^H_{0;1}=\psi_2$.
	\end{enumerate}
	A priori those two conditions are not equivalent. However, in \cite{ogata2021h3gmathbb} it is proven that the first equivalence implies the second one. The proof that equivalent states have the same index only uses the second condition. This means that one could also have taken the second equivalence condition as the starting point. This is exactly what we do in this paper. One should also be able to show that the second condition implies the first one using a similar construction as in the above remark without the group and averaging the resulting interaction over the group (where one then has to use that a finite group has a normalized Haar measure).\\\\
	There were two things that are in theorems \ref{thrm:ExistenceFirstIndex} and \ref{thrm:ExistenceSecondIndex} but were not discussed in \cite{ogata2021h3gmathbb}. This is the part that proves that the index is a group homomorphism under stacking and the part that the index is independent of the choice of basis for the on-site Hilbert space. However, on inspection of the arguments used in this paper one can conclude that these also hold for the $H^3$-valued index.\\\\
	All in all, we have provided a proof that the work done in \cite{ogata2021h3gmathbb} also implies the following theorem:
	\begin{theorem}\label{thrm:ExistenceOriginalIndex}
		For any $\AA$ satisfying the construction from section \ref{sec:QuasiLocalC*Algebra} and any choice of on-site group action $U\in\hom(G,U(\CC^d))$, there exists a map
		\begin{equation}
			\textrm{Index}^{\AA,U}_{\text{no trans}}:\{\omega\in\PP(\AA)| \omega\text{ is SRE and }G\text{-invariant}\}\rightarrow H^3(G,\TT)
		\end{equation}
		that is well-defined (doesn't depend on the choice of disentangler). This map will satisfy:
		\begin{enumerate}
			\item For any $G$ invariant family of interactions $\Phi\in\BB_{F_\phi}([0,1])$ we have that
			\begin{equation}
				\textrm{Index}^{\AA,U}_{\text{no trans}}(\omega)=\textrm{Index}^{\AA,U}_{\text{no trans}}(\omega\circ\gamma^\Phi_{0;1}).
			\end{equation}
			\item This index is a group homomorphism under the stacking operator by which we mean that
			\begin{equation}
				\textrm{Index}_{\text{no trans}}^{\AA^2,U\otimes U}(\omega_1\otimes_{\text{stack}}\omega_2)=\textrm{Index}_{\text{no trans}}^{\AA,U}(\omega_1)+\textrm{Index}_{\text{no trans}}^{\AA,U}(\omega_2).
			\end{equation}
			\item The index is independent of the choice of basis for the on-site Hilbert space by which we mean that for any on-site unitary $V$ we get that
			\begin{equation}
				\textrm{Index}^{\AA,\Ad{V^\dagger}(U)}_{\text{no trans}}(\omega\circ i_V) =\textrm{Index}^{\AA,U}_{\text{no trans}}(\omega).
			\end{equation}
		\end{enumerate}
	\end{theorem}
	\section{Translation invariance in one dimension}\label{sec:OneDimensionalIndices}
	Let $\AA_{L_0}$ be the quasi-local $C^*$ algebra defined on the one-dimensional lattice. Let $U_i\in\hom(G,\UU(\AA_i))$ (for $i\in\ZZ$) be the on-site group action. Let $\beta_g$ be again so that for each finite $I\subset \ZZ$, $\beta_g^I=\Ad{\bigotimes_{i\in I}U_i(g)}$. In this section, we will look at states satisfying the split property. We say that $\omega\in\PP(\AA_{L_0})$ satisfies the split property iff. it is quasi equivalent to $\omega|_{L}\otimes\omega|_{R}$. This implies that there exists a GNS triple of $\omega$, $(\HH,\pi,\Omega)$ that is of the form $\HH=\HH_L\otimes\HH_R$, $\pi=\pi_L\otimes\pi_R$. We will denote the space of pure states satisfying the split property by $\mathcal{S}\PP(\AA)$. Now let us additionally ask that $\omega$ is invariant under the group action ($\omega\circ\beta_g=\omega$). The following now holds:
	\begin{lemma}
		There exist maps $u_\sigma:G\rightarrow \UU(\HH_\sigma)$ (for all $\sigma\in\{L,R\}$) satisfying
		\begin{align}
			\Ad{u_\sigma(g)}\circ\pi_\sigma&=\pi_\sigma\circ\beta_g^\sigma&u_L(g)\otimes u_R(g)\Omega&=\Omega.
		\end{align}
		They are unique up to a $G$-dependent phase. Furthermore, there exists a 2-cochain $C:G^2\rightarrow U(1)$ that satisfies
		\begin{align}
			u_R(g)u_R(h)u_R(gh)^{\dagger}&=C(g,h)\id_{H_R}&u_L(g)u_L(h)u_L(gh)^{\dagger}&=C(g,h)^{-1}\id_{H_L}.
		\end{align}
		Its second group cohomology class is independent of the choice of GNS triple or the choice of (a $G$-dependent) phase in $u_R(g)$. This index is invariant under locally generated automorphisms generated by $G$-invariant interactions.
	\end{lemma}
	\begin{proof}
		See \cite{ogata2019classification}.
	\end{proof}
	The cohomology class of this 2-cochain is called the one-dimensional SPT index. Now let us assume additionally (on top of the split property and the $G$-invariance) that $\omega$ be translation invariant. To this end, let $\nu$ be the automorphism that translates $\BB$ by one site to the right. We assume that $\omega\circ\nu=\omega$. The following now holds:
	\begin{lemma}
		There exists a unique $w\in\UU(\HH)$ such that
		\begin{align}
			\Ad{w}\circ\pi&=\pi\circ\nu&w\Omega&=\Omega.
		\end{align}
	\end{lemma}
	\begin{proof}
		Since $\omega\circ\nu=\omega$, we get that $(\HH,\pi\circ\nu,\Omega)\ue (\HH,\pi,\Omega)$. The uniqueness of the GNS triple now implies the existence of this $w$. Its uniqueness follows from the irreducibility of $\pi$.
	\end{proof}
	We can use this to define an additional 1d index for translation invariant states. To do this, we will need to use the finiteness of the group explicitly. We will use one particular property of this, namely that finite groups have discrete U(1) representations. By this we mean the following:
	\begin{lemma}\label{lem:FiniteGroupsHaveDiscreteU(1)Representations}
		Let $G$ be a finite group. Let $\alpha:[0,1]\rightarrow\hom(G,U(1)):\lambda\mapsto \alpha_\lambda$ be a map such that $\lambda\mapsto\alpha_\lambda(g)$ is continuous for all $g\in G$. If $\alpha_0(g)=1$ for all $g\in G$ then $\alpha_\lambda(g)=1$ for all $g\in G$ and all $\lambda\in[0,1]$.
	\end{lemma}
	\begin{proof}
		 Take $\alpha$ as above with $\alpha_0(g)=1$ for all $g\in G$, we will prove that $\alpha_\lambda(g)=1$ for all $g\in G$ and all $\lambda\in[0,1]$. To do this, let $n$ be the order of $G$. It follows from Lagrange's theorem that $g^n=\id_G$ for all $g\in G$. This implies that there exists a unique ($\lambda$ and $g$ dependent) $k\in\NN\cap [0,n-1]$ such that $\alpha_\lambda(g)=\exp(2\pi ik/n)$. Since for any fixed $g\in G$,  $\alpha_\lambda(g)$ has to be continuously connected to $1$, we deduce that $k=0$ and hence the result follows.
	\end{proof}
	\begin{lemma}
		There exists a (unique) $\alpha\in\hom(G,U(1))$ satisfying that
		\begin{equation}
			\pi(U_{-1}(g)) (\id_{\HH_L}\otimes u_R(g))=\alpha(g)w^\dagger(\id_{\HH_L}\otimes u_R(g)) w.
		\end{equation}
		This index is invariant under locally generated automorphisms generated by one-parameter families of interactions in $\BB_{F_\phi}([0,1])$ (with $F_\phi$ as presented in \eqref{eq:OurFFunction}) that are both $G$ and translation invariant.
	\end{lemma}
	\begin{proof}
		The fact that $\alpha$ exists follows from the fact that $\pi$ is irreducible and combined with the fact that
		\begin{equation}
			\Ad{\pi(U_{-1}(g)) (\id_{\HH_L}\otimes u_R(g))}\circ\pi=\Ad{w^\dagger(\id_{\HH_L}\otimes u_R(g)) w}\circ\pi=\pi\circ\beta_g^{[-1,\infty[}.
		\end{equation}
		The fact that $\alpha$ is a $U(1)-$representation follows from the fact that $u_R$ is the lift of a projective representation. Suppose that $u_R(g)u_R(h)=C(g,h)u_R(gh)$ then
		\begin{equation}
			\pi(U_{-1}(g)) (\id_{\HH_L}\otimes u_R(g))\pi(U_{-1}(h)) (\id_{\HH_L}\otimes u_R(h))=C(g,h)\pi(U_{-1}(gh)) (\id_{\HH_L}\otimes u_R(gh))
		\end{equation}
		on the one hand whereas
		\begin{equation}
			w^\dagger(\id_{\HH_L}\otimes u_R(g)) ww^\dagger(\id_{\HH_L}\otimes u_R(h)) w=C(g,h)w^\dagger(\id_{\HH_L}\otimes u_R(gh)) w.
		\end{equation}
		These two equations can only be consistent if $\alpha(g)\alpha(h)=\alpha(gh)$. The proof that this is independent of the choice of GNS triple is straightforward. Now we need to show that this index is invariant under LGA's generated by $G$-invariant interactions. To this end let $F_\phi$ be our F-function from equation \eqref{eq:OurFFunction}. Let $\Phi\in\BB_{F_\phi}([0,1])$ be a $G$-invariant, translation invariant interaction. For simplicity of notation let $\Phi_{\text{split}}=\Phi_{\nu^{-1}(L)}+\Phi_{\nu(R)}$. Let $\lambda\in[0,1]\mapsto A_\lambda\in\UU(\AA)$ be a norm continuous one-parameter family such that
		\begin{align}\label{eq:DefinitionSplitting1DAppendix}
			\gamma^\Phi_{0;\lambda}&=\Ad{A_\lambda}\circ\gamma^{\Phi_{\text{split}}}_{0;\lambda}&A_0&=\id_\AA.
		\end{align}
		and define $A\defeq A_1$. This exists because of lemma \versionDifference{\ref{lem:PropertiesLocallyGeneratedAutomorphisms1d}}{B.1. of \cite{jappens2023spt}}. By construction $(\HH,\pi\circ\gamma^{\Phi_{\text{split}}}_{0;1},\pi(A^\dagger)\Omega)$ is a GNS triple of $\omega\circ\gamma^\Phi_{0;1}$. Let $\tilde{w}\defeq \pi(A^\dagger)w\pi(A)$ then $\tilde{\alpha}(g)\defeq \pi(U_{-1}(g))u_R(g)\tilde{w}^\dagger u_R(g)^\dagger \tilde{w}$ is the index of $\omega\circ\gamma^\Phi_{0;1}$. Putting the $A$'s to the front and inserting the old index gives
		\begin{equation}
			\tilde{\alpha}(g)=\alpha(g)w^\dagger \pi\circ\beta_g^R(\nu(A^\dagger)A)\pi(A^\dagger\nu(A))w.
		\end{equation}
		This shows that we conclude the proof if we can show that $\beta_g^R(\nu(A^\dagger)A)=\nu(A^\dagger)A$. We will do this proof in two steps. The first step is that we prove that $\Ad{\beta_g^R(\nu(A^\dagger)A)}=\Ad{\nu(A^\dagger)A}$. We use the first equation of \eqref{eq:DefinitionSplitting1DAppendix} and obtain
		\begin{align}
			\Ad{\beta_g^R(\nu(A^\dagger)A)}=\stkout{\beta_g^R}\circ\gamma^{\nu(\Phi_{\text{split}})}_{0;1}\circ\stkout{\nu\circ\nu^{-1}}\circ\gamma^{\Phi_{\text{split}}}_{1;0}\circ\stkout{(\beta_g^R)^{-1}}=\Ad{\nu(A^\dagger)A}.
		\end{align}
		This shows that there exists some $\alpha'\in\hom(G,U(1))$ such that $\alpha'(g)=\beta_g^R(\nu(A^\dagger)A)(A^\dagger \nu(A))$. However since this has to hold for each $\lambda\in[0,1]$, we can define for each $\lambda$ some $\alpha_\lambda\in\hom(G,U(1))$ through $\alpha_\lambda(g)=\beta_g^R(\nu(A^\dagger_\lambda)A_\lambda)(A^\dagger_\lambda \nu(A_\lambda))$. This implies that $\alpha'(g)=\alpha_1(g)$ has to be connected to $\alpha_0(g)=1$. Now we will use that the $U(1)$ representations of a finite group are discrete. More specifically we will use lemma \ref{lem:FiniteGroupsHaveDiscreteU(1)Representations}. This implies that $\alpha'(g)=1$ for all $g\in G$.
	\end{proof}
	All together we constructed a proof of the following theorem:
	\begin{theorem}
		There exist maps
		\begin{align}
			\textrm{Index}_{1d}&:\{\omega\in\mathcal{S}\PP(\AA)|\omega\circ\beta_g=\omega\}\rightarrow H^2(G,\TT)\\
			\nonumber
			\textrm{Index}_{1d\text{ trans}}&:\{\omega\in\mathcal{S}\PP(\AA)|\omega\circ\beta_g=\omega\text{ and }\omega\circ\nu=\omega\}\rightarrow H^1(G,\TT)\cong \hom(G,\TT).
		\end{align}
		For any $G-$invariant family of interactions $\Phi\in\BB_{F}([0,1])$, we have that $\textrm{Index}_{1d}(\omega)=\textrm{Index}_{1d}(\omega\circ\gamma^\Phi_{0;1})$. For any $G-$invariant family of interactions $\Phi'\in\BB_{F}([0,1])$ that is also translation invariant, we have that $\textrm{Index}_{1d\text{ trans}}(\omega)=\textrm{Index}_{1d\text{ trans}}(\omega\circ\gamma^{\Phi'}_{0;1})$.
	\end{theorem}
	\section{Cone and translation operators}\label{sec:ConeAndTranslationOperators}
	\begin{figure}
		\centering
		\def\s{0.4}
		\resizebox{0.21\textwidth}{!}{%
			\begin{tikzpicture}
\draw[draw=white,line width=0mm] (1*\s,1*\s) coordinate (a) -- (0,1*\s) coordinate (b) -- (1*\s,2*\s) coordinate (c);

\fill[fill=green!30!white] (0,1*\s) -- (-4*\s,5*\s) -- (-5*\s,5*\s) -- (-5*\s,-5*\s) -- (-4*\s,-5*\s) -- (0,-1*\s);
\fill[fill=green!30!white] (0,1*\s) -- (4*\s,5*\s) -- (5*\s,5*\s) -- (5*\s,-5*\s) -- (4*\s,-5*\s) -- (0,-1*\s);
\fill[fill=red!30!white] (0,1*\s) -- (4*\s,5*\s) -- (-4*\s,5*\s);
\fill[fill=red!30!white] (0,-1*\s) -- (4*\s,-5*\s) -- (-4*\s,-5*\s);

\draw[draw=black,line width=0.3mm]
plot[smooth,samples=2,domain=0:4*\s] (\x,\x+1*\s);
\draw[draw=black,line width=0.5mm] (0,-\s) -- (0,\s);
\draw[draw=black,line width=0.3mm]
plot[smooth,samples=2,domain=-4*\s:0] (\x,-\x+1*\s);
\draw[draw=black,line width=0.3mm]
plot[smooth,samples=2,domain=0:4*\s] (\x,-\x-1*\s);
\draw[draw=black,line width=0.3mm]
plot[smooth,samples=2,domain=-4*\s:0] (\x,\x-1*\s);

\draw[draw=black,line width=0.3mm] (0,1*\s) -- (2*\s,1*\s);
\pic [draw, ->, "$\theta$", angle eccentricity=1.5, angle radius=\s*1.5cm,line width=0.3mm] {angle=a--b--c};

\node at (-3*\s,0) {$\xi_L$};
\node at (3.5*\s,0) {$\xi_R$};
\node at (0,3*\s) {$\Theta$};
\end{tikzpicture}}
		$\quad$
		\def\s{0.4}
		\resizebox{0.21\textwidth}{!}{%
			\begin{tikzpicture}
\draw[draw=white,line width=0mm] (1,0) coordinate (a) -- (0,0) coordinate (b) -- (1,1) coordinate (c);

\fill[fill=green!30!white] (0,0) -- (-5*\s,-5*\s) -- (-5*\s,5*\s);
\fill[fill=green!30!white] (0,0) -- (5*\s,-5*\s) -- (5*\s,5*\s);
\fill[fill=red!30!white] (0,1*\s) -- (4*\s,5*\s) -- (-4*\s,5*\s);
\fill[fill=red!30!white] (0,-1*\s) -- (4*\s,-5*\s) -- (-4*\s,-5*\s);

\draw[draw=black,line width=0.3mm]
plot[smooth,samples=2,domain=0:4*\s] (\x,\x+1*\s);
\draw[draw=black,line width=0.3mm]
plot[smooth,samples=2,domain=-4*\s:0] (\x,-\x+1*\s);
\draw[draw=black,line width=0.3mm]
plot[smooth,samples=2,domain=0:4*\s] (\x,-\x-1*\s);
\draw[draw=black,line width=0.3mm]
plot[smooth,samples=2,domain=-4*\s:0] (\x,\x-1*\s);
\draw[draw=black,line width=0.3mm]
plot[smooth,samples=2,domain=-5*\s:5*\s] (\x,\x);
\draw[draw=black,line width=0.3mm]
plot[smooth,samples=2,domain=-5*\s:5*\s] (\x,-\x);

\draw[draw=black,line width=0.3mm] (0,0) -- (2*\s,0);
\pic [draw, ->, "$\theta$", angle eccentricity=1.5, angle radius=\s*1.5cm,line width=0.3mm] {angle=a--b--c};

\node at (-3*\s,0) {$\xi_L$};
\node at (3.5*\s,0) {$\xi_R$};
\node at (0,3*\s) {$\Theta$};
\end{tikzpicture}}
		$\quad$
		\def\s{0.4}
		\resizebox{0.21\textwidth}{!}{%
			\begin{tikzpicture}
\draw[draw=white,line width=0mm] (2*\s,1*\s) coordinate (a) -- (1*\s,1*\s) coordinate (b) -- (2*\s,2*\s) coordinate (c);

\fill[fill=green!30!white] (-1*\s,1*\s) -- (-5*\s,5*\s) -- (-5*\s,-5*\s) -- (-1*\s,-1*\s);
\fill[fill=green!30!white] (1*\s,1*\s) -- (5*\s,5*\s) -- (5*\s,-5*\s) -- (1*\s,-1*\s);
\fill[fill=red!30!white] (0,1*\s) -- (4*\s,5*\s) -- (-4*\s,5*\s);
\fill[fill=red!30!white] (0,-1*\s) -- (4*\s,-5*\s) -- (-4*\s,-5*\s);

\draw[draw=black,line width=0.3mm]
plot[smooth,samples=2,domain=0:4*\s] (\x,\x+1*\s);
\draw[draw=black,line width=0.5mm] (-\s,-\s) -- (-\s,\s);
\draw[draw=black,line width=0.5mm] (\s,-\s) -- (\s,\s);
\draw[draw=black,line width=0.3mm]
plot[smooth,samples=2,domain=-4*\s:0] (\x,-\x+1*\s);
\draw[draw=black,line width=0.3mm]
plot[smooth,samples=2,domain=0:4*\s] (\x,-\x-1*\s);
\draw[draw=black,line width=0.3mm]
plot[smooth,samples=2,domain=-4*\s:0] (\x,\x-1*\s);
\draw[draw=black,line width=0.5mm] (\s,\s) -- (5*\s,5*\s);
\draw[draw=black,line width=0.5mm] (\s,-\s) -- (5*\s,-5*\s);
\draw[draw=black,line width=0.5mm] (-\s,\s) -- (-5*\s,5*\s);
\draw[draw=black,line width=0.5mm] (-\s,-\s) -- (-5*\s,-5*\s);

\draw[draw=black,line width=0.3mm] (1*\s,1*\s) -- (3*\s,1*\s);
\pic [draw, ->, "$\theta$", angle eccentricity=1.5, angle radius=\s*1.5cm,line width=0.3mm] {angle=a--b--c};

\node at (-3*\s,0) {$\xi_L$};
\node at (3.5*\s,0) {$\xi_R$};
\node at (0,3*\s) {$\Theta$};
\end{tikzpicture}}
		$\quad$
		\def\s{0.4}
		\resizebox{0.24\textwidth}{!}{%
			\begin{tikzpicture}
	\draw[draw=white,line width=0mm] (1+\s,0) coordinate (a) -- (\s,0) coordinate (b) -- (1+\s,1) coordinate (c);
	
	\fill[fill=green!30!white] (-1*\s,0) -- (-6*\s,-5*\s) -- (-6*\s,5*\s);
	\fill[fill=green!30!white] (1*\s,0) -- (6*\s,-5*\s) -- (6*\s,5*\s);
	\fill[fill=red!30!white] (0,1*\s) -- (4*\s,5*\s) -- (-4*\s,5*\s);
	\fill[fill=red!30!white] (0,-1*\s) -- (4*\s,-5*\s) -- (-4*\s,-5*\s);
		
	\draw[draw=black,line width=0.3mm]
	plot[smooth,samples=2,domain=0:4*\s] (\x,\x+1*\s);
	\draw[draw=black,line width=0.3mm]
	plot[smooth,samples=2,domain=-4*\s:0] (\x,-\x+1*\s);
	\draw[draw=black,line width=0.3mm]
	plot[smooth,samples=2,domain=0:4*\s] (\x,-\x-1*\s);
	\draw[draw=black,line width=0.3mm]
	plot[smooth,samples=2,domain=-4*\s:0] (\x,\x-1*\s);
	\draw[draw=black,line width=0.3mm]
	plot[smooth,samples=2,domain=0:5*\s] (\x+1*\s,\x);
	\draw[draw=black,line width=0.3mm]
	plot[smooth,samples=2,domain=0:5*\s] (\x+1*\s,-\x);
	\draw[draw=black,line width=0.3mm]
	plot[smooth,samples=2,domain=-5*\s:0] (\x-1*\s,-\x);
	\draw[draw=black,line width=0.3mm]
	plot[smooth,samples=2,domain=-5*\s:0] (\x-1*\s,\x);
		
	\draw[draw=black,line width=0.3mm] (1*\s,0) -- (3*\s,0);
	\pic [draw, ->, "$\theta$", angle eccentricity=1.5, angle radius=\s*1.5cm,line width=0.3mm] {angle=a--b--c};
		
	\node at (-4*\s,0) {$\xi_L$};
	\node at (4.5*\s,0) {$\xi_R$};
	\node at (0,3*\s) {$\Theta$};
\end{tikzpicture}}
		\caption{This is a graphical depiction of the support of the automorphisms $\xi_L$ and $\xi_R$ with respect to $\Theta$ in the definitions of $\textrm{Cone}_{\sigma}(\alpha_0,\theta)$, $\textrm{Cone}_{\sigma}^W(\alpha_0,\theta)$, $\textrm{Cone}_{\nu^\sigma(\sigma)}(\alpha_0,\theta)$ and $\textrm{Cone}_{\nu^\sigma(\sigma)}^W(\alpha_0,\theta)$ respectively.}
		\label{fig:ConeOperators}
	\end{figure}
	In this section, we will fix certain objects that we will later on use to define an $H^2(G,\TT)-$valued index for states satisfying assumption \ref{assumption} and an $H^1(G,\TT)-$valued index for states satisfying assumption \ref{assumption:2Translations}. To this end, let $\omega\in\PP(\AA)$ satisfy assumption \ref{assumption}. Fix a product state $\omega_0$ and an $\alpha\in\QAut{\AA}$ such that $\omega=\omega_0\circ\alpha$. Fix $(\HH_0=\HH_L\otimes\HH_R,\pi_0=\pi_L\otimes\pi_R,\Omega_0)$, a GNS triple (see theorem 2.3.16 in \cite{bratteli1979operator}) of $\omega_0$ where ($\forall\sigma\in\{L,R\}$) $\pi_\sigma:\AA_\sigma\rightarrow\BB(\HH_\sigma)$ is a GNS representation of the restriction of $\omega_0$ to $\AA_\sigma$. We also fix, $V_1$, $\alpha_0=\alpha_L\otimes\alpha_R$ and $\Theta$ as a choice of decomposition for $\alpha$ such that $\Theta$ and $\eta_g$ commute (see figure \ref{fig:SetupWithQAutomorphism}). Because we will be working with states that are translation invariant, it makes sense to define the translation action on the GNS space of $\omega_0$:
	\begin{lemma}\label{lem:Definition_v_And_w}
		There exists a unique $v\in\UU(\HH_0)$ such that
		\begin{align}
			\Ad{v}\circ\pi_0&=\pi_0\circ\alpha_0\circ\Theta\circ\tau\circ\Theta^{-1}\circ\alpha_0^{-1}&&\text{and}&\pi_0(V_1)v\pi_0(V_1^\dagger)\Omega_0&=\Omega_0.
		\end{align}
		If additionally, $\omega$ satisfies assumption \ref{assumption:2Translations}, there exists a unique $w\in\UU(\HH_0)$ such that
		\begin{align}
			\Ad{w}\circ\pi_0&=\pi_0\circ\alpha_0\circ\Theta\circ\nu\circ\Theta^{-1}\circ\alpha_0^{-1}&&\text{and}&\pi_0(V_1)w\pi_0(V_1^\dagger)\Omega_0&=\Omega_0.
		\end{align}
	\end{lemma}
	\begin{proof}
		We only do the first part as the second part is analogous. Since $\omega\circ\tau=\omega$ we get by the uniqueness of the GNS triple (see Corollary 2.3.17 and Theorem 2.3.19 from \cite{bratteli1979operator}) that there exists a unique $\tilde{v}\in\UU(\HH_0)$ satisfying
		\begin{align}
			\Ad{\tilde{v}}\circ\pi_0\circ\alpha&=\pi_0\circ\alpha\circ\tau&\tilde{v}\Omega_0&=\Omega_0.
		\end{align}
		Since $\alpha=\Ad{V_1}\circ\alpha_0\circ\Theta$ we get that
		\begin{align}
			\Ad{\pi_0(V_1^\dagger)\tilde{v}\pi_0(V_1)}\circ\pi_0\circ\alpha_0\circ\Theta&=\pi_0\circ\alpha_0\circ\Theta\circ\tau&\tilde{v}\Omega_0&=\Omega_0.
		\end{align}
		Choosing $v=\pi_0(V_1^\dagger)\tilde{v}\pi_0(V_1)$  concludes the proof.
	\end{proof}
	We now define a subgroup of $\UU(\HH_0)$ that includes the representations of cone automorphisms (see figure \ref{fig:ConeOperators} for a graphical depiction of the supports of $\xi$):
	\begin{definition}\label{def:ConeOperators}
		Take $0<\theta<\pi/2$, $\alpha_L\in\Aut{\AA_{L}}$, $\alpha_R\in\Aut{\AA_{R}}$ and $\alpha_0\defeq \alpha_L\otimes\alpha_R$. Take $\sigma\in\{L,R\}$ and $x\in\UU(\HH_{\sigma})\otimes\id_{\HH_{\ZZ^2/\sigma}}$ then we say that $x$ is a cone operator on $\sigma$ (or in short $x\in\textrm{Cone}_{\sigma}(\alpha_0,\theta)$) if and only if there exists a $\xi\in\Aut{\AA_{W(C_\theta)\cap\sigma}}$ such that
		\begin{equation}
			\Ad{x}\circ\pi_0\circ\alpha_0=\pi_0\circ\alpha_0\circ\xi.
		\end{equation}
		$\textrm{Cone}_\sigma(\alpha_0,\theta)$ is a subgroup of $\UU(\HH_\sigma)\otimes\id_{\ZZ^2/\sigma}$. We have that
		\begin{align}\label{eq:CommutantPropertyCones}
			\textrm{Cone}_R(\alpha_0,\theta)&\subseteq\textrm{Cone}_L(\alpha_0,\theta)'&\textrm{Cone}_L(\alpha_0,\theta)&\subseteq\textrm{Cone}_R(\alpha_0,\theta)'.
		\end{align}
		If additionally $\xi\in\Aut{\AA_{C_\theta\cap\sigma}}$, $\xi\in\Aut{\AA_{\nu^\sigma(W(C_\theta)\cap\sigma)}}$ or $\xi\in\Aut{\AA_{\nu^\sigma(C_\theta\cap\sigma)}}$ respectively, we will say that $x\in\textrm{Cone}_\sigma^W(\alpha_0,\theta)$, $x\in\textrm{Cone}_{\nu^\sigma(\sigma)}(\alpha_0,\theta)$ or $x\in\textrm{Cone}_{\nu^\sigma(\sigma)}^W(\alpha_0,\theta)$ respectively.
	\end{definition}
	\begin{proof}
		The proof of equation \eqref{eq:CommutantPropertyCones} just follows from the fact that we defined $\textrm{Cone}_\sigma(\alpha_0,\theta)$ such that
		\begin{align}
			\textrm{Cone}_L(\alpha_0,\theta)&\subseteq \UU(\HH_L)\otimes\id_{\HH_R}&\textrm{Cone}_R(\alpha_0,\theta)&\subseteq \id_{\HH_L} \otimes \UU(\HH_R).
		\end{align}
	\end{proof}
	We will also define a generalization of this. We will define a subgroup of $\UU(\HH_0)$ that includes both the group $\pi_0\circ\alpha_0\circ\Theta(\UU(\AA_R))$ and the representations of cone automorphisms:
	\begin{definition}
		Take $0<\theta<\pi/2$, $\alpha_L\in\Aut{\AA_{L}}$, $\alpha_R\in\Aut{\AA_{R}}$ and $\Theta\in\Aut{\AA_{W(C_\theta)^c}}$. Take $\sigma\in\{L,R\}$ and take $x\in\UU(\HH_0)$ then we say that $x$ is an inner after cone operator on $\sigma$ (or in short $x\in \IAC_\sigma(\alpha_0,\theta,\Theta)$) if there exists an $A\in\UU(\AA_\sigma)$ and a $y\in\textrm{Cone}_\sigma(\alpha_0,\theta)$ such that
		\begin{equation}
			x=\pi_0\circ\alpha_0\circ\Theta(A)y.
		\end{equation}
		$\IAC_\sigma$ is a subgroup of $\UU(\HH_0)$. We get that
		\begin{align}\label{eq:CommutantProperty}
			\IAC_R(\alpha_0,\theta,\Theta)&\subseteq\IAC_L(\alpha_0,\theta,\Theta)'&\IAC_L(\alpha_0,\theta,\Theta)&\subseteq\IAC_R(\alpha_0,\theta,\Theta)'.
		\end{align}
		If, additionally, $y\in \textrm{Cone}_\sigma^W(\alpha_0,\theta)$ we will say that $x\in \IAC_\sigma^W(\alpha_0,\theta,\Theta)$. If $A\in\UU(\AA_{\nu^\sigma(\sigma)})$ and $y\in \textrm{Cone}_{\nu^\sigma(\sigma)}(\alpha_0,\theta)$, we say that $x\in\IAC_{\nu^\sigma(\sigma)}(\alpha_0,\theta,\Theta)$. If $A\in\UU(\AA_{\nu^\sigma(\sigma)})$ and $y\in \textrm{Cone}_{\nu^\sigma(\sigma)}^W(\alpha_0,\theta)$, we say that $x\in\IAC_{\nu^\sigma(\sigma)}^W(\alpha_0,\theta,\Theta)$.
	\end{definition}
	\begin{proof}
		Take $u_L\in\IAC_L(\alpha_0,\theta,\Theta)$ and $u_R\in\IAC_R(\alpha_0,\theta,\Theta)$ arbitrary. Take $A_\sigma\in\UU(\AA_\sigma)$ and $v_\sigma\in\textrm{Cone}_\sigma(\alpha_0,\theta)$ (for $\sigma\in\{L,R\}$) such that $u_\sigma=\pi_0\circ\alpha_0\circ\Theta(A_\sigma)v_\sigma$. We have to prove that $[u_L,u_R]=0$. To do this, we need three things. First of all, by construction, we get that $[A_L,A_R]=0$. Secondly, we showed in \ref{eq:CommutantPropertyCones} that $[v_L,v_R]=0$. We now only have to show that $[\pi_0\circ\alpha_0\circ\Theta(A_L),v_R]$ (and vice versa) as using these three properties repeatedly on the expression $[u_L,u_R]$ shows that it vanishes. This property is clearly equivalent to showing that $\Ad{v_R}\circ \pi_0\circ\alpha_0\circ\Theta(A_L)=\pi_0\circ\alpha_0\circ\Theta(A_L)$. Using definition \ref{def:ConeOperators} and the fact that $\xi_R$ and $\Theta$ commute, this reduces to showing that $\xi_R(A_L)=A_L$. This is true by construction.
	\end{proof}
	\begin{remark}
		The main reason for the introduction of the $\textrm{Cone}_\sigma^W(\alpha_0,\theta)$ and $\IAC^W_\sigma(\alpha_0,\theta,\Theta)$ subgroups is that they are still (inner after) cone operators when translated. More specifically, let $y\in\textrm{Cone}_\sigma^W(\alpha_0,\theta)$ then both $vyv^\dagger$ and $v^\dagger y v$ are elements of $\textrm{Cone}_\sigma(\alpha_0,\theta)$. The analogous statement is also true for the $\IAC$ operators. Similarly, in the case where there are two translation symmetries, we get that for $x\in\IAC_{\nu^\sigma(\sigma)}(\alpha_0,\theta)$ or $x\in\IAC_{\nu^\sigma(\sigma)}^W(\alpha_0,\theta,\Theta)$, both $wx w^\dagger$ and $w^\dagger x w$ are in $\IAC_{\sigma}(\alpha_0,\theta,\Theta)$ or $\IAC_{\sigma}^W(\alpha_0,\theta,\Theta)$ respectively.
	\end{remark}
	There is a standard lemma we will often refer to:
	\begin{lemma}\label{lem:SplittingOfUnitary}
		Let $\AA$ and $\BB$ be arbitrary unital $C^*$-algebras. Let $(\HH_\AA,\pi_\AA)$ and $(\HH_\BB,\pi_\BB)$ be arbitrary irreducible $*$-representations on $\AA$ and $\BB$ respectively. Let $U\in\UU(\HH_\AA\otimes\HH_\BB)$ be such that there exists an $\alpha_\AA\in\Aut{\AA}$ and an $\alpha_\BB\in\Aut{\BB}$ satisfying
		\begin{equation}
			\Ad{U}\circ(\pi_\AA\otimes\pi_\BB)=\pi_\AA\circ\alpha_\AA\otimes\pi_\BB\circ\alpha_\BB
		\end{equation}
		then there exists a $U_\AA\in\UU(\HH_\AA)$ and a $U_\BB\in\UU(\HH_\BB)$ such that $U=U_\AA\otimes U_\BB$.
	\end{lemma}
	\begin{proof}
		From the assumptions we see that $\forall A\in\AA$
		\begin{equation}
			\Ad{U}\circ\pi_\AA\otimes\pi_\BB(A\otimes\id)=\pi_\AA\circ\alpha_\AA(A)\otimes\id\subset \BB(\HH_\AA)\otimes\id.
		\end{equation}
		Since $\Ad{U}$ is continuous in weak operator topology (on $\BB(\HH_{\AA}\otimes\HH_{\BB})$), it follows that we can extend the map
		\begin{equation}
			\Ad{U}|_{\textrm{Im}(\pi_\AA)}:\textrm{Im}(\pi_\AA)\otimes\id\rightarrow \BB(\HH_\AA)\otimes\id
		\end{equation}
		to the closure in weak operator topology. By irreducibility (and the Von Neumann bicommutant theorem), we get that $\pi_\AA(\AA)''=\BB(\HH_\AA)$ and therefore we get a map
		\begin{equation}
			\Ad{U}|_{\BB(\HH_\AA)\otimes\id}:\BB(\HH_\AA)\otimes\id\rightarrow \BB(\HH_\AA)\otimes\id.
		\end{equation}
		By restriction, this gives rise to an automorphism $\tau_\AA:\UU(\HH_\AA)\rightarrow\UU(\HH_\AA)$. By Wigners theorem any automorphism of $\BB(\HH_\AA)$ is inner and therefore there exists a $U_\AA$ such that $\tau_\AA=\Ad{U_\AA}$. Doing the same on $\BB$ gives us similarly a $\tau_\BB$ and a $U_\BB$. We now get
		\begin{equation}
			\Ad{U}\circ(\pi_\AA\otimes\pi_\BB)=\tau_\AA\circ\pi_\AA\otimes\tau_\BB\circ\pi_\BB=\Ad{U_\AA\otimes U_\BB}\circ(\pi_\AA\otimes\pi_\BB).
		\end{equation}
		By the irreducibility of $\pi_\AA\otimes\pi_\BB$ we get that (up to some irrelevant phase) we need that $U=U_\AA\otimes U_\BB$ concluding the proof.
	\end{proof}
	Among other things, it shows that if a unitary on the GNS space of $\omega_0$ has an adjoint action that is the product of cones then it is a product of cone operators:
	\begin{lemma}\label{lem:UsingIrreducibilityAndWignerTheorem}
		Take $x\in\UU(\HH_0)$ such that there exist $\xi_\sigma\in\Aut{\AA_{C_\theta\cap\sigma}}$ (for $\sigma\in\{L,R\}$) satisfying
		\begin{equation}
			\Ad{x}\circ\pi_0\circ\alpha_0=\pi_0\circ\alpha_0\circ\xi_L\otimes\xi_R
		\end{equation}
		then there exist $x_\sigma\in\textrm{Cone}_\sigma(\alpha_0,\theta)$ such that $x=x_L\otimes x_R$. These are unique up to a phase.
	\end{lemma}
	\begin{proof}
		Using lemma \ref{lem:SplittingOfUnitary}, the result follows.
	\end{proof}
	Following Ogata \cite{ogata2021h3gmathbb}, we now define objects using the following lemma:
	\begin{lemma}\label{lem:Definition_W_And_u}
		There unitaries $W_g\in\UU(\HH_0)$ and $u_{\sigma}(g,h)\in\UU(\HH_{\sigma})$ for all $g,h\in G$ and for all $\sigma\in\{L,R\}$ satisfying
		\begin{align}
			\nonumber
			\Ad{W_g}\circ\pi_0&=\pi_0\circ\alpha_0\circ\Theta\circ\eta_g\circ\beta_g^U\circ\Theta^{-1}\circ\alpha_0^{-1}\\
			\Ad{u_\sigma(g,h)}\circ\pi_\sigma&=\pi_\sigma\circ\alpha_\sigma\circ\eta_g^\sigma\circ\beta_g^{\sigma U}\circ\eta_h^{\sigma}\circ(\beta_g^{\sigma U})^{-1}\circ(\eta^\sigma_{gh})^{-1}\circ\alpha_\sigma^{-1}\\
			\nonumber
			u_L(g,h)\otimes u_R(g,h)&=W_gW_hW_{gh}^{-1}.
		\end{align}
	\end{lemma}
	\begin{proof}
		The result follows from the fact that assumption \ref{assumption} is sufficient to define the objects used in lemma 2.1 and definition 2.2 of \cite{ogata2021h3gmathbb}.
	\end{proof}
	These objects will be the starting point for our definition. Clearly the $u_\sigma(g,h)$ are elements of $\textrm{Cone}_{\nu^\sigma(\sigma)}^W(\alpha_0,\theta)$. The $W_g$ have the following property:
	\begin{lemma}\label{lem:AdjointOverConeIsInCone}
		Take $\Xi^{\sigma}\in\textrm{Cone}_\sigma(\alpha_0,\theta)$ (for all $\sigma\in\{L,R\}$), then $\Ad{W_g}(\Xi^\sigma)\in\textrm{Cone}_\sigma(\alpha_0,\theta)$ (for all $g\in G$). If, more generally, we take $\Xi^\sigma\in\IAC_\sigma(\alpha_0,\Theta)$, then $\Ad{W_g}(\Xi^\sigma)\in\IAC_\sigma(\alpha_0,\Theta)$. Similarly if $\Xi^{\sigma}\in\textrm{Cone}_\sigma^W(\alpha_0,\theta)$, then $\Ad{W_g}(\Xi^\sigma)\in\textrm{Cone}_\sigma^W(\alpha_0,\theta)$. If additionally, $\Xi^{\sigma}\in\textrm{Cone}_{\nu^\sigma(\sigma)}(\alpha_0,\theta)$ or $\Xi^{\sigma}\in\textrm{Cone}_{\nu^\sigma(\sigma)}^W(\alpha_0,\theta)$, $\Ad{W_g}(\Xi^\sigma)\in\textrm{Cone}_{\nu^\sigma(\sigma)}(\alpha_0,\theta)$ and $\textrm{Cone}_{\nu^\sigma(\sigma)}^W(\alpha_0,\theta)$ respectively.
	\end{lemma}
	\begin{proof}
		Take $\tilde{\Xi}^\sigma$ such that $\Xi^\sigma=\tilde{\Xi}^\sigma\otimes\id_{\HH_{\sigma^c}}$. Take $\xi_\sigma\in\Aut{\AA_{W(C_\theta)\cap\sigma}}$ such that
		\begin{equation}
			\Ad{\tilde\Xi^\sigma}\circ\pi_\sigma\circ\alpha_\sigma=\pi_\sigma\circ\alpha_\sigma\circ \xi_\sigma.
		\end{equation}
		We have that
		\begin{equation}
			\Ad{\Ad{W_g}(\Xi^R)}\circ\pi_0\circ\alpha_0\circ\Theta=\pi_0\circ\alpha_0\circ\Theta\circ\zeta_{g,R}
		\end{equation}
		where
		\begin{equation}
			\zeta_{g,\sigma}\defeq \eta_g\circ\beta_g^{U}\circ\xi_\sigma\circ(\beta_g^{U})^{-1}\circ\eta_g^{-1}.
		\end{equation}
		Since $\zeta_{g,\sigma}\in\Aut{\AA_{C_\theta}\cap\AA_{\sigma}}$ we get that
		\begin{equation}
			\Ad{\Ad{W_g}(\Xi^R)}\circ\pi_0\circ\alpha_0=\pi_L\circ\alpha_L\otimes\pi_R\circ\alpha_R\circ\zeta_{g,R}.
		\end{equation}
		Using lemma \ref{lem:UsingIrreducibilityAndWignerTheorem} there exists a $Z_{g,R}\in\UU(\HH_R)$ satisfying that $\Ad{W_g}(\Xi^R)=\id_{\HH_L}\otimes Z_{g,R}$. To show the second result we only need to observe that for any $A\in\UU(\AA_R)$ we get that
		\begin{align}
			\Ad{W_g}\left(\pi_0\circ\alpha_0\circ\Theta(A)\Xi_R\right)&=\Ad{W_g}\circ\pi_0\circ\alpha_0\circ\Theta(A)\Ad{W_g}(\Xi_R)\\
			\nonumber
			&=\pi_0\circ\alpha_0\circ\Theta\circ\xi_R(A)\Ad{W_g}(\Xi_R).
		\end{align}
		By the previous result, this is clearly in $\IAC_R(\alpha_0,\theta,\Theta)$ concluding the proof.
	\end{proof}
	We now have to define two more objects (remember that $v$ was defined in lemma \ref{lem:Definition_v_And_w}):
	\begin{lemma}\label{lem:Definition_K}
		There exist $K_g^{L}\in\textrm{Cone}_{\nu^L(L)}(\alpha_0,\theta)$ and $K_g^{R}\in\textrm{Cone}_{\nu^R(R)}(\alpha_0,\theta)$ such that
		\begin{equation}
			v^\dagger W_g v W_g^\dagger=K_g^L\otimes K_g^R=K_g
		\end{equation}
		and they are unique (up to a phase). They satisfy the identity
		\begin{equation}
			\Ad{K^R_g}\circ\pi_0=\pi_0\circ \alpha_0\circ \tau^{-1}\circ \eta_g^R\circ\beta_g^{RU}\circ\tau\circ(\beta_g^{RU})^{-1}\circ(\eta_g^{R})^{-1}\circ\alpha_0^{-1}.\quad\footnote{Observe in particular that $\Theta$ has cancelled out}
		\end{equation}
	\end{lemma}
	\begin{proof}
		By some straightforward calculation, we have that
		\begin{align}
			\Ad{K_g}\circ\pi_0&=\Ad{v^\dagger W_g v W_g^\dagger}\circ\pi_0\\
			\nonumber
			&=\bigotimes_{\sigma=L,R}\pi_\sigma\circ\alpha_\sigma\circ\tau^{-1}\circ \eta_g^\sigma\circ\beta_g^{\sigma U}\circ\tau\circ(\beta_g^{\sigma U})^{-1}\circ(\eta_g^{\sigma})^{-1}\circ\alpha_\sigma^{-1}.
		\end{align}
		Using lemma \ref{lem:UsingIrreducibilityAndWignerTheorem} concludes the proof that these unitaries exist and are unique up to a $G-$dependent phase.
	\end{proof}
	If, additionally, $\omega$ satisfies assumption \ref{assumption1dWithTranslation}, we can use these objects to define $b^\sigma_g$:
	\begin{lemma}
		There exist a $b_g^L\in\textrm{Cone}_{\nu^{-1}(L)}^W(\alpha_0,\theta)$ and a $b_g^R\in\textrm{Cone}_{R}^W(\alpha_0,\theta)$ (both unique up to an exchange of a $G-$dependent phase) satisfying that
		\begin{equation}
			w^\dagger W_g w W_g^\dagger=b_g:=b_g^L\otimes b_g^R.
		\end{equation}
		It will satisfy
		\begin{equation}\label{eq:automorphismBelongingTo_b}
			\Ad{b_g^\sigma\otimes\id_{\HH_{\ZZ^2/\sigma}}}\circ\pi_0\circ\alpha_0=\pi_0\circ\alpha_0\circ\nu^{-1}\circ\eta_g^\sigma\circ\nu\circ(\eta_g^\sigma)^{-1}.
		\end{equation}
	\end{lemma}
	\begin{proof}
		It is easy enough to show that
		\begin{equation}
			\Ad{w^\dagger W_g w W_g^\dagger}\circ\pi_0=\pi_0\circ (\alpha_L\circ\nu^{-1}\circ\eta_g^L\circ\nu\circ(\eta_g^L)^{-1}\circ\alpha_L^{-1})\otimes(\alpha_R\circ\nu^{-1}\circ\eta_g^R\circ\nu\circ(\eta_g^R)^{-1}\circ\alpha_R^{-1}).
		\end{equation}
		To do this one only has to use that $\beta_g^U$ commutes with $\nu$. By lemma \ref{lem:UsingIrreducibilityAndWignerTheorem} the result now follows.
	\end{proof}
	\section{Defining the indices}
	In this section, we will use the objects defined in section \ref{sec:ConeAndTranslationOperators} to define the $H^2(G,\TT)$-valued index introduced in theorem \ref{thrm:ExistenceFirstIndex} and the $H^1(G,\TT)$-valued index introduced in theorem \ref{thrm:ExistenceSecondIndex}.
	\subsection{The \texorpdfstring{$H^2(G,\TT)$}{H2}-valued index}\label{sec:DefinitionH2Index}
	\begin{lemma}\label{lem:Definition2Cochain}
		There exists a $C:G^2\rightarrow U(1)$ such that 
		\begin{equation}\label{eq:Definition2Cochain}
			K_g^R\Ad{W_g}\left(K_h^R\Ad{W_hW_{gh}^\dagger}\left((K_{gh}^R)^\dagger\right)\right)u_
			R(g,h)=C(g,h)v^\dagger u_R(g,h)v
		\end{equation}
		for all $g,h\in G.$
	\end{lemma}
	\begin{proof}
		Since the GNS representation is irreducible this is equivalent to showing that the left and righthand side of equation \eqref{eq:Definition2Cochain} have the same adjoint action on the GNS representation. We first prove the result for the full tensor product. By using the definition of $u(g,h)$ we get
		\begin{align}\label{eq:ProofOf:lem:Definition2CochainFirstEquation}
			v^\dagger u_L(g,h)v v^\dagger u_R(g,h) v&=v^\dagger (u_L(g,h)\otimes u_R(g,h)) v =v^\dagger W_g W_h W_{gh}^{-1}v.
		\end{align}
		Now using the definition of $K_g$ gives:
		\begin{align}
			\eqref{eq:ProofOf:lem:Definition2CochainFirstEquation}&=K_gW_g v^\dagger W_h W_{gh}^{-1}v=K_gW_g K_h W_h v^\dagger W_{gh}^{-1}v\\
			\nonumber
			&=K_gW_g K_h W_h W_{gh}^\dagger K_{gh}^\dagger=K_gW_g K_h W_h W_{gh}^\dagger K_{gh}^\dagger W_{gh}W_h^\dagger W_g^\dagger W_gW_hW_{gh}^\dagger\\
			\nonumber
			&=K_g\Ad{W_g}\left(K_h\Ad{W_hW_{gh}^\dagger}\left(K_{gh}^\dagger\right)\right)u_L(g,h)\otimes u_
			R(g,h).
		\end{align}
		Using lemma \ref{lem:AdjointOverConeIsInCone} we get that
		\begin{align}
			&(K_g^L\otimes K_g^R)\Ad{W_g}\left((K_h^L\otimes K_h^R)\Ad{W_hW_{gh}^\dagger}\left((K_{gh}^L\otimes K_{gh}^R)^{-1}\right)\right)\\
			\nonumber
			&=K_g^L\Ad{W_g}\left(K_h^L\Ad{W_hW_{gh}^\dagger}\left((K_{gh}^L)^\dagger\right)\right)\otimes K_g^R\Ad{W_g}\left(K_h^R\Ad{W_hW_{gh}^\dagger}\left((K_{gh}^R)^\dagger\right)\right)
		\end{align}
		concluding the proof.
	\end{proof}
	We the next goal is to prove that this function $C$ is a $2$-cochain. We will do this by proving that the $3$-cochain constructed in \cite{ogata2021h3gmathbb} is invariant under some substitution. More precisely, we will use:
	\begin{lemma}\label{lem:3cochainIsInvariant}
		The three-cochain $C'(g,h,k)$ defined through
		\begin{equation}\label{eq:defintion3CochainProof2Cochain}
			u_R(g,h)u_R(gh,k)u_R(g,hk)^\dagger\Ad{W_g}(u_R(h,k)^\dagger)=C'(g,h,k)\id.
		\end{equation}
		is invariant under the substitution
		\begin{align}\label{eq:SubstitutionForProofCochain}
			W_g&\rightarrow K_g W_g&u_R(g,h)&\rightarrow K_g^R\Ad{W_g}\left(K_h^R\Ad{W_hW_{gh}^\dagger}\left((K_{gh}^R)^\dagger\right)\right)u_
			R(g,h)
		\end{align}
	\end{lemma}
	\begin{proof}
		Inserting substitution \eqref{eq:SubstitutionForProofCochain} into \eqref{eq:defintion3CochainProof2Cochain} gives
		\begin{align}
		&u_R(g,h)u_R(gh,k)u_R(g,hk)^\dagger\Ad{W_g}\left(u_R(h,k)^\dagger\right)\\
		\rightarrow&K_g^R\Ad{W_g}\left(K_h^R\Ad{W_hW_{gh}^\dagger}\left((K_{gh}^R)^\dagger\right)\right)u_
		R(g,h)\\
		\nonumber
		&K_{gh}^R\Ad{W_{gh}}\left(K_k^R\Ad{W_kW_{ghk}^\dagger}\left((K_{ghk}^R)^\dagger\right)\right)u_
		R(gh,k)\\
		\nonumber
		&u_R(g,hk)^\dagger \Ad{W_g}\left( \Ad{W_{hk}W_{ghk}^\dagger}(K^R_{ghk})(K^R_{hk})^{\dagger} \right)(K^R_g)^\dagger\\
		\nonumber
		&\Ad{K_gW_g}\left(u_R(h,k)^\dagger \Ad{W_h}\left(\Ad{W_kW_{hk}^\dagger}\left(K_{hk}^R\right)(K^R_k)^\dagger\right)(K^R_h)^\dagger\right).
		\end{align}
		Using the fact that $W_gW_hW_{gh}^\dagger=u_L\otimes u_R(g,h)$ and lemma \ref{lem:AdjointOverConeIsInCone} (for the substitution $K_g\rightarrow K_g^R$ in the last line) one now gets
		\begin{align}
		=&K_g^R\Ad{W_g}\left(K_h^R\Ad{W_hW_{gh}^\dagger}\left((K_{gh}^R)^\dagger\right)\right)\underline{W_gW_hW_{gh}^\dagger u_L(g,h)^\dagger}\\
		\nonumber
		&K_{gh}^R\Ad{W_{gh}}\left(K_k^R\Ad{W_kW_{ghk}^\dagger}\left((K_{ghk}^R)^\dagger\right)\right) \underline{W_{gh}W_kW_{ghk}^\dagger u_L(gh,k)^\dagger}\\
		\nonumber
		&\underline{W_{ghk}W_{hk}^\dagger W_{g}^\dagger u_L(g,hk)} \Ad{W_g}\left( \Ad{W_{hk}W_{ghk}^\dagger}(K^R_{ghk})(K^R_{hk})^{\dagger} \right)\stkout{(K^R_g)^\dagger}\\
		\nonumber
		&\:\stkout{K^R_g}\Ad{W_g}\left(u_R(h,k)^\dagger \Ad{W_h}\left(\Ad{W_kW_{hk}^\dagger}\left(K_{hk}^R\right)(K^R_k)^\dagger\right)(K^R_h)^\dagger\right)(K^R_g)^\dagger.
		\end{align}
		We will now use the fact that the $u_L(g,hk)$ commutes with everything that has support only on the right. Combining this with lemma \ref{lem:AdjointOverConeIsInCone} gives
		\begin{align}
		=&K_g^R\Ad{W_g}\left(K_h^R\Ad{W_hW_{gh}^\dagger}\left((K_{gh}^R)^\dagger\right)\right)W_gW_hW_{gh}^\dagger u_L(g,h)^\dagger\\
		\nonumber
		&K_{gh}^R\Ad{W_{gh}}\left(K_k^R\Ad{W_kW_{ghk}^\dagger}\left((K_{ghk}^R)^\dagger\right)\right) W_{gh}W_kW_{ghk}^\dagger u_L(gh,k)^\dagger\\
		\nonumber
		&W_{ghk}W_{hk}^\dagger W_{g}^\dagger \Ad{W_g}\left( \Ad{W_{hk}W_{ghk}^\dagger}(K^R_{ghk})(K^R_{hk})^{\dagger} \right)\underline{u_L(g,hk)}\\
		\nonumber
		&\:\Ad{W_g}\left(u_R(h,k)^\dagger \Ad{W_h}\left(\Ad{W_kW_{hk}^\dagger}\left(K_{hk}^R\right)(K^R_k)^\dagger\right)(K^R_h)^\dagger\right)(K^R_g)^\dagger\\
		=&K_g^R\Ad{W_g}\left(K_h^R\Ad{W_hW_{gh}^\dagger}\left((K_{gh}^R)^\dagger\right)\right)W_gW_hW_{gh}^\dagger u_L(g,h)^\dagger\\
		\nonumber
		&K_{gh}^R\Ad{W_{gh}}\left(K_k^R\Ad{W_kW_{ghk}^\dagger}\left((K_{ghk}^R)^\dagger\right)\right) W_{gh}W_kW_{ghk}^\dagger u_L(gh,k)^\dagger\\
		\nonumber
		&\Ad{W_{ghk}W_{hk}^\dagger W_{g}^\dagger}  \left(\Ad{W_g}\left( \Ad{W_{hk}W_{ghk}^\dagger}(K^R_{ghk})(K^R_{hk})^{\dagger} \right)\right)\\
		\nonumber
		&\underline{u_R(g,hk)^\dagger}\Ad{W_g}\left(u_R(h,k)^\dagger \Ad{W_h}\left(\Ad{W_kW_{hk}^\dagger}\left(K_{hk}^R\right)(K^R_k)^\dagger\right)(K^R_h)^\dagger\right)(K^R_g)^\dagger.
		\end{align}
		Doing the same with $u_L(gh,k)^\dagger$ now gives
		\begin{align}
		=&K_g^R\Ad{W_g}\left(K_h^R\Ad{W_hW_{gh}^\dagger}\left((K_{gh}^R)^\dagger\right)\right)W_gW_hW_{gh}^\dagger u_L(g,h)^\dagger\\
		\nonumber
		&K_{gh}^R\Ad{W_{gh}}\left(K_k^R\Ad{W_kW_{ghk}^\dagger}\left((K_{ghk}^R)^\dagger\right)\right)  \\
		\nonumber
		&\Ad{W_{gh}W_k\stkout{W_{ghk}^\dagger W_{ghk}}W_{hk}^\dagger W_{g}^\dagger}  \left(\Ad{W_g}\left( \Ad{W_{hk}W_{ghk}^\dagger}(K^R_{ghk})(K^R_{hk})^{\dagger} \right)\right)\\
		\nonumber
		&\underline{u_R(gh,k)}u_R(g,hk)^\dagger\Ad{W_g}\left(u_R(h,k)^\dagger \Ad{W_h}\left(\Ad{W_kW_{hk}^\dagger}\left(K_{hk}^R\right)(K^R_k)^\dagger\right)(K^R_h)^\dagger\right)(K^R_g)^\dagger.
		\end{align}
		When also applying this to $u_L(g,h)^\dagger$ one gets
		\begin{align}
		=&K_g^R\Ad{W_g}\left(K_h^R\Ad{W_hW_{gh}^\dagger}\left((K_{gh}^R)^\dagger\right)\right)W_gW_hW_{gh}^\dagger (K^R_g)^\dagger\\
		\nonumber
		&K_{gh}^R\Ad{W_{gh}}\left(K_k^R\Ad{W_kW_{ghk}^\dagger}\left((K_{ghk}^R)^\dagger\right)\right)\\
		\nonumber
		&\Ad{W_{gh}W_kW_{hk}^\dagger W_{g}^\dagger}  \left(\Ad{W_g}\left( \Ad{W_{hk}W_{ghk}^\dagger}(K^R_{ghk})(K^R_{hk})^{\dagger} \right)\right)\\
		\nonumber
		&W_{gh}W_h^\dagger W_g^\dagger \underline{u_R(g,h)} u_R(gh,k)u_R(g,hk)^\dagger\Ad{W_g}\left(u_R(h,k)^\dagger\right)\\
		\nonumber
		&\Ad{W_g}\left(\Ad{W_h}\left(\Ad{W_kW_{hk}^\dagger}\left(K_{hk}^R\right)(K^R_k)^\dagger\right)(K^R_h)^\dagger\right)(K^R_g)^\dagger
		\end{align}
		Filling in the equation for the 3-cochain (equation \eqref{eq:defintion3CochainProof2Cochain}) now gives
		\begin{align}
		=&\underline{C'(g,h,k)}K_g^R\Ad{W_g}\left(K_h^R\Ad{W_hW_{gh}^\dagger}\left((K_{gh}^R)^\dagger\right)\right)W_gW_hW_{gh}^\dagger \\
		\nonumber
		&K_{gh}^R\Ad{W_{gh}}\left(K_k^R\Ad{W_kW_{ghk}^\dagger}\left((K_{ghk}^R)^\dagger\right)\right)\\
		\nonumber
		&\Ad{W_{gh}W_kW_{hk}^\dagger \stkout{W_{g}^\dagger}}  \left(\underline{\id}\stkout{\Ad{W_g}}\left( \Ad{W_{hk}W_{ghk}^\dagger}(K^R_{ghk})(K^R_{hk})^{\dagger} \right)\right)\\
		\nonumber
		&W_{gh}W_h^\dagger W_g^\dagger\Ad{W_g}\left(\Ad{W_h}\left(\Ad{W_kW_{hk}^\dagger}\left(K_{hk}^R\right)(K^R_k)^\dagger\right)(K^R_h)^\dagger\right)(K^R_g)^\dagger.
		\end{align}
		Fully writing out the adjoints now gives:
		\begin{align}
		=&C'(g,h,k)K_g^RW_gK_h^RW_hW_{gh}^\dagger (K_{gh}^R)^\dagger W_{gh}W_h^\dagger W_g^\dagger W_gW_hW_{gh}^\dagger \\
		\nonumber
		&K_{gh}^RW_{gh}K_k^RW_kW_{ghk}^\dagger(K_{ghk}^R)^\dagger W_{ghk}W_k^\dagger W_{gh}^\dagger  \\
		\nonumber
		&W_{gh}W_kW_{hk}^\dagger W_{hk}W_{ghk}^\dagger K^R_{ghk} W_{ghk}W_{hk}^\dagger(K^R_{hk})^{\dagger} W_{hk}W_k^\dagger W_{gh}^\dagger\\
		\nonumber
		&W_{gh}W_h^\dagger W_g^\dagger W_gW_h W_kW_{hk}^\dagger K_{hk}^RW_{hk}W_k^\dagger(K^R_k)^\dagger W_h^\dagger (K^R_h)^\dagger W_g^\dagger (K^R_g)^\dagger\\
		=&\id C'(g,h,k)
		\end{align}
		concluding the proof.
	\end{proof}
	\begin{lemma}
		The function $C$ as defined in lemma \ref{lem:Definition2Cochain} is a 2-cochain.
	\end{lemma}
	\begin{proof}
		Take the 3-cochain $C'$ introduced in the last lemma. It is clear that this 3-cochain is invariant under the substitution
		\begin{align}
			W_g&\rightarrow v^\dagger W_g v&u_R(g,h)&\rightarrow v^\dagger u_R(g,h)v.
		\end{align}
		If we now prove that it is also invariant under the substitution in equation \ref{eq:SubstitutionForProofCochain}, we have proved our result because using equation \eqref{eq:Definition2Cochain} we then get that
		\begin{equation}
			\stkout{C'(g,h,k)}=C(h,k)C(gh,k)^{-1}C(g,hk)C(g,h)^{-1}\stkout{C'(g,h,k)}
		\end{equation}
		proving the result. The invariance of the three cochain under this substitution is proven in lemma \ref{lem:3cochainIsInvariant}.
	\end{proof}
	Our translation index is now defined as
	\begin{definition}\label{def:DefinitionOfTheH2ValuedIndex}
		Let $C:G^2\rightarrow U(1)$ be the 2-cochain defined in \ref{lem:Definition2Cochain}. Take $\phi:G^2\rightarrow\TT$ such that $C(g,h)=\exp(i\phi(g,h))$ then we define the index as
		\begin{equation}
			\textrm{Index}^{\AA,U}(\theta,\tilde{\beta}_g,\eta_g,\alpha_0,\Theta,\omega,\omega_0)\defeq\expval{\phi}\in H^2(G,\TT)
		\end{equation}
		and (as advertised) it is only a function of the automorphisms (and the product state) not on the choice of the GNS triple of $\omega_0$ or on the choice of phases in $W_g,u_L(g,h),u_R(g,h),v,K_g^L$ and $K_g^R$.\footnote{Even though we explicitly wrote down the on-site group action $U$, this index is only dependent on $\Ad{U}$.}
	\end{definition}
	\begin{proof}
		The construction is invariant under the choice of GNS triple since this simply amounts to an adjoint action by some unitary on every operator. Now we will show that it is invariant under the choice of phases of our operators. Clearly the 2-cochain $C$ is invariant under
		\begin{align}
			u_L(g,h)&\rightarrow \alpha(g)\alpha(g)\alpha(gh)^{-1}\beta(g,h)^{-1} u_L(g,h)&u_R(g,h)&\rightarrow \beta(g,h)u_R(g,h)\\
			\nonumber
			W_g&\rightarrow\alpha(g)W_g&v&\rightarrow \gamma v.
		\end{align}
		Under the transformation
		\begin{align}
			K_g^L&\rightarrow \delta(g)^{-1}K_g^L&K_g^R&\rightarrow \delta(g)K_g^R
		\end{align}
		we get $C(g,h)\rightarrow \delta(g)\delta(h)\delta(gh)^{-1}C(g,h)$ which is still in the same equivalence class concluding the proof.
	\end{proof}
	\subsection{The \texorpdfstring{$H^1(G,\TT)$}{H1}-valued index}\label{sec:DefinitionH1Index}
	\begin{figure}
	\centering
	\def\s{0.4}
	\resizebox{0.35\textwidth}{!}{%
		\begin{tikzpicture}
			\draw[draw=white,line width=0mm] (1+\s,0) coordinate (a) -- (\s,0) coordinate (b) -- (1+\s,1) coordinate (c);
			
			\fill[fill=green!30!white] (-1*\s,0) -- (-6*\s,-5*\s) -- (-6*\s,5*\s);
			\fill[fill=green!30!white] (1*\s,0) -- (6*\s,-5*\s) -- (6*\s,5*\s);
			\fill[fill=red!30!white] (0,1*\s) -- (4*\s,5*\s) -- (-4*\s,5*\s);
			\fill[fill=red!30!white] (0,-1*\s) -- (4*\s,-5*\s) -- (-4*\s,-5*\s);
			
			\draw[draw=black,line width=0.3mm]
			plot[smooth,samples=2,domain=0:4*\s] (\x,\x+1*\s);
			\draw[draw=black,line width=0.3mm]
			plot[smooth,samples=2,domain=-4*\s:0] (\x,-\x+1*\s);
			\draw[draw=black,line width=0.3mm]
			plot[smooth,samples=2,domain=0:4*\s] (\x,-\x-1*\s);
			\draw[draw=black,line width=0.3mm]
			plot[smooth,samples=2,domain=-4*\s:0] (\x,\x-1*\s);
			\draw[draw=black,line width=0.3mm]
			plot[smooth,samples=2,domain=0:5*\s] (\x+1*\s,\x);
			\draw[draw=black,line width=0.3mm]
			plot[smooth,samples=2,domain=0:5*\s] (\x+1*\s,-\x);
			\draw[draw=black,line width=0.3mm]
			plot[smooth,samples=2,domain=-5*\s:0] (\x-1*\s,-\x);
			\draw[draw=black,line width=0.3mm]
			plot[smooth,samples=2,domain=-5*\s:0] (\x-1*\s,\x);
			
			\draw[draw=black,line width=0.3mm] (1*\s,0) -- (3*\s,0);
			\pic [draw, ->, "$\theta$", angle eccentricity=1.5, angle radius=\s*1.5cm,line width=0.3mm] {angle=a--b--c};
			
			\node at (-4*\s,0) {$\eta^L_g$};
			\node at (4.5*\s,0) {$\eta^R_g$};
			\node at (0,3*\s) {$\Theta$};
			\node at (-1*\s,-1*\s) {$\bullet$};
			\node at (-1.45*\s,-1.5*\s) {$\mathsmaller{\mathsmaller{\left(\scalebox{0.5}{-}1,\scalebox{0.5}{-}1\right)}}$};
			
	\end{tikzpicture}}
	\caption{This figure indicates the support area of the different automorphisms when there are two translation directions. The angle $\theta$ still has to be smaller then or equal to what was indicated here so that the $\Theta$ and the $\eta_g$ commute but now we must be able to do both a horizontal and a vertical widening of the support of $\eta_g$. Also the coordinate $(-1,-1)$ which will be of particular interest is indicated.}
	\label{fig:SetupForTwoTranslationsNearProof}
\end{figure}
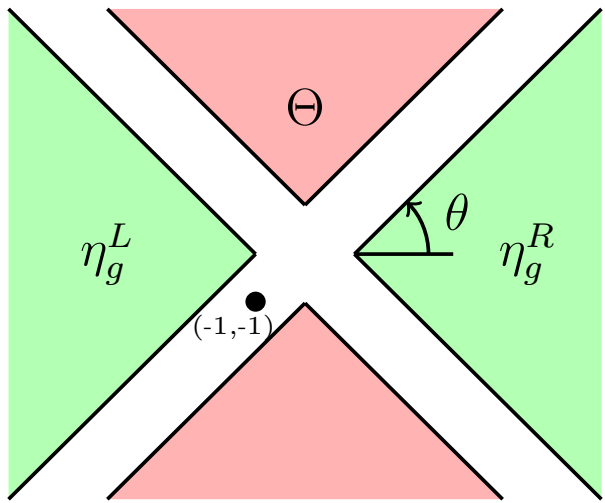
	\begin{lemma}\label{lem:DefinitionAlpha}
		There exists an $\alpha:G\rightarrow U(1)$ satisfying
		\begin{align}\label{eq:DefinitionOfTheH_1Index}
			v^\dagger b_g^R vh_g K_g^R (b_g^R)^\dagger&=\alpha(g)w^\dagger K_g^R w&&\text{where}&h_g&\defeq\pi_0\circ\alpha_0\circ\Theta(U_{(-1,-1)}(g)).
		\end{align}
	\end{lemma}
	\begin{proof}
		We have that
		\begin{align}
			&\Ad{w^\dagger K_g^R w}\circ\pi_0\circ\alpha_0\circ\Theta=\pi_0\circ\alpha_0\circ\Theta\circ\nu^{-1}\circ\tau^{-1}\circ\eta_g^R\circ\beta_g^{RU}\circ\tau\circ(\beta_g^{RU})^{-1}\circ(\eta_g^R)^{-1}\circ\nu.
		\end{align}
		Now we will use the fact that because of translation invariance of the group action, we have that $\tau^{-1}\circ\beta_g^{RU}\circ\tau\circ(\beta_g^{RU})^{-1}=\beta_g^{RL_{-1}}$ with $L_j=\{(x,y)\in\ZZ^2|y=j\}$. This leads to
		\begin{align}
			&=\pi_0\circ\alpha_0\circ\Theta\circ\nu^{-1}\circ\underline{\tau^{-1}}\circ\eta_g^R\circ\tau\circ\beta_g^{R,L_{-1}}\circ(\eta_g^R)^{-1}\circ\nu\\
			\nonumber
			&=\Ad{\underline{v^\dagger}}\circ \pi_0\circ\alpha_0\circ\Theta\circ\nu^{-1}\circ\eta_g^R\circ\tau\circ\beta_g^{R,L_{-1}}\circ(\eta_g^R)^{-1}\circ\nu.
		\end{align}
		Inserting $b^R_g v v^\dagger (b^R_g)^\dagger$ now gives
		\begin{equation}
			=\Ad{v^\dagger\underline{b^R_g v v^\dagger (b^R_g)^\dagger}}\circ \pi_0\circ\alpha_0\circ\Theta\circ\nu^{-1}\circ\eta_g^R\circ\tau\circ\beta_g^{R,L_{-1}}\circ(\eta_g^R)^{-1}\circ\nu.
		\end{equation}
		Using the inverse of equation \eqref{eq:automorphismBelongingTo_b} we now get
		\begin{align}
			&=\Ad{v^\dagger b^R_g v }\circ \pi_0\circ\alpha_0\circ\Theta\circ\underline{\tau^{-1}\circ\eta_g^R\circ\nu^{-1}\circ\stkout{(\eta_g^R)^{-1}\circ\nu}}\circ\stkout{\nu^{-1}\circ\eta_g^R}\circ\tau\circ\beta_g^{R,L_{-1}}\circ(\eta_g^R)^{-1}\circ\nu\\
			\nonumber
			&=\Ad{v^\dagger b_g^R v}\circ\pi_0\circ\alpha_0\circ\Theta\circ\tau^{-1}\circ\eta_g^R\circ\tau\circ\nu^{-1}\circ\beta_g^{R,L_{-1}}\circ(\eta_g^R)^{-1}\circ\nu.
		\end{align}
		Now we will use the fact that $\nu^{-1}\circ\beta_g^{R,L_{-1}}\circ\nu=\Ad{U_{(-1,-1)}(g)}\circ\beta_g^{R,L_{-1}}$ to obtain
		\begin{equation}
			=\Ad{v^\dagger b_g^R v}\circ\pi_0\circ\alpha_0\circ\Theta\circ\tau^{-1}\circ\eta_g^R\circ\tau\circ\Ad{U_{(-1,-1)}(g)}\circ\beta_g^{R,L_{-1}}\circ\nu^{-1}\circ(\eta_g^R)^{-1}\circ\nu.
		\end{equation}
		Since (as you can check in figure \ref{fig:SetupForTwoTranslationsNearProof}) $\tau^{-1}(\eta_g^R(\tau(U_{(-1,-1)}(g))))=U_{(-1,-1)}(g)$ this gives
		\begin{align}
			&=\Ad{v^\dagger b_g^R v}\circ\pi_0\circ\alpha_0\circ\Theta\circ\Ad{U_{(-1,-1)}(g)}\circ\tau^{-1}\circ\eta_g^R\circ\tau\circ\beta_g^{R,L_{-1}}\circ\nu^{-1}\circ(\eta_g^R)^{-1}\circ\nu\\
			\nonumber
			&=\Ad{v^\dagger b_g^R vh_g}\circ\pi_0\circ\alpha_0\circ\Theta\circ\tau^{-1}\circ\eta_g^R\circ\tau\circ\beta_g^{R,L_{-1}}\circ\nu^{-1}\circ(\eta_g^R)^{-1}\circ\nu.
		\end{align}
		Now using again that $\tau\circ\beta_g^{R,L_{-1}}=\beta_{g}^{RU}\circ\tau\circ(\beta_g^{RU})^{-1}$ we get:
		\begin{align}
			&=\Ad{v^\dagger b_g^R vh_g}\circ\pi_0\circ\alpha_0\circ\Theta\circ\tau^{-1}\circ\eta_g^R\circ\beta_{g}^{RU}\circ\tau\circ(\beta_g^{RU})^{-1}\circ\nu^{-1}\circ(\eta_g^R)^{-1}\circ\nu\\
			\nonumber
			&=\Ad{v^\dagger b_g^R vh_gK_g^R}\circ\pi_0\circ\alpha_0\circ\Theta\circ\eta_g^R\circ\nu^{-1}\circ(\eta_g^R)^{-1}\circ\nu.
		\end{align}
		After again using \eqref{eq:automorphismBelongingTo_b} we obtain:
		\begin{align}
			=\Ad{v^\dagger b_g^R vh_gK_g^R(b_g^R)^\dagger}\circ\pi_0\circ\alpha_0\circ\Theta.
		\end{align}
		By the irreducibility of $\pi_0$, this implies that there indeed exists such an $\alpha$.
	\end{proof}
	We have that $h_g\in\IAC_L(\theta,\alpha_0,\Theta)$. On top of that, we also have that
	\begin{lemma}\label{lem:W_g_And_h_g_Commute}
		For all $g,h\in G$ we have that
		\begin{align}
			[h_g,W_h]&=0&[h_g,b_h^L]&=0.
		\end{align}
	\end{lemma}
	\begin{proof}
		We have that (see figure \ref{fig:SetupForTwoTranslationsNearProof})
		\begin{align}
			\Ad{W_h}(h_g)&=\pi_0\circ\alpha_0\circ\Theta\circ\eta_h\circ\beta_h^U(U_{(-1,-1)}(g))=\pi_0\circ\alpha_0\circ\Theta(U_{(-1,-1)}(g))=h_g
		\end{align}
		concluding the proof of the first item. The second proof is analogous.
	\end{proof}
	Finally, before we can show that the $\alpha$ is a $U(1)$ representation we need to prove an analog of the equation in the 2-cochain lemma \ref{lem:Definition2Cochain}:
	\begin{lemma}\label{lem:translating_u_R_To_The_Right_identity}
		The equality
		\begin{equation}
			\Ad{b_g^R W_g b_h^R W_h W_{gh}^\dagger (b_{gh}^R)^\dagger W_{gh}W_h^\dagger W_g^\dagger u_R(g,h)}\circ\pi_0=\Ad{w^\dagger u_R(g,h) w}\circ\pi_0
		\end{equation}
		holds.
	\end{lemma}
	\begin{proof}
		The proof is analogous to the proof of \ref{lem:Definition2Cochain}. We now show that
		\begin{equation}
			\Ad{b_g W_g b_h W_h W_{gh}^\dagger (b_{gh})^\dagger W_{gh}W_h^\dagger W_g^\dagger u_L(g,h)\otimes u_R(g,h)}\circ\pi_0=\Ad{w^\dagger u_R(g,h)\otimes u_R(g,h) w}\circ\pi_0
		\end{equation}
		holds. This part is completely the same. Afterwards we have to prove that: $b_g W_g b_h W_h W_{gh}^\dagger (b_{gh})^\dagger W_{gh}W_h^\dagger W_g^\dagger$, is split. This is a straightforward calculation.
	\end{proof}
	We now need to prove the analogue of \ref{lem:3cochainIsInvariant}.
	\begin{lemma}\label{lem:2cochainIsInvariant}
		The two-cochain as defined through
		\begin{equation}\label{eq:Definition2Cochain2TranslationSectionAppendix}
			K_g^RW_gK_h^RW_hW_{gh}^\dagger(K_{gh}^R)^\dagger W_{gh}W_{h}^\dagger W_g^\dagger u_R(g,h)=C(g,h)v^\dagger u_R(g,h)v.
		\end{equation}
		is invariant under the substitution
		\begin{align}
			K_g^R&\mapsto v^\dagger b_g^R v h_g K_g^R (b_g^R)^\dagger&W_g&\mapsto w^\dagger W_g w&u_R(g,h)&\mapsto w^\dagger u_R(g,h)w.
		\end{align}
	\end{lemma}
	\begin{proof}
		To show this, notice that we have by construction and by lemma \ref{lem:translating_u_R_To_The_Right_identity} that
		\begin{align}
			\Ad{b_g W_g}\circ\pi_0&=\Ad{w^\dagger W_g w}\circ\pi_0\\
			\Ad{b_g^R W_g b_h^R W_h W_{gh}^\dagger (b_{gh}^R)^\dagger W_{gh}W_h^\dagger W_g^\dagger u_R(g,h)}\circ\pi_0&=\Ad{w^\dagger u_R(g,h) w}\circ\pi_0.
		\end{align}
		This shows that there exists a $\beta:G\rightarrow U(1)$ and a $\gamma:G\times G\rightarrow U(1)$ such that
		\begin{align}
			w^\dagger W_g w&=\beta(g)b_g^Lb_g^RW_g&w^\dagger u_R(g,h) w&=\gamma(g,h)b_g^R W_g b_h^R W_h W_{gh}^\dagger (b_{gh}^R)^\dagger W_{gh}W_h^\dagger W_g^\dagger u_R(g,h).
		\end{align}
		Since the equation is invariant under the substitution with these two phases all that is left to do is to show that equation \eqref{eq:Definition2Cochain2TranslationSectionAppendix} is invariant under the substitution:
		\begin{align}
			K_g^R&\mapsto v^\dagger b_g^R v h_g K_g^R (b_g^R)^\dagger=v^\dagger b_g^R v K_g^R (b_g^R)^\dagger h_g&W_g&\mapsto b_g^Lb_g^RW_g
		\end{align}
		\begin{equation}
			u_R(g,h)\mapsto b_g^R W_g b_h^R W_h W_{gh}^\dagger (b_{gh}^R)^\dagger W_{gh}W_h^\dagger W_g^\dagger u_R(g,h).
		\end{equation}
		To show this first notice that by using the fact that $h_g\in\IAC_R(\theta,\alpha_0,\Theta)'$ and lemma \ref{lem:W_g_And_h_g_Commute} we get that
		\begin{align}
			&K_g^R W_g K_h^R W_ hW_{gh}^\dagger(K_{gh}^R)^\dagger\\
			\nonumber
			&\mapsto v^\dagger b_g^R v \underline{h_g} K_g^R (b_g^R)^\dagger b_g^Lb_g^RW_g v^\dagger b_h^R v \underline{h_h} K_h^R (b_h^R)^\dagger b_h^Lb_h^RW_h W_{gh}^\dagger (b_{gh}^Lb_{gh}^R)^\dagger (v^\dagger b_{gh}^R v \underline{h_{gh}} K_{gh}^R (b_{gh}^R)^\dagger)^\dagger\\
			\nonumber
			&=\underline{h_gh_hh_{gh}^\dagger} v^\dagger b_g^R v  K_g^R (b_g^R)^\dagger b_g^Lb_g^RW_g v^\dagger b_h^R v  K_h^R (b_h^R)^\dagger b_h^Lb_h^RW_h W_{gh}^\dagger (b_{gh}^Lb_{gh}^R)^\dagger (v^\dagger b_{gh}^R v  K_{gh}^R (b_{gh}^R)^\dagger)^\dagger
		\end{align}
		(all the $h-$operators can be put in front). Since $h_g$ is an honest representation we get that the $h-$operators cancel out. This leaves us with checking that equation \eqref{eq:Definition2Cochain2TranslationSectionAppendix} is invariant under
		\begin{align}
			K_g^R&\mapsto v^\dagger b_g^R v K_g^R (b_g^R)^\dagger&W_g&\mapsto b_g^Lb_g^RW_g
		\end{align}
		\begin{equation*}
			u_R(g,h)\mapsto b_g^R W_g b_h^R W_h W_{gh}^\dagger (b_{gh}^R)^\dagger W_{gh}W_h^\dagger W_g^\dagger u_R(g,h).
		\end{equation*}
		Using the fact that
		\begin{align}
			(b_{gh}^R)^\dagger W_{gh}W_h^\dagger W_g^\dagger u_R(g,h)&=(b_{gh}^R)^\dagger u_L(g,h)^\dagger=u_L(g,h)^\dagger(b_{gh}^R)^\dagger=W_{gh}W_h^\dagger W_g^\dagger u_R(g,h) (b_{gh}^R)^\dagger
		\end{align}
		we see that we can prove that our equation is invariant under the substitution using lemma \ref{lem:TransformationUnderDelta}. This concludes the proof.
	\end{proof}
	Using this we can show that:
	\begin{lemma}
		The $\alpha$ defined previously is a $U(1)$-representation.
	\end{lemma}
	\begin{proof}
		Take the definition of the 2-cochain (lemma \ref{lem:Definition2Cochain}). This equation is invariant under the substitution
		\begin{align}
			K_g^R&\mapsto w^\dagger K_g^R w&W_g&\mapsto w^\dagger W_g w&u_R(g,h)&\mapsto w^\dagger u_R(g,h)w.
		\end{align}
		If we now also show that it is invariant under the substitution
		\begin{align}
			K_g^R&\mapsto v^\dagger b_g^R v h_g K_g^R (b_g^R)^\dagger&W_g&\mapsto w^\dagger W_g w&u_R(g,h)&\mapsto w^\dagger u_R(g,h)w
		\end{align}
		then we get using the definition of $\alpha$ that
		\begin{equation}
			\stkout{C(g,h)}=\stkout{C(g,h)}\frac{\alpha(g)\alpha(h)}{\alpha(gh)}
		\end{equation}
		which would conclude the proof. This invariance is proven in lemma \ref{lem:2cochainIsInvariant}.
	\end{proof}
	\begin{definition}\label{def:DefinitionOfTheH1ValuedIndex}
		Let $\alpha$ be the $U(1)$-representation defined in lemma \ref{lem:DefinitionAlpha}. Take $\phi\in\hom(G,\TT)$ such that $\alpha(g)=\exp(i\phi(g))$. We define the 2 translation index as
		\begin{equation}
			\textrm{Index}_{2\:\text{trans}}^{\AA,U}(\theta,\tilde{\beta}_g,\eta_g,\alpha_0,\Theta,\omega,\omega_0)\defeq \phi\in H^1(G,\TT)
		\end{equation}
		and (as advertised) it is only a function of the automorphisms (and the product state) not of the choice of the GNS triple of $\omega_0$ or the choice of phases in $W_g,u_L(g,h),u_R(g,h),v,w,K^L_g,K^R_g,b^L_g$ and $b^R_g$.\footnote{It is however explicitly dependent on the on-site group action $U$, and not only $\Ad{U}$.}
	\end{definition}
	\begin{proof}
		The construction is invariant under the choice of GNS triple since this simply amounts to an adjoint action by some unitary on every operator. The proof that it is independent of the choice of phases is just as trivial.
	\end{proof}
	One can also define this index starting from operators acting on the left. As the following lemma shows this gives us the opposite index (the complex conjugate of the $U(1)$-representation):
	\begin{lemma}
		Define $\tilde{\alpha}(g)$ such that
		\begin{equation}\label{eq:DefinitionOfTheH_1IndexAlternative}
			v^\dagger b_g^L v h_g^\dagger K_g^L (b_g^L)^\dagger=\tilde{\alpha}(g)w^\dagger K_g^L w
		\end{equation}
		then $\alpha(g)\tilde{\alpha}(g)=1$.
	\end{lemma}
	\begin{proof}
		By multiplying the left hand side of equation \eqref{eq:DefinitionOfTheH_1IndexAlternative} with the left hand side of equation \eqref{eq:DefinitionOfTheH_1Index} we obtain
		\begin{equation}
			v^\dagger b_g v K_g b_g^\dagger=v^\dagger b_g v v^\dagger W_g v W_g^\dagger b_g^\dagger=w^\dagger v^\dagger W_g v W_g^\dagger w=w^\dagger K_g w
		\end{equation}
		whereas by multiplying the right-hand sides of equations \eqref{eq:DefinitionOfTheH_1IndexAlternative} and \eqref{eq:DefinitionOfTheH_1Index} we would obtain $\alpha(g)\tilde{\alpha}(g)w^\dagger K_g w$. These two results can only be consistent if the lemma holds.
	\end{proof}
	\section{The indices are well-defined}\label{sec:IndexIsInvariantUnderChoices}
	In this section, we will show that the index from definition \ref{def:DefinitionOfTheH2ValuedIndex} and the index from definition \ref{def:DefinitionOfTheH1ValuedIndex} are only dependent on $\omega$ and not on the choices of our automorphisms nor on $\omega_0$. On this last item, we remark:
	\begin{remark}\label{rem:NontrivialProductState}
		A product state can have a non-trivial $H^1(G,\TT)$ index and the index is not defined relative to $\omega_0$ so in particular
		\begin{equation}
			\textrm{Index}_{2\:\text{trans}}^{\AA,U}(\theta,\beta_g^U,\textrm{Id},\textrm{Id},\textrm{Id},\omega_0,\omega_0)
		\end{equation}
		can still be non-trivial. In fact in this case we get $\alpha(g)=\omega_0(U_{(-1,-1)}(g))$. To show this, notice that we can take $b_g=\id_{\HH}$ and we can choose $K_g^R$ so that it leaves the cyclic vector of $\omega_0$, $\Omega_0$ invariant. Applying the definition of the index on this cyclic vector then gives us $h_g\Omega_0=\alpha(g)\Omega_0$. This obviously implies that $\bra{\Omega_0}\ket{h_g\Omega_0}=\alpha(g)\bra{\Omega_0}\ket{\Omega_0}$ giving us indeed the advertised equation.
	\end{remark}
	We now show independence on $\alpha$ and the choice of its decomposition.
	\begin{lemma}
		Take $\omega_{01},\omega_{02}\in\PP(\AA)$ product states and let $\alpha_1,\alpha_2\in\Aut{\AA}$ be such that $\omega_{01}\circ\alpha_1=\omega_{02}\circ\alpha_2=\omega$. Let $V_{11},V_{12}\in\UU(\AA)$, $\alpha_{L/R,1},\alpha_{L/R,2}\in\Aut{\AA_{L/R}}$ and $\Theta_1,\Theta_2\in \Aut{\AA_{W(C_\theta)^c}}$ be such that $\alpha_1=\Ad{V_{11}}\circ\alpha_{01}\circ\Theta_1$ and $\alpha_2=\Ad{V_{12}}\circ\alpha_{02}\circ \Theta_2$ with $\alpha_{0,i}=\alpha_{L,i}\otimes\alpha_{R,i}$, then
		\begin{equation}
			\textrm{Index}^{\AA,U}(\theta,\tilde{\beta}_g,\eta_g,\alpha_{0,1},\Theta_1,\omega,\omega_{01})=\textrm{Index}^{\AA,U}(\theta,\tilde{\beta}_g,\eta_g,\alpha_{0,2},\Theta_2,\omega,\omega_{02}).
		\end{equation}
		Additionally, when $\omega$ satisfies assumption \ref{assumption:2Translations},
		\begin{equation}
			\textrm{Index}^{\AA,U}_{\text{2 trans}}(\theta,\tilde{\beta}_g,\eta_g,\alpha_{0,1},\Theta_1,\omega,\omega_{01})=\textrm{Index}^{\AA,U}_{\text{2 trans}}(\theta,\tilde{\beta}_g,\eta_g,\alpha_{0,2},\Theta_2,\omega,\omega_{02}).
		\end{equation}
	\end{lemma}
	\begin{proof}
		We will first prove the result in the case that $\omega_0=\omega_{01}=\omega_{02}$ and then generalize this result. Since $\omega_0\circ\alpha_2\circ\alpha_1^{-1}=\omega_0$ there exists a $\tilde{x}\in\UU(\HH_0)$ such that $\pi_0\circ\alpha_2\circ\alpha_1^{-1}=\Ad{\tilde{x}}\circ\pi_0$. Now define $x\in\UU(\HH_0)$ to be $x\defeq \pi_0(V_{12}^\dagger)\tilde{x} \pi_0(V_{11})$ then
		\begin{equation}
			\pi_0\circ\alpha_{02}\circ\Theta_2=\Ad{x}\circ\pi_0\circ\alpha_{01}\circ\Theta_1.
		\end{equation}
		Now take $W_{g,1},u_{R,1}(g,h), v$ and $K^R_{g,1}$ to be the operators belonging to the first choice (with arbitrary phases). When, $\omega$ satisfies assumption \ref{assumption:2Translations}, also take $b^R_{g,1}$ and $w$, the additional operators belonging to the first choice.
		We have (see \cite{ogata2021h3gmathbb} lemma 2.11)
		\begin{align}
			\Ad{xW_{g,1}x^\dagger}\circ\pi_0&=\pi_0\circ \alpha_{02}\circ\Theta_2\circ\eta_g\circ\beta_g^U\circ\Theta_2^{-1}\circ\alpha_{02}^{-1},\\
			\nonumber
			\Ad{xu_{R,1}(g,h)x^\dagger}\circ\pi_0&=\pi_0\circ \alpha_{02}\circ\eta_g^R\circ\beta_g^{RU}\circ\eta_h^R\circ\beta_{h}^{RU}\circ(\beta_{gh}^{RU})^{-1}\circ(\eta_{gh}^R)^{-1}\circ\alpha_{02}^{-1}.
		\end{align}
		Through similar arguments we get
		\begin{align}
			\Ad{xvx^\dagger}\circ\pi_0&=\pi_0\circ\alpha_{02}\circ\Theta_2\circ\tau\circ\Theta_2^{-1}\circ\alpha_{02}^{-1}\\
			\nonumber
			\Ad{xK^R_{g,1}x^\dagger}\circ\pi_0&=\pi_0\circ \alpha_{02}\circ\tau^{-1}\circ\eta_g^R\circ\beta_g^{RU}\circ\tau\circ(\beta_g^{RU})^{-1}\circ(\eta^R_g)^{-1}\circ\alpha_{02}^{-1}
		\end{align}
		and
		\begin{align}
			\Ad{xwx^\dagger}\circ\pi_0&=\pi_0\circ\alpha_{02}\circ\Theta_2\circ\nu\circ\Theta_2^{-1}\circ\alpha_{02}^{-1}\\
			\nonumber
			\Ad{xb^R_{g,1}x^\dagger}\circ\pi_0&=\pi_0\circ \alpha_{02}\circ\nu^{-1}\circ\eta_g^R\circ\nu\circ(\eta_g^R)^{-1}\circ\alpha_{02}^{-1}
		\end{align}
		in the case where $\omega$ satisfies assumption \ref{assumption:2Translations}. This shows that $xW_{g,1}x^\dagger$, $xu_{R,1}(g,h)x^\dagger$, $xvx^\dagger$, $xK^R_{g,1}x^\dagger$, $xwx^\dagger$ and $x b^R_{g,1}$ are operators belonging to the second choice. Since our index is invariant under this substitution this concludes the proof when $\omega_0=\omega_{01}=\omega_{02}$. Now suppose that $\omega_{01}\neq\omega_{02}$. Since they are both product states there exists a $\gamma\in\Aut{\AA}$ satisfying $\omega_{02}=\omega_{01}\circ\gamma$ that is of the form $\gamma=\gamma^L\otimes\gamma^R$. We now have
		\begin{equation}
			\textrm{Index}(\theta,\tilde{\beta}_g,\eta_g,\alpha_{0,2},\Theta_2,\omega,\omega_{02})=\textrm{Index}(\theta,\tilde{\beta}_g,\eta_g,\alpha_{0,2},\Theta_2,\omega,\omega_{01}\circ\gamma)=\textrm{Index}(\theta,\tilde{\beta}_g,\eta_g,\gamma\circ\alpha_{0,2},\Theta_2,\omega,\omega_{01})
		\end{equation}
		and similarly for the second index.
	\end{proof}
	We now show that the $H^2(G,\TT)$ and $H^1(G,\TT)$-valued indices are invariant under some transformation parametrized by the group $\IAC_R^W$:
	\begin{lemma}\label{lem:TransformationUnderDelta}
		Take $\omega_0,\theta,\alpha_0,\Theta$ and $\eta_g$ as usual, take $v,W_{g,1},u_{\sigma,1}(g,h)$ and $K_{g,1}^\sigma$ the operators corresponding to these automorphisms and take $\delta^\sigma_g\in\IAC_\sigma^W(\alpha_0,\theta,\Theta)$ (for all $\sigma\in\{L,R\}$ and all $g\in G$). Define
		\begin{align}
			W_{g,2}&\defeq\delta_g W_{g,1}&
			u_{\sigma,2}(g,h)&\defeq \delta_g^\sigma W_{g,1}\delta_h^\sigma W_{g,1}^\dagger u_{\sigma,1}(g,h)(\delta_{gh}^\sigma)^\dagger&
			K^\sigma_{g,2}&\defeq v^\dagger \delta_g^\sigma v K_{g,1}^\sigma (\delta_g^\sigma)^\dagger
		\end{align}
		then equation \eqref{eq:Definition2Cochain}, defining the index of $v,W_{g,1},u_{\sigma,1}(g,h)$ and $K_{g,1}^\sigma$, is still valid when we replace these operators by $v,W_{g,2},u_{\sigma,2}(g,h)$ and $K_{g,2}^\sigma$ respectively. The two cochain also remains unchanged.
	\end{lemma}
	\begin{lemma}\label{lem:TransformationUnderDeltaTwoTranslations}
		Additionally, on top of the assumptions and definitions of lemma \ref{lem:TransformationUnderDelta}, let $\omega$ satisfy assumption \ref{assumption:2Translations} and take $w$ and $b_{g,1}^\sigma$ the additional operators belonging to the automorphisms. Define $b_{g,2}^\sigma\defeq w^\dagger\delta^\sigma_g w b_{g,1}^\sigma (\delta^\sigma_g)^\dagger$ then equation \ref{eq:DefinitionOfTheH_1Index}, defining the index of $v,w,W_{g,1},u_{\sigma,1}(g,h),K_{g,1}^\sigma,h_g$ and $b_{g,1}^\sigma$ is still valid when we replace these operators by $v,w,W_{g,2},u_{\sigma,2}(g,h),h_g,K_{g,2}^\sigma$ and $b_{g,2}^\sigma$. The one cochain also remains unchanged.
	\end{lemma}
	\begin{proof}
		We start by proving lemma \ref{lem:TransformationUnderDelta}. First, we will prove that we still have that
		\begin{equation}\label{eq:TransformationUnderDeltaProofNewUIsSplitAppendix}
			u_{L,2}(g,h)u_{R,2}(g,h)=W_{gh,2}W_{h,2}^\dagger W_{g,2}^\dagger.
		\end{equation}
		To do this first insert the definition to obtain
		\begin{align}
			u_{L,2}(g,h)u_{R,2}(g,h)=\delta_g^L W_{g,1}\delta_h^L W_{g,1}^\dagger u_{L,1}(g,h)(\delta_{gh}^L)^\dagger \delta_g^R W_{g,1}\delta_h^R W_{g,1}^\dagger u_{R,1}(g,h)(\delta_{gh}^R)^\dagger.
		\end{align}
		Since by lemma \ref{lem:AdjointOverConeIsInCone}, $ W_{g,1}\delta_g^\sigma W_{g,1}^\dagger\in\IAC_\sigma(\alpha_0,\theta,\Theta)$ (for $\sigma\in\{L,R\}$ and $g\in G$) we get by repetitive use of equation \eqref{eq:CommutantProperty} that
		\begin{align}
			&=\delta_g^L W_{g,1}\delta_h^L W_{g,1}^\dagger \underline{\delta_g^R W_{g,1}\delta_h^R W_{g,1}^\dagger} u_{L,1}(g,h)(\delta_{gh}^L)^\dagger  u_{R,1}(g,h)(\delta_{gh}^R)^\dagger\\
			\nonumber
			&=\delta_g^L \delta_g^R W_{g,1}\delta_h^L \stkout{ W_{g,1}^\dagger W_{g,1}}\delta_h^R W_{g,1}^\dagger u_{L,1}(g,h)  u_{R,1}(g,h)\underline{(\delta_{gh}^L)^\dagger}(\delta_{gh}^R)^\dagger\\
			\nonumber
			&=\delta_g W_{g,1}\delta_h \stkout{W_{g,1}^\dagger W_{g,1}} W_{h,1} W_{gh,1}^\dagger(\delta_{gh})^\dagger=W_{g,2}W_{h,2}W_{gh,2}^\dagger
		\end{align}
		concluding the proof of equation \eqref{eq:TransformationUnderDeltaProofNewUIsSplitAppendix}. Using this we obtain:
		\begin{align}
			&K_{g,2}^RW_{g,2}K_{h,2}^{R}W_{h,2}W_{gh,2}^\dagger(K_{gh,2}^R)^\dagger W_{gh,2}W_{h,2}^\dagger W_{g,2}^\dagger u_{R,2}(g,h)=K_{g,2}^RW_{g,2}K_{h,2}^{R}W_{h,2}W_{gh,2}^\dagger(K_{gh,2}^R)^\dagger u_{L,2}(g,h)^\dagger\\
			=&K_{g,2}^RW_{g,2}K_{h,2}^{R}W_{h,2}W_{gh,2}^\dagger u_{L,2}(g,h)^\dagger (K_{gh,2}^R)^\dagger=K_{g,2}^RW_{g,2}K_{h,2}^{R}\stkout{W_{h,2}W_{gh,2}^\dagger W_{gh,2}W_{h,2}^\dagger} W_{g,2}^\dagger u_{R,2}(g,h) (K_{gh,2}^R)^\dagger.
		\end{align}
		Filling this in the definition of the 2-cochain gives
		\begin{align}
			\nonumber
			C_2(g,h)&=K_{g,2}^RW_{g,2}K_{h,2}^{R}W_{g,2}^\dagger u_{R,2}(g,h) (K_{gh,2}^R)^\dagger v^\dagger u_{R,2}(g,h)^\dagger v\\
			&=v^\dagger \delta_g^R v K_{g,1}^R (\delta_g^R)^\dagger \delta_g W_{g,1}v^\dagger \delta_h^R v K_{h,1}^R (\delta_h^R)^\dagger W_{g,1}^\dagger \delta_g^\dagger\\
			\nonumber
			&\qquad \delta_g^R W_{g,1}\delta_h^R W_{g,1}^\dagger u_{R,1}(g,h)\stkout{(\delta_{gh}^R)^\dagger \delta_{gh}^R} (K_{gh,1}^R)^\dagger v^\dagger \stkout{(\delta_{gh}^R)^\dagger v  v^\dagger \delta^R_{gh}} u_{R,1}(g,h)^\dagger W_{g,1}(\delta^R_h)^\dagger W_{g,1}^\dagger(\delta_g^R)^\dagger v.
		\end{align}
		We now insert $(K_{h,1}^R)^\dagger W_{g,1}^\dagger (K_{g,1}^R)^\dagger K_{g,1}^R W_{g,1}K_{h,1}^R=\id:$
		\begin{align}
			\nonumber
			C_2(g,h)&=v^\dagger \delta_g^R v K_{g,1}^R (\delta_g^R)^\dagger \delta_g W_{g,1}v^\dagger \delta_h^R v K_{h,1}^R (\delta_h^R)^\dagger W_{g,1}^\dagger \delta_g^\dagger\delta_g^R W_{g,1}\delta_h^R \left((K_{h,1}^R)^\dagger W_{g,1}^\dagger (K_{g,1}^R)^\dagger K_{g,1}^R W_{g,1}K_{h,1}^R\right)\\	
			&\qquad  W_{g,1}^\dagger u_{R,1}(g,h) (K_{gh,1}^R)^\dagger v^\dagger u_{R,1}(g,h)^\dagger v v^\dagger W_{g,1}(\delta^R_h)^\dagger W_{g,1}^\dagger(\delta_g^R)^\dagger v.
		\end{align}
		Filling in the old index now gives:
		\begin{align}
			C_2(g,h)&=\underline{C_1(g,h)}v^\dagger \delta_g^R v K_{g,1}^R (\delta_g^R)^\dagger \delta_g W_{g,1}v^\dagger \delta_h^R v K_{h,1}^R (\delta_h^R)^\dagger W_{g,1}^\dagger \delta_g^\dagger\\
			\nonumber
			&\qquad \delta_g^R W_{g,1}\delta_h^R (K_{h,1}^R)^\dagger W_{g,1}^\dagger (K_{g,1}^R)^\dagger \underline{\id} v^\dagger W_{g,1}(\delta^R_h)^\dagger W_{g,1}^\dagger(\delta_g^R)^\dagger v
		\end{align}
		Using equation \eqref{eq:CommutantProperty} we now get:
		\begin{align}
			C_2(g,h)&=C_1(g,h)v^\dagger \delta_g^R v K_{g,1}^R \stkout{\delta_g^L} W_{g,1}v^\dagger \delta_h^R v K_{h,1}^R (\delta_h^R)^\dagger W_{g,1}^\dagger \stkout{(\delta_g^L)^\dagger}\\
			\nonumber
			&\qquad  W_{g,1}\delta_h^R (K_{h,1}^R)^\dagger W_{g,1}^\dagger (K_{g,1}^R)^\dagger v^\dagger W_{g,1}(\delta^R_h)^\dagger W_{g,1}^\dagger(\delta_g^R)^\dagger v\\
			\nonumber
			&=C_1(g,h)v^\dagger \delta_g^R v K_{g,1}^\stkout{R} W_{g,1}v^\dagger \delta_h^R v \stkout{K_{h,1}^R (\delta_h^R)^\dagger W_{g,1}^\dagger W_{g,1}\delta_h^R (K_{h,1}^R)^\dagger} W_{g,1}^\dagger (K_{g,1}^\stkout{R})^\dagger v^\dagger W_{g,1}(\delta^R_h)^\dagger W_{g,1}^\dagger(\delta_g^R)^\dagger v\\
			\nonumber
			&=C_1(g,h)v^\dagger \delta_g^R v K_{g,1} W_{g,1}v^\dagger \delta_h^R v W_{g,1}^\dagger (K_{g,1})^\dagger v^\dagger W_{g,1}(\delta^R_h)^\dagger W_{g,1}^\dagger(\delta_g^R)^\dagger v.
		\end{align}
		Using the fact that by definition $K_{g,1}W_{g,1}=v^\dagger W_{g,1}v$ we obtain
		\begin{align}
			C_2(g,h)&=C_1(g,h)v^\dagger \delta_g^R \underline{W_{g,1}} \delta_h^R \underline{W_{g,1}^\dagger} W_{g,1}(\delta^R_h)^\dagger W_{g,1}^\dagger(\delta_g^R)^\dagger v=C_1(g,h)
		\end{align}
		concluding the proof of lemma \ref{lem:TransformationUnderDelta}. We now proceed with the proof of lemma \ref{lem:TransformationUnderDeltaTwoTranslations}. We define the index in function of the new variables:
		\begin{align}
			\nonumber
			\alpha_2(g)&\defeq v^\dagger b^R_{g,2}v h_g K^R_{g,2}(b^R_{2,g})^\dagger w^\dagger (K^R_{2,g})^\dagger w\\
			&=v^\dagger \left(w^\dagger\delta^R_g w b_{g,1}^R (\delta^R_g)^\dagger\right) v h_g \left(v^\dagger \delta_g^R v K_{g,1}^R (\delta_g^R)^\dagger\right) \left(w^\dagger\delta^R_g w b_{g,1}^R (\delta^R_g)^\dagger\right)^\dagger w^\dagger (v^\dagger \delta_g^R v K_{g,1}^R (\delta_g^R)^\dagger)^\dagger w\\
			\nonumber
			&=v^\dagger w^\dagger\delta^R_g w b_{g,1}^R (\delta^R_g)^\dagger v h_g v^\dagger \delta_g^R v K_{g,1}^R \stkout{(\delta_g^R)^\dagger \delta_g^R} (b_{1,g}^R)^\dagger w^\dagger \stkout{(\delta_g^R)^\dagger w w^\dagger \delta_g^R} (K_{g,1}^R)^\dagger v^\dagger (\delta_g^R)^\dagger v  w.
		\end{align}
		If we now insert $vv^\dagger,$ we obtain
		\begin{align}
			\alpha_2(g)&=v^\dagger w^\dagger\delta^R_g w b_{g,1}^R v\stkout{v^\dagger (\delta^R_g)^\dagger v} h_g \stkout{v^\dagger \delta_g^R v} K_{g,1}^R (b_{1,g}^R)^\dagger w^\dagger (K_{g,1}^R)^\dagger v^\dagger (\delta_g^R)^\dagger v  w
		\end{align}
		where the cancelation is due to the fact that $v^\dagger \delta^R_g v\in\textrm{IAC}_{R}(\alpha_0,\theta,\Theta)$ whereas $h_g\in \textrm{IAC}_{L}(\alpha_0,\theta,\Theta)$. Again after inserting $vv^\dagger$ and $ww^\dagger$ we now get
		\begin{align}
			\alpha_2(g)&=v^\dagger w^\dagger\delta^R_g w \underline{v v^\dagger} b_{g,1}^R v h_g K_{g,1}^R (b_{1,g}^R)^\dagger w^\dagger (K_{g,1}^R)^\dagger \underline{ww^\dagger} v^\dagger (\delta_g^R)^\dagger v  w\\
			\nonumber
			&=\stkout{v^\dagger w^\dagger\delta^R_g w v} \alpha_1(g) \stkout{w^\dagger v^\dagger (\delta_g^R)^\dagger v  w}=\alpha_1(g)
		\end{align}
		concluding the proof.
	\end{proof}
	We will now show that the index is independent of the choice of $\tilde{\beta}_g$ and its decomposition.
	\begin{lemma}\label{lem:InvarianceUnderChoiceBeta}
		Take $\tilde{\beta}_{g,1},\tilde{\beta}_{g,2}\in\Aut{\AA},V_{g,21},V_{g,22}\in\UU(\AA),\eta_{g,1}^L,\eta_{g,2}^L\in \Aut{\AA_{\nu^{-1}(L\cap C_\theta)}}$ and $\eta_{g,1}^R,\eta_{g,2}^R\in \Aut{\AA_{\nu(R\cap C_\theta)}}$ such that there exist $V_{g,21},V_{g,22}\in\UU(\AA)$ satisfying
		\begin{align}\label{eq:InvarianceUnderChoiceBetaTwoBetas}
			\tilde{\beta}_{g,1}&=\Ad{V_{g,21}}\circ\eta_{g,1}\circ\beta^U_g&\tilde{\beta}_{g,2}&=\Ad{V_{g,22}}\circ\eta_{g,2}\circ\beta^U_g&&\text{and}&\omega\circ\tilde{\beta}_{g,1}&=\omega\circ\tilde{\beta}_{g,2}=\omega
		\end{align}then
		\begin{equation}
			\textrm{Index}(\theta,\tilde{\beta}_{g,1},\eta_{g,1},\alpha_{0},\Theta,\omega,\omega_0)=\textrm{Index}(\theta,\tilde{\beta}_{g,2},\eta_{g,2},\alpha_{0},\Theta,\omega,\omega_0).
		\end{equation}
		If additionally, $\omega$ satisfies assumption \ref{assumption:2Translations},	then
		\begin{equation}
			\textrm{Index}_{\text{2 trans}}(\theta,\tilde{\beta}_{g,1},\eta_{g,1},\alpha_{0},\Theta,\omega,\omega_0)=\textrm{Index}_{\text{2 trans}}(\theta,\tilde{\beta}_{g,2},\eta_{g,2},\alpha_{0},\Theta,\omega,\omega_0).
		\end{equation}
	\end{lemma}
	\begin{proof}
		Take $\alpha=\Ad{V_1}\circ\alpha_{0}\circ\Theta$ the usual decomposition. Since $\omega_0\circ\alpha\circ\tilde{\beta}_{g,1}\circ(\tilde{\beta}_{g,2})^{-1}=\omega_0\circ\alpha$, there exist $\tilde{\delta}_g\in\UU(\HH_0)$ such that: $\Ad{\tilde{\delta}_g}\circ\pi_0\circ\alpha=\pi_0\circ\alpha\circ\tilde{\beta}_{g,2}\circ(\tilde{\beta}_{g,1})^{-1}$. Inserting equation \eqref{eq:InvarianceUnderChoiceBetaTwoBetas} and the decomposition of $\alpha$ in this gives
		\begin{equation}
			\Ad{\tilde{\delta}_g}\circ\pi_0\circ\Ad{V_1}\circ\alpha_0\circ\Theta=\pi_0\circ\Ad{V_1}\circ\alpha_0\circ\Theta\circ\Ad{V_{g,21}}\circ\eta_{g,1}\circ\eta_{g,2}^{-1}\circ\Ad{V_{g,22}^\dagger}.
		\end{equation}
		Putting the $V_1$'s to the other side gives:
		\begin{equation}
			\Ad{\pi_0(V_1^\dagger) \tilde{\delta}_g\pi_0(V_1)}\circ\pi_0\circ\alpha_0\circ\Theta=\pi_0\circ\alpha_0\circ\Theta\circ\Ad{V_{g,21}}\circ\eta_{g,1}\circ\eta_{g,2}^{-1}\circ\Ad{V_{g,22}^\dagger}.
		\end{equation}
		Doing the same for the $V_{g,21}$ and $V_{g,22}$ now gives
		\begin{align}
			\Ad{\pi_0\circ\alpha_0\circ\Theta(V_{g,21}^\dagger)\pi_0(V_1^\dagger)\tilde{\delta}_g\pi_0(V_1)}\circ\pi_0\circ\alpha_0\circ\Theta\circ\Ad{V_{g,22}}&=\pi_0\circ\alpha_0\circ\Theta\circ\eta_{g,2}\circ(\eta_{g,1})^{-1}\\
			\Ad{\pi_0\circ\alpha_0\circ\Theta(V_{g,21}^\dagger)\pi_0(V_1^\dagger)\tilde{\delta}_g\pi_0(V_1)\pi_0\circ\alpha_0\circ\Theta(V_{g,22})}\circ\pi_0&=\pi_0\circ\alpha_0\circ\eta_{g,2}\circ(\eta_{g,1})^{-1}\circ\alpha_0^{-1}.
		\end{align}
		Since the last equation is split we have by lemma \ref{lem:UsingIrreducibilityAndWignerTheorem} that we can take $\delta_g^L\in\textrm{Cone}_{\nu^{-1}(L)}^W(\alpha_0,\theta)$ and $\delta_g^R\in\textrm{Cone}_{\nu(R)}^W(\alpha_0,\theta)$ such that
		\begin{equation}
			\delta_g^L\otimes\delta_g^R=\pi_0\circ\alpha_0\circ\Theta(V_{g,21}^\dagger)\pi_0(V_1^\dagger)\tilde{\delta}_g\pi_0(V_1)\pi_0\circ\alpha_0\circ\Theta(V_{g,22}).
		\end{equation}
		Take $W_{g,1},u_{R,1}(g,h)$ and $K^R_{g,1}$ to be the operators belonging to the first choice (with arbitrary phases). Define
		\begin{align}
			W_{g,2}&\defeq\delta_g W_{g,1}&u_{R,2}(g,h)&\defeq \delta_g^R W_{g,1}\delta_h^R W_{g,1}^\dagger u_{R,1}(g,h)(\delta_{gh}^R)^\dagger&K^R_{g,2}&\defeq v^\dagger \delta_g^R v K_{g,1}^R (\delta_g^R)^\dagger
		\end{align}
		then by construction $W_{g,2},u_{R,2}(g,h)$ and $K^R_{g,2}$ are operators belonging to the second choice. Now using lemma \ref{lem:TransformationUnderDelta} concludes the proof. The proof of the second part is completely analogous, only we now define, $b_{g,2}^\sigma\defeq w^\dagger\delta^\sigma_g w b_{g,1}^\sigma (\delta^\sigma_g)^\dagger$ as well and conclude the proof by using lemma \ref{lem:TransformationUnderDeltaTwoTranslations}.
	\end{proof}
	\begin{lemma}
		Both indices are independent of the choice of angle $\theta$.
	\end{lemma}
	\begin{proof}
		Take $0<\theta_1<\theta_2<\pi/2$. Take $\alpha\in \QAut{\AA}$ and (for all $\sigma\in\{L,R\}$) take $\eta_{g,1}^\sigma\in\Aut{\AA_{\nu^{\sigma}(\sigma\cap C_{\theta_1})}}$ and $\eta_{g,2}^\sigma\in\Aut{\AA_{\nu^{\sigma}(\sigma\cap C_{\theta_2})}}$ to be operators belonging to $\tilde{\beta}_{g,1}$ and $\tilde{\beta}_{g,2}$ leaving $\omega$ invariant. Since $\Aut{\AA_{\nu^{\sigma}(\sigma\cap C_{\theta_1})}}\subset \Aut{\AA_{\nu^{\sigma}(\sigma\cap C_{\theta_2})}}$ the result now follows from lemma \ref{lem:InvarianceUnderChoiceBeta}.
	\end{proof}
	Due to all these considerations, from here onward, we will write the $H^2(G,\TT)$ and $H^1(G,\TT)$-valued indices as $\textrm{Index}^{\AA,U}(\omega)$ and $\textrm{Index}^{\AA,U}_{\text{2 trans}}(\omega)$ respectively.
	\section{Examples}\label{sec:examples}
	In subsections \ref{sec:ExampleOneTranslation} and \ref{sec:ExampleTwoTranslations} we will give a special class of translation invariant states that satisfy assumptions \ref{assumption} and \ref{assumption:2Translations} respectively. We start our construction by taking a state $\phi\in\PP(\AA_{L_0})$ that satisfies assumption \ref{assumption1d}. This state satisfies the following lemma:
	\begin{lemma}\label{lem:phiExampleConsistentWithConjectureSplitProperty}
		This $\phi$ satisfies the split property.
	\end{lemma}
	\begin{proof}
		Take $(\HH_L\otimes\HH_R,\pi_L\otimes\pi_R,\Omega_L\otimes\Omega_R)$ a GNS triple of $\phi_0$. By construction $(\HH_L\otimes\HH_R,\pi_L\circ\tilde\alpha_L\otimes\pi_R\circ\tilde\alpha_R,\pi_L\otimes\pi_R(b^\dagger)\Omega_L\otimes\Omega_R)$ is a GNS triple of $\phi$. This concludes the proof that $\phi$ satisfies the split property.
	\end{proof}
	For any $j\in\ZZ$ and $n\in\NN$, take
	\begin{align}
		L_j&\defeq \{(x,y)\in\ZZ^2|y=j\}&L_{j}^{n}&\defeq \{(x,y)\in\ZZ^2|y=j,x\in[-n,n]\}\\
		L_{<j}&\defeq \{(x,y)\in\ZZ^2|y<j\}&L_{>j}&\defeq \{(x,y)\in\ZZ^2|y>j\}.
	\end{align}
	We introduce some additional notation, for any automorphism $\alpha\in\Aut{\AA_{L_0}}$ we define $\alpha_{(\cdot,j)}\defeq \tau^j\circ\alpha\circ\tau^{-j}$. We will now define the infinite tensor product state.
	\begin{definition}\label{def:InfiniteTensorProductState}
		Take $\phi\in\PP(\AA_{L_0})$ arbitrary. There exists a unique state $\omega$ satisfying that for any $i\in\ZZ$ we have that for any $A\in\AA_{L_i}$ and $B\in\AA_{L_i^c}$, $\omega(A\otimes B)=\phi\circ\tau^{-i}(A)\omega(B)$.
	\end{definition}
	\begin{proof}
		This condition gives a construction by induction for $\omega(A)$ for each $A\in\AA_{\text{loc}}$. For general elements we define $\omega(A)\defeq \lim_{n\rightarrow\infty}\omega(A_n)$ where $A_n$ is a sequence in $\AA_{\text{loc}}$ converging in $\AA$.
	\end{proof}
	In what follows, we will fix a GNS triple for $\omega_0$, $(\HH_0,\pi_0,\Omega_0)$ that is of the form: $\HH_0=\bigotimes_{\rho\sigma}\HH_{\rho,\sigma}$, $\pi_0=\bigotimes_{\rho\sigma}\pi_{\rho,\sigma}$ and $\Omega_0=\bigotimes_{\rho\sigma}\Omega_{\rho,\sigma}$, where the $\pi_{\rho,\sigma}:\AA_{\rho\cap\sigma}\rightarrow\BB(\HH_{\rho,\sigma})$ are all irreducible representations for all $\rho\in \{L_{>-1},L_{-1},L_{<-1}\}$ and $\sigma\in\{L,R\}$.
	\begin{remark}\label{rem:GNS_One_Dimensional}
		We have (by construction) that $\pi_{L_{-1},L}\circ\alpha_L|_{L_{-1}}\otimes \pi_{L_{-1},R}\circ\alpha_R|_{L_{-1}}$ is a GNS representation of $\phi$. 
	\end{remark}
	\subsection{Example with non-trivial \texorpdfstring{$H^2(G,\TT)$}{}-valued index}\label{sec:ExampleOneTranslation}
	By the result of \cite{ogata2019classification}, lemma \ref{lem:phiExampleConsistentWithConjectureSplitProperty} implies that $\phi$ has a well-defined 1d SPT index. The following lemma says that the infinite tensor product state also has a well-defined $H^2(G,\TT)$-valued index.
	\begin{lemma}\label{lem:TensorProductStateHasWellDefinedH^2Index}
		Let $\omega$ be the infinite tensor product of a state $\phi$ that satisfies assumption \ref{assumption1d} then $\omega$ satisfies assumption \ref{assumption}.
	\end{lemma}
	\begin{proof}
		To show that $\tilde{\beta}_g$ exists note that we can take $\tilde{\beta}_g$ to be simply $\beta_g^{U}$. The translation invariance was true by construction. We now only have to find an $\alpha\in\QAut{\AA}$ that satisfies $\omega=\omega_0\circ\alpha$. We define $\alpha$ as the infinite product automorphism of $\tilde{\alpha}_L^{-1}\otimes\tilde{\alpha}_R^{-1}\circ\Ad{b^\dagger}$. This Automorphism indeed satisfies that $\omega_0\defeq \omega\circ\alpha^{-1}$ is a product state. This is because for any $A\in\AA_{L_{i}}$, $B\in\AA_{L_{j}}$ with $i\neq j$ we have that $\omega_0(A\otimes B)=\phi_0(A)\phi_0(B)$. We will now show that $\alpha\in\QAut{\AA}$. To do this, take $0<\theta<\pi/2$ and let $b_j$ be a summable sequence converging to $b$ such that $b_j\in\UU(\AA_{L_0^{n(j)}})$ where $n(j)\in\NN$ is the largest number such that $L_{j}^{n(j)}\subset W(C_\theta)^c$. Define $\Theta\in\Aut{\AA_{W(C_\theta)^c}}$ through $\Theta=\bigotimes_{j\in\ZZ}\Ad{b_\abs{j}^\dagger}_{(\cdot,j)}$. Let $\alpha_\sigma\in\Aut{\AA_\sigma}$ be $\alpha_\sigma\defeq \bigotimes_{j\in\ZZ}(\tilde\alpha_\sigma^{-1})_{(\cdot,j)}$. Now define $\tilde V_{1,m}\defeq\bigotimes_{j\in\ZZ\cap[-m,m]}\tau^j (b^\dagger b_\abs{j})$. We will now show that the limit $\tilde V_1\defeq\lim_{m\rightarrow\infty}\tilde V_{1,m}$ exists (is an element of $\AA$). By construction we have that for any $\epsilon$ there exists an $n_0>0$ such that $\epsilon_n\defeq\sum_{i=n}^\infty\norm{b-b_i}<\epsilon$ for all $n>n_0$. We will now show that our sequence is a Cauchy sequence. Let $\epsilon>0$ and take $n_0$ accordingly. For any $n,m\geq n_0$ we have that
		\begin{align}
			\norm{\tilde{V}_{1,n}-\tilde{V}_{1,m}}&\leq \norm{\tilde{V}_{1,n}-\tilde{V}_{1,n_0}}+\norm{\tilde{V}_{1,m}-\tilde{V}_{1,n_0}}=\norm{\tilde{V}_{1,n}\tilde{V}_{1,n_0}^\dagger-\id}+\norm{\tilde{V}_{1,m}\tilde{V}_{1,n_0}^\dagger-\id}.
		\end{align}
		We will find a bound on the first term as the bound of the second term is analogous. First notice that we have for any tensor product that
		\begin{equation}
			A\otimes B-C\otimes D=(A-C+C)\otimes B-C\otimes D=(A-C)\otimes B+C\otimes (B-D).
		\end{equation}
		By using this property and the triangle inequality recursively we get
		\begin{align}
			&\norm{\bigotimes_{i=-n}^{-n_0}(b^\dagger b_\abs{i})_i\bigotimes_{i=n_0}^{n}(b^\dagger b_i)_i-\id_{\HH_{L_{-n,\ldots,-n_0}}}\otimes\id_{\HH_{L_{n_0,\ldots,n}}}}\leq 2\sum_{i=n_0}^{n}\norm{b^\dagger b_i-\id_{\HH_{L_i}}}<2\epsilon.
		\end{align}
		This proves that the sequence is a Cauchy sequence and hence the convergence follows. To conclude the proof, we merely have to define $V_1\defeq \alpha_0(\tilde{V}_1)$. We then get
		\begin{equation}
			\Ad{V_1}\circ \alpha_0\circ \Theta= \alpha_0\circ\Ad{\tilde V_1}\circ=\bigotimes_{j\in\ZZ}\tilde\alpha_0^{-1}\circ\Ad{b^\dagger b_\abs{j}}\circ \Ad{b_\abs{j}^\dagger}_{(\cdot,j)}=\bigotimes_{j\in\ZZ}\tilde\alpha_0^{-1}\circ\Ad{b^\dagger}_{(\cdot,j)}=\alpha
		\end{equation}
		concluding the proof.
	\end{proof}
	This shows that $\omega$ has a well-defined 2D translation index. We will now show that $\textrm{Index}(\omega)=\textrm{Index}_{1d}(\phi)$.
	\begin{remark}
		Remark \ref{rem:GNS_One_Dimensional} implies that if we take any $k^R:G\rightarrow \UU(\HH_{L_{-1},R})$ such that
		\begin{equation}
			\Ad{k^R(g)}\circ\pi_{L_{-1},R}\circ\alpha_R|_{L_{-1}}=\pi_{L_{-1},R}\circ\alpha_R|_{L_{-1}}\circ\beta_{g}^{L_{-1}\cap R}
		\end{equation}
		then there exists a 2 cochain $C$ satisfying that $k^R(g)k^R(h)k^R(gh)^{-1}=C(g,h)\id_{\HH_{L_{-1},R}}$. Furthermore, the equivalence class this cochain is in, is given by $\textrm{Index}_{1d}(\phi)$.
	\end{remark}
	Let $\tilde k:G\rightarrow \UU(\HH_{L_{-1}})$ be the unique operator that satisfies
	\begin{align}
		\Ad{\tilde{k}_g}\circ\pi_{L_{-1}}&=\pi_{L_{-1}}\circ \circ\alpha|_{L_{-1}}\circ\beta_g^{L_{-1}}\circ \alpha^{-1}|_{L_{-1}}&\tilde k_g\Omega_{L_{-1}}&=\Omega_{L_{-1}}.
	\end{align}
	Similarly, let $\tilde{w}:G\rightarrow\UU(\HH_{L_{>-1},L}\otimes\HH_{L_{>-1},R})$ be the unique operator such that
	\begin{align}
		\Ad{\tilde w_{g}}\circ\pi_{L_{>-1}}&=\pi_{L_{>-1}}\circ\alpha|_{L_{>-1}}\circ\beta_g^{U}\circ \alpha^{-1}|_{L_{>-1}}&\tilde w_{g}\Omega_{L_{>-1}}&=\Omega_{L_{>-1}}.
	\end{align}
	Clearly, both functions are (linear) representations of the group. Now define
	\begin{equation}\label{eq:ChoiceOfW_g_ExampleH2Index}
		W_g=\pi_{0}(V_1)^{\dagger}(\tilde{w}_g\otimes\id_{H_{L_{-1}}}\otimes\id_{\HH_{L_{<-1}}}) \pi_{0}(V_1).
	\end{equation}
	This operator-valued function is also a linear representation and satisfies the conditions indicated in lemma \ref{lem:Definition_W_And_u}. Since $W_g$ is a linear representation we get that $W_gW_hW_{gh}^{-1}=\id_{\HH_0}$. This implies that we can take $u_\sigma(g,h)=\bigotimes_{\rho\in\{L_{>-1},L_{-1},L_{<-1}\}}\id_{\HH_{\sigma,\rho}}$ (for any $\sigma\in\{L,R\}$). All that is now left is to calculate $K^\sigma_g\in\UU(\HH_\sigma)$. We get that
	\begin{align}
		K_g&=v^\dagger W_g v W_g^\dagger=v^\dagger \pi_{0}(V_1)^{\dagger}(\tilde{w}_g\otimes\id_{\HH_{L_{-1}}}\otimes\id_{\HH_{L_{<-1}}}) \pi_{0}(V_1) v W_g^\dagger\\
		&=\pi_{0}(V_1)^{\dagger}\tilde v^\dagger(\tilde{w}_g\otimes\id_{\HH_{L_{-1}}}\otimes\id_{\HH_{L_{<-1}}}) \tilde v \pi_{0}(V_1) W_g^\dagger.
	\end{align}
	Notice however that $\tilde v^\dagger(\tilde{w}_g\otimes\id_{\HH_{L_{-1}}}\otimes\id_{\HH_{L_{<-1}}}) \tilde v$ and $\tilde{w}_g\otimes\tilde{k}_g\otimes\id_{\HH_{L_{<-1}}}$ both leave the cyclic vector invariant and have the same adjoint action on the GNS representation. By the uniqueness of such operators, we get that
	\begin{equation}
		K_g=\pi_{0}(V_1)^{\dagger}(\tilde{w}_g\otimes\tilde{k}_g\otimes\id_{\HH_{L_{<-1}}}) \pi_{0}(V_1) W_g^\dagger.
	\end{equation}
	Using the adjoint of equation \eqref{eq:ChoiceOfW_g_ExampleH2Index} this gives
	\begin{align}
		K_g&=\pi_{0}(V_1)^{\dagger}(\id_{H_{L_{>-1}}}\otimes \tilde k_g \otimes\id_{\HH_{L_{<-1}}}) \pi_{0}(V_1)=\id_{H_{L_{>-1}}}\otimes \pi_{L_{-1}}(b)^\dagger \tilde k_g \pi_{L_{-1}}(b) \otimes\id_{\HH_{L_{<-1}}}
	\end{align}
	where $\tilde{v}=\pi_0(V_1)^\dagger v\pi_0(V_1)$. Now there exists a $k^L:G\rightarrow \UU(\HH_{L_{-1},L})$ and a $k^R:G\rightarrow \UU(\HH_{L_{-1},R})$ such that
	\begin{equation}
		k^L_g\otimes k^R_g=\pi_{L_{-1}}(b)^\dagger \tilde k_g \pi_{L_{-1}}(b)
	\end{equation}
	and these $k^\sigma_g$ will satisfy what was written in remark \ref{rem:GNS_One_Dimensional}. Define $K^\sigma_g=k^\sigma_g\otimes\id_{\HH_{L_{>-1},\sigma}}\otimes\id_{\HH_{L_{<-1},\sigma}}$. Notice that we have (by construction) that $\forall g,h\in G$, $[K^\sigma_g\otimes\id_{\HH_{\ZZ^2/\sigma}},W_h]=0$. We now get (using this equation and the fact that $W_g$ is a representation)
	\begin{align}
		&K_g^R W_g K_h^R W_h W_{gh}^\dagger (K_{gh}^R)^\dagger \stkout{W_{gh}W_h^\dagger W_g^\dagger u_R(g,h)}=\stkout{W_g W_h W_{gh}^\dagger} K_g^R K_h^R (K_{gh}^R)^\dagger\\
		\nonumber
		&=k_g^R k_h^R (k_{gh}^R)^\dagger\otimes \id_{\HH_{L_{>-1},\sigma}}\otimes\id_{\HH_{L_{<-1},\sigma}}
	\end{align}
	and by remark \ref{rem:GNS_One_Dimensional} this gives that $C(g,h)=\tilde{C}(g,h)$.
	\subsection{Example with non-trivial \texorpdfstring{$H^1(G,\TT)$}{}-valued index}\label{sec:ExampleTwoTranslations}
	This example will be very similar to what was done in subsection \ref{sec:ExampleOneTranslation}. In this case, however, we will require that the translation morphism $\nu$ acts as an automorphism on the $C^*$ algebra $\AA_{L_i}$ (for all $i$) and that our one-dimensional state is invariant under this automorphism. In this part, we let $\phi\in\PP(\AA)$ satisfy assumption \ref{assumption1dWithTranslation}. Since this state satisfies the split property (see lemma \ref{lem:phiExampleConsistentWithConjectureSplitProperty}) we can define an $H^1(G,\TT)$-valued index for it (see section \ref{sec:OneDimensionalIndices} for a construction of this index). Now we will take again the infinite tensor product (see definition \ref{def:InfiniteTensorProductState}) of such states then we get:
	\begin{lemma}
		Let $\omega_\phi$ be the infinite tensor product of a state $\phi$ that satisfies assumption \ref{assumption1dWithTranslation} then $\omega_\phi$ satisfies assumption \ref{assumption:2Translations}.
	\end{lemma}
	\begin{proof}
		This lemma follows from lemma \ref{lem:TensorProductStateHasWellDefinedH^2Index}.
	\end{proof}
	This shows that $\omega_\phi$ indeed has a well-defined $H^1(G,\TT)$ index. We will now show that $\textrm{Index}_{\text{2 trans}}(\omega_\phi)=\textrm{Index}_{\text{1d trans}}(\phi)$.
	\begin{remark}\label{rem:GNS_One_DimensionalTwoTranslations}
		Remark \ref{rem:GNS_One_Dimensional} implies that if we take any $k^R:G\rightarrow \HH_{L_{-1},R}$ such that
		\begin{equation}
			\Ad{k^R(g)}\circ\pi_{L_{-1},R}\circ\alpha_R|_{L_{-1}}=\pi_{L_{-1},R}\circ\alpha_R|_{L_{-1}}\circ\beta_{g}^{L_{-1}\cap R}
		\end{equation}
		and $w_{L_{-1}}$ to be such that
		\begin{equation}
			\Ad{w_{L_{-1}}}\circ\pi_{L_{-1},L}\otimes \pi_{L_{-1},R}=\pi_{L_{-1},L}\otimes \pi_{L_{-1},R}\circ\nu
		\end{equation}
		then there exists a $U(1)$ representation $\alpha$ satisfying that
		\begin{equation}
			(\pi_{L_{-1},L}(U_{(-1,-1)}(g))\otimes \id_{\HH_{L_{-1},R}}) (\id_{\HH_{L_{-1},L}}\otimes k^R(g))=\alpha(g) w_{L_{-1}}^\dagger(\id_{\HH_{L_{-1},L}}\otimes k^R(g))w_{L_{-1}}
		\end{equation}
		and that it is given by $\textrm{Index}_{\text{1d trans}}(\phi)$.
	\end{remark}
	Just like in subsection \ref{sec:ExampleOneTranslation}, let $\tilde k:G\rightarrow \UU(\HH_{L_{-1}})$ be the unique operator that satisfies
	\begin{align}
		\Ad{\tilde{k}_g}\circ\pi_{L_{-1}}&=\pi_{L_{-1}}\circ \circ\alpha|_{L_{-1}}\circ\beta_g^{L_{-1}}\circ \alpha^{-1}|_{L_{-1}}&\tilde k_g\Omega_{L_{-1}}&=\Omega_{L_{-1}}.
	\end{align}
	We can take
	\begin{equation}
		K_g=\id_{H_{L_{>-1}}}\otimes \pi_{L_{-1}}(b)^\dagger \tilde k_g \pi_{L_{-1}}(b) \otimes\id_{\HH_{L_{<-1}}} =\id_{H_{L_{>-1}}}\otimes (k_g^L\otimes k_g^R) \otimes\id_{\HH_{L_{<-1}}}.
	\end{equation}
	Similarly we can take $w_\rho\in\HH_\rho$ (where $\rho\in\{L_{>-1},L_{-1},L_{<-1}\}$) satisfying
	\begin{equation}
		\Ad{w_\rho}\circ\pi_\rho\circ\alpha_{0}|_\rho\circ\Theta|_\rho=\pi_\rho\circ\alpha_{0}|_\rho\circ\Theta|_\rho\circ\nu
	\end{equation}
	and then we get that $w=w_{L_{>-1}}\otimes w_{L_{-1}}\otimes w_{L_{<-1}}$. Since $\eta_g=\textrm{Id}_{\AA}$ we can simply take $b_g^\sigma=\id_{\HH_\sigma}$. We now have that
	\begin{equation}
		\pi_0\circ\alpha_0\circ\Theta(U_{(-1,-1)})\id_{\HH_L}\otimes K_g^R=\id_{\HH_{L_{>-1}}}\otimes(\pi_{L_{-1},L}\circ\tilde\alpha_L(U_{(-1,-1)})\otimes k^R(g))\otimes\id_{\HH_{L_{>-1}}}.
	\end{equation}
	By what was written in remark \ref{rem:GNS_One_DimensionalTwoTranslations} we get that
	\begin{align}
		&=\id_{H_{L_{>-1}}}\otimes\alpha(g) w_{L_{-1}}(\id_{\HH_{L_{-1},L}}\otimes k^R(g))w_{L_{-1}}^\dagger\otimes \id_{H_{L_{<-1}}}\\
		\nonumber
		&=\alpha(g)w^\dagger (\id_{H_{L_{>-1}}}\otimes \id_{\HH_{L_{-1},L}}\otimes k^R(g)\otimes \id_{H_{L_{<-1}}}) w=\alpha(g)w^\dagger (\id_{\HH_L}\otimes K^R_g) w
	\end{align}
	showing that indeed $\textrm{Index}_{\text{2 trans}}(\omega_\phi)=\textrm{Index}_{\text{1d trans}}(\phi)$.
	\section{Invariance of the indices under LGAs}\label{sec:AllIndicesInvariantUnderLGA}
	The goal of this section is now to show that the indices we constructed are invariant under locally generated automorphisms. More specifically, if we let $H\in\BB_{F_\phi}([0,1])$, be a $G$-invariant, translation invariant (in the vertical direction) one-parameter family of interactions, then we will show that $\textrm{Index}(\omega\circ\gamma^H_{0;1})=\textrm{Index}(\omega)$. If additionally, $\omega$ and $H$ are translation invariant in both directions then we will show that $\textrm{Index}_{\text{2 trans}}(\omega\circ\gamma^H_{0;1})=\textrm{Index}_{\text{2 trans}}(\omega)$.
	\subsection{The morphisms \texorpdfstring{$\phi_1$}{} and \texorpdfstring{$\phi_2$}{}}
	The goal of this section is to define a pair of morphisms, $\phi_1$ and $\phi_2$, that will be used for defining the operators belonging to $\omega\circ\gamma^H_{0;1}$. For notational simplicity in this section $\Hsplit=H_{\nu^{-1}(L)}+H_{\nu(R)}$ (with $H_{\nu^{-1}(L)}$ and $H_{\nu(R)}$ the restrictions of $H$ as defined in \ref{sec:Interactions}). Clearly we have that $\gamma^{\Hsplit}_{t;s}=\gamma^{H_{\nu^{-1}(L)}}_{t;s}\otimes\gamma^{H_{\nu(R)}}_{t;s}$ and we still have that $\gamma^{\Hsplit}_{t;s}\circ\tau=\tau\circ\gamma^{\Hsplit}_{t;s}$ (just like we have for $H$). Additionally, in the case where there are two translation symmetries, we get that $\nu(\Hsplit)=H_L+H_{\nu\circ\nu(R)}$ while $\nu^{-1}(\Hsplit)=H_{\nu^{-1}\circ\nu^{-1}(L)}+H_{R}$.\\\\
	For the remainder of the section, we will use a decomposition using the following lemma:
	\begin{lemma}\label{lem:DefinitionOfGroupMorphismVAutEquation}
		Take $0<\theta<\pi/2$ and $\Theta\in \Aut{\AA_{W(C_\theta)^c}}$. There exist $\Phi^U_{0;1}\in\Aut{\AA_{W(C_\theta)^c\cap U}}$ and $\Phi^D_{0;1}\in\Aut{\AA_{W(C_\theta)^c\cap D}}$, both commuting with $\beta_g$, such that there exists some $a\in\UU(\AA)$ satisfying that
		\begin{equation}\label{eq:DefinitionOfGroupMorphismVAutEquation}
			\gamma^H_{0;1}\circ\gamma^{\Hsplit}_{1;0}=\Ad{a}\circ \Phi^U_{0;1}\otimes\Phi^D_{0;1}.
		\end{equation}
	\end{lemma}
	\begin{proof}
		This follows from the fact that $\gamma^H_{0;1}\circ\gamma^{\Hsplit}_{1;0}\in \textrm{GVAut}_1(\AA)$ (see lemma \versionDifference{\ref{lem:PropertiesLocallyGeneratedAutomorphisms}}{C.4. of \cite{jappens2023spt}} part 2 which was based on similar statements made in \cite{ogata2021h3gmathbb}).
	\end{proof}
	In what follows, we will define $\tilde{a}\defeq \pi_0\circ\alpha_0\circ\Theta(a)$. The main property we will use of this operator is that it satisfies the following lemma:
	\begin{lemma}\label{lem:PropertyTilde_a}
		Let $A\in\UU(\AA_{W(C_\theta)\cap\sigma})$ then
		\begin{equation}
			\Ad{\tilde{a}}\circ\pi_0\circ\alpha_0(A)=\Ad{\tilde{a}}\circ\pi_0\circ\alpha_0\circ\Theta(A)=\pi_0\circ\alpha_0\circ\Theta\circ\gamma^H_{0;1}\circ\gamma^{\Hsplit}_{1;0}(A).
		\end{equation}
	\end{lemma}
	\begin{proof}
		Using lemma \ref{lem:DefinitionOfGroupMorphismVAutEquation} we get that
		\begin{align}
			\Ad{\tilde{a}}\circ\pi_0\circ\alpha_0\circ\Theta(A)&=\pi_0\circ\alpha_0\circ\Theta\circ\Ad{a}(A)=\pi_0\circ\alpha_0\circ\Theta\circ\gamma^H_{0;1}\circ\gamma^{\Hsplit}_{1;0}\circ\Phi^U_{1;0}\otimes\Phi^D_{1;0}(A)\\
			\nonumber
			&=\pi_0\circ\alpha_0\circ\Theta\circ\gamma^H_{0;1}\circ\gamma^{\Hsplit}_{1;0}(A)
		\end{align}
		here $\Phi_{1;0}=\Phi_{0;1}^{-1}$. This concludes the proof.
	\end{proof}
	We will then use this property to define a morphism:
	\begin{lemma}\label{lem:DefinitionOfGroupMorphism}
		Let $\tilde{a}=\pi_0\circ\alpha_0\circ\Theta(a)$ and define the map
		\begin{align}
			\phi_{1,\sigma}&:\IAC_\sigma(\alpha_0,\theta,\Theta) \rightarrow \UU(\HH):x=\pi_0\circ\alpha_0\circ\Theta(A)y\mapsto \phi_1(x)=\pi_0\circ\alpha_0\circ\Theta\circ\gamma^H_{0;1}\circ\gamma^{\Hsplit}_{1;0}(A)\tilde{a}y\tilde{a}^\dagger
		\end{align}
		where $A\in\UU(\AA_\sigma)$ and $y\in\textrm{Cone}_\sigma(\alpha_0,\theta)$. This map satisfies that
		\begin{enumerate}
			\item  for (the unique) $\xi\in\Aut{\AA_{W(C_\theta)\cap\sigma}}$, such that $\Ad{y}\circ\pi_0\circ\alpha_0=\pi_0\circ\alpha_0\circ\xi$, we get that
			\begin{equation}\label{eq:ConditionDefinitionOfGroupMorphism}
				\Ad{\phi_{1,\sigma}(x)}\circ\pi_0\circ\alpha_0\circ\Theta\circ\gamma^H_{0;1}\circ\gamma^{\Hsplit}_{1;0}=\pi_0\circ\alpha_0\circ\Theta\circ\gamma^H_{0;1}\circ\gamma^{\Hsplit}_{1;0}\circ\Ad{A}\circ\xi.
			\end{equation}
			\item it is well-defined (independent of the choice of $A$ and $y$).
			\item it is a group homomorphism.
			\item $\tilde{a}^\dagger \phi_1(x)\tilde{a}\in\IAC_\sigma(\alpha_0,\theta,\Theta\circ \Phi^U_{0;1}\otimes\Phi^D_{0;1})$.
			\item if additionally, $x\in \IAC_\sigma^W(\alpha_0,\theta,\Theta)$, $\IAC_{\nu^{\sigma}(\sigma)}(\alpha_0,\theta,\Theta)$ or $\IAC_{\nu^{\sigma}(\sigma)}^W (\alpha_0,\theta,\Theta)$ then  $\tilde{a}^\dagger \phi_1(x)\tilde{a}\in\IAC_\sigma^W(\alpha_0,\theta,\Theta\circ \Phi^U_{0;1}\otimes\Phi^D_{0;1})$, $\IAC_{\nu^{\sigma}(\sigma)}(\alpha_0,\theta,\Theta\circ \Phi^U_{0;1}\otimes\Phi^D_{0;1})$ or $\IAC_{\nu^{\sigma}(\sigma)}^W (\alpha_0,\theta,\Theta\circ \Phi^U_{0;1}\otimes\Phi^D_{0;1})$ respectively.
		\end{enumerate}
	\end{lemma}
	\begin{proof}
		In this proof, we take $\sigma=R$, the proof when $\sigma=L$ is equivalent. Take $x\in \IAC_R(\alpha_0,\theta,\Theta)$ arbitrary. Take $A\in\UU(\AA_R)$ and $y\in\textrm{Cone}_R(\alpha_0,\theta,\Theta)$ such that $x=\pi_0\circ\alpha_0\circ\Theta(A)y$ and take $\xi\in\Aut{\AA_{W(C_\theta)}\cap \AA_R}$ the automorphism such that
		\begin{equation}
			\Ad{y}\circ\pi_0\circ\alpha_0\circ\Theta=\pi_0\circ\alpha_0\circ\Theta\circ\xi.
		\end{equation}
		Take $a$ and $\Phi=\Phi^U\otimes\Phi^D$ as given in equation \eqref{eq:DefinitionOfGroupMorphismVAutEquation}. We now define
		\begin{equation}
			\phi_{1,R}(x)\defeq \pi_0\circ\alpha_0\circ\Theta\circ\gamma^H_{0;1}\circ\gamma^{\Hsplit}_{1;0}(A)\tilde{a}y\tilde{a}^\dagger.
		\end{equation}
		We now have to prove three things:
		\begin{enumerate}
			\item It satisfies equation \eqref{eq:ConditionDefinitionOfGroupMorphism}.
			\item This map is well-defined (independent of our choices).
			\item This is a group homomorphism.
		\end{enumerate}
		To show the first item just observe that
		\begin{align}
			\nonumber
			&\Ad{\phi_{1,R}(x)}\circ\pi_0\circ\alpha_0\circ\Theta\circ\gamma^H_{0;1}\circ\gamma^{\Hsplit}_{1;0}\\
			&=\Ad{\pi_0\circ\alpha_0\circ\Theta\circ\gamma^H_{0;1}\circ\gamma^{\Hsplit}_{1;0}(A)\pi_0\circ\alpha_0\circ\Theta(a)y\pi_0\circ\alpha_0\circ\Theta(a)^\dagger}\circ\pi_0\circ\alpha_0\circ\Theta\circ\gamma^H_{0;1}\circ\gamma^{\Hsplit}_{1;0}\\
			\nonumber
			&=\Ad{\pi_0\circ\alpha_0\circ\Theta\circ\gamma^H_{0;1}\circ\gamma^{\Hsplit}_{1;0}(A)}\circ\pi_0\circ\alpha_0\circ\Theta\circ\Ad{a}\circ\xi\circ\Ad{a^\dagger}\circ\gamma^H_{0;1}\circ\gamma^{\Hsplit}_{1;0}.
		\end{align}
		Now we insert $\gamma^H_{0;1}\circ\gamma^{\Hsplit}_{1;0}\circ\gamma^{\Hsplit}_{0;1}\circ\gamma^H_{1;0}=\text{Id}$ and obtain:
		\begin{align}
			&=\Ad{\pi_0\circ\alpha_0\circ\Theta\circ\gamma^H_{0;1}\circ\gamma^{\Hsplit}_{1;0}(A)}\circ\pi_0\circ\alpha_0\circ\Theta\circ(\gamma^H_{0;1}\circ\gamma^{\Hsplit}_{1;0}\circ\gamma^{\Hsplit}_{0;1}\circ\gamma^H_{1;0})\\
			\nonumber
			&\quad\circ\Ad{a}\circ\xi\circ\Ad{a^\dagger}\circ\gamma^H_{0;1}\circ\gamma^{\Hsplit}_{1;0}\\
			&=\pi_0\circ\alpha_0\circ\Theta\circ\gamma^H_{0;1}\circ\gamma^{\Hsplit}_{1;0}\circ\Ad{A}\circ\gamma^{\Hsplit}_{0;1}\circ\gamma^H_{1;0}\circ\Ad{a}\circ\xi\circ\Ad{a^\dagger}\circ\gamma^H_{0;1}\circ\gamma^{\Hsplit}_{1;0}.
		\end{align}
		Using equation \eqref{eq:DefinitionOfGroupMorphismVAutEquation} this gives
		\begin{align}
			&=\pi_0\circ\alpha_0\circ\Theta\circ\gamma^H_{0;1}\circ\gamma^{\Hsplit}_{1;0}\circ\Ad{A}\circ\Phi^{-1}\circ\xi\circ\Phi=\pi_0\circ\alpha_0\circ\Theta\circ\gamma^H_{0;1}\circ\gamma^{\Hsplit}_{1;0}\circ\Ad{A}\circ\xi
		\end{align}
		concluding the proof of the first item. To show the second item, suppose we have two different representatives of $x$, say
		\begin{equation}
			x=\pi_0\circ\alpha_0\circ\Theta(A_1)y_1=\pi_0\circ\alpha_0\circ\Theta(A_2)y_2.
		\end{equation}
		By construction, this implies that $\pi_0\circ\alpha_0\circ\Theta(A_2^\dagger A_1)=y_2 y_1^\dagger$. This in turn implies that $\Ad{A_2^\dagger A_2}\in\Aut{\AA_{W(C_\theta)\cap R}}$ and hence we require that $A_2^\dagger A_1\in\UU(\AA_{(W(C_\theta)\cap R)^c}')=\UU(\AA_{W(C_\theta)\cap R})$ (this last equality follows from section 3 of \cite{NaScWe_2013}). By using this equation combined with lemma \ref{lem:PropertyTilde_a} we get that
		\begin{align}
			\Ad{\tilde{a}}(y_2y_1^\dagger)&=\Ad{\tilde{a}}\circ\pi_0\circ\alpha_0\circ\Theta(A_2^\dagger A_1)=\pi_0\circ\alpha_0\circ\Theta\circ\gamma^H_{0;1}\circ\gamma^{\Hsplit}_{1;0}(A_2^\dagger A_1).
		\end{align}
		Reworking this equation gives the desired
		\begin{equation}
			\phi_{1,R}(\pi_0\circ\alpha_0\circ\Theta(A_1)y_1)=\phi_{1,R}(\pi_0\circ\alpha_0\circ\Theta(A_2)y_2).
		\end{equation}
		To then show the third item, take $x_1,x_2\in \IAC_R(\alpha_0,\Theta)$ arbitrary. Take $A_1,A_2\in\UU(\AA_R),y_1,y_2\in \textrm{Cone}_R(\alpha_0,\Theta)$ and $\xi_1,\xi_2\in\Aut{\AA_{W(C_\theta)}\cap\AA_R}$ such that
		\begin{align}
			x_i&=\pi_0\circ\alpha_0\circ\Theta(A_i)y_i&\Ad{y_i}\circ\pi_0\circ\alpha_0\circ\Theta&=\pi_0\circ\alpha_0\circ\Theta\circ\xi_i.
		\end{align}
		We have
		\begin{align}
			x_1x_2&=\pi_0\circ\alpha_0\circ\Theta(A_1)y_1\pi_0\circ\alpha_0\circ\Theta(A_2)y_2=\pi_0\circ\alpha_0\circ\Theta(A_1)y_1\pi_0\circ\alpha_0\circ\Theta(A_2)y_1^{\dagger}y_1y_2\\
			\nonumber
			&=\pi_0\circ\alpha_0\circ\Theta(A_1\xi_1(A_2))y_1y_2
		\end{align}
		giving
		\begin{equation}
			\phi_{1,R}(x_1x_2)=\pi_0\circ\alpha_0\circ\Theta\circ\gamma^H_{0;1}\circ\gamma^{\Hsplit}_{1;0}(A_1\xi_1(A_2))\tilde{a}y_1y_2\tilde{a}^\dagger.
		\end{equation}
		where $\tilde{a}=\pi_0\circ\alpha_0\circ\Theta(a)$. Filling this in gives
		\begin{align}
			\nonumber
			&\phi_{1,R}(x_1x_2)\phi_{1,R}(x_2)^\dagger\phi_{1,R}(x_1)^\dagger\\
			\label{eq:HomomorphismCondition}
			&=\pi_0\circ\alpha_0\circ\Theta\circ\gamma^H_{0;1}\circ\gamma^{\Hsplit}_{1;0}(A_1\xi_1(A_2))\tilde{a}y_1\stkout{y_2\tilde{a}^\dagger \tilde{a}y_2^\dagger}\tilde{a}^\dagger\\
			\nonumber
			&\quad\pi_0\circ\alpha_0\circ\Theta\circ\gamma^H_{0;1}\circ\gamma^{\Hsplit}_{1;0}(A_2^\dagger)\tilde{a}y_1^\dagger\tilde{a}^\dagger \pi_0\circ\alpha_0\circ\Theta\circ\gamma^H_{0;1}\circ\gamma^{\Hsplit}_{1;0}(A_1^\dagger).
		\end{align}
		Writing out a part of this gives:
		\begin{align}
			&\Ad{\tilde{a}y_1\tilde{a}^\dagger}\circ\pi_0\circ\alpha_0\circ\Theta\circ\gamma^H_{0;1}\circ\gamma^{\Hsplit}_{1;0}(A_2^\dagger)\\
			\nonumber
			&=\pi_0\circ\alpha_0\circ\Theta\circ\Ad{a}\circ\xi_1\circ\Ad{a^\dagger}\circ\gamma^H_{0;1}\circ\gamma^{\Hsplit}_{1;0}(A_2^\dagger).
		\end{align}
		Using equation \eqref{eq:DefinitionOfGroupMorphismVAutEquation} now yields:
		\begin{align}
			&=\pi_0\circ\alpha_0\circ\Theta\circ\Ad{a}\circ\xi_1\circ\Phi(A_2^\dagger)=\pi_0\circ\alpha_0\circ\Theta\circ\Ad{a}\circ\Phi\circ\xi_1(A_2^\dagger)=\pi_0\circ\alpha_0\circ\Theta\circ\gamma^H_{0;1}\circ\gamma^{\Hsplit}_{1;0}\circ\xi_1(A_2^\dagger).
		\end{align}
		This shows that
		\begin{equation}
			\Ad{\tilde{a}y_1\tilde{a}^\dagger}\circ\pi_0\circ\alpha_0\circ\Theta\circ\gamma^H_{0;1}\circ\gamma^{\Hsplit}_{1;0}(A_2^\dagger)=\pi_0\circ\alpha_0\circ\Theta\circ\gamma^H_{0;1}\circ\gamma^{\Hsplit}_{1;0}\circ\xi_1(A_2^\dagger).
		\end{equation}
		Inserting this in equation \eqref{eq:HomomorphismCondition} shows that $\phi_{1,R}(x_1x_2)\phi_{1,R}(x_1)^\dagger\phi_{1,R}(x_2)^\dagger=\id$, concluding the proof of the third item. To show the last two items one only has to work out $\Ad{\tilde{a}^\dagger \phi_{1,R}(x)\tilde{a}}\circ\pi_0\circ\alpha_0\circ\Theta$, in both situations. Doing this concludes the proof.
	\end{proof}
	We now have to extend the above definition to the group
	\begin{equation}
		\IAC_{L\times R}(\alpha_0,\theta,\Theta)=\{ab=ba|a\in\IAC_{L}(\alpha_0,\theta,\Theta),b\in\IAC_{R}(\alpha_0,\theta,\Theta)\}.
	\end{equation}
	\begin{lemma}\label{lem:extensionOfPhi1Definition}
		Define
		\begin{equation}
			\phi_1:\IAC_{L\times R}(\alpha_0,\theta,\Theta)\rightarrow\UU(\HH):ab\mapsto \phi_{1,L}(a)\phi_{1,R}(b)
		\end{equation}
		where $a\in \IAC_{L}(\alpha_0,\theta,\Theta)$ and $b\in \IAC_{R}(\alpha_0,\theta,\Theta)$. This satisfies two points.
		\begin{enumerate}
			\item It is well-defined, by which we mean that for any $a,\tilde{a}\in \IAC_{L}(\alpha_0,\theta,\Theta)$ and for any $b,\tilde{b}\in \IAC_{R}(\alpha_0,\theta,\Theta)$ such that $ab=\tilde{a}\tilde{b}$ we have that $\phi_{1,L}(a)\phi_{1,R}(b)=\phi_{1,L}(\tilde a)\phi_{1,R}(\tilde b)$.
			\item It is still a group homomorphism.
		\end{enumerate}
	\end{lemma}
	\begin{proof}
		The first item is trivial, this is because the fact that $ab=\tilde{a}\tilde{b}$ already implies that $(a,b)$ and $(\tilde{a},\tilde{b})$ can only differ up to an exchange in phase. To show the second item, we only have to show that $\phi_{1,L}(a)\phi_{1,R}(b)=\phi_{1,R}(b)\phi_{1,L}(a)$ for any $a\in \IAC_{L}(\alpha_0,\theta,\Theta)$ and for any $b\in \IAC_{R}(\alpha_0,\theta,\Theta)$. To show this, we first write out $b=\pi_0\circ\alpha_0\circ\Theta(B)y_b$ where $B\in\UU(\AA_R)$ and where $y_b\in\textrm{Cone}_R(\alpha_0,\theta)$. We can now split the proof into two parts. First, we want to prove that
		\begin{equation}
			\Ad{\phi_{1,L}(a)}\circ\pi_0\circ\alpha_0\circ\Theta\circ\gamma_{0;1}^H\circ\gamma^\Hsplit_{1;0}(B)=\pi_0\circ\alpha_0\circ\Theta\circ\gamma_{0;1}^H\circ\gamma^\Hsplit_{1;0}(B).
		\end{equation}
		This can be done using equation \eqref{eq:ConditionDefinitionOfGroupMorphism}. Secondly, we want to prove that $\Ad{\phi_{1,L}(a)}(\phi_{1,R}(y_b))=(\phi_{1,R}(y_b))$. We can rewrite this equation to
		\begin{equation}
			\Ad{\phi_{1,R}(y_b)}(\phi_{1,L}(a))=(\phi_{1,L}(a)).
		\end{equation}
		We now write out $a=\pi_0\circ\alpha_0\circ\Theta(A)y_a$ with $A\in\UU(\AA_L)$ and $y_a\in\textrm{Cone}_L(\alpha_0,\theta)$. We again split the proof of this equality into two parts. The equality with the invariance of the $A$ part again uses equation \eqref{eq:ConditionDefinitionOfGroupMorphism}. The equality with the invariance of the $y_a$ part can be done by inserting the definition and observing that the $\tilde{a}$ cancels.
	\end{proof}
	\begin{lemma}\label{lem:DefinitionOfWgMap}
		Define again $\tilde{a}$ through equation \eqref{eq:DefinitionOfGroupMorphismVAutEquation}. Let
		\begin{equation}
			\phi_2:\{\Ad{v^{-p}}(W_g)|g\in G,p\in\{-1,0,1\}\}\rightarrow \UU(\HH):\Ad{v^{-p}} (W_g)\mapsto \Ad{\tilde{a}v^{-p}} (W_g)
		\end{equation}
		for the case where there is one translation symmetry, let
		\begin{align}
			\phi_2&:\{\Ad{w^{-s}v^{-p}}(W_g)|g\in G\text{ and }s,p\in\{-1,0,1\}\}\rightarrow \UU(\HH)\\
			\nonumber
			&:\Ad{w^{-s}v^{-p}} (W_g)\mapsto \Ad{\tilde{a}w^{-s}v^{-p}} (W_g)
		\end{align}
		in the case where $\omega$ and $H$ are translation invariant in both directions. This satisfies
		\begin{align}\label{eq:lemDefinitionOfWgMapCondition}
			&\Ad{\phi_2(w^{-s}v^{-p}W_g v^{p}w^{s})}\circ\pi_0\\
			\nonumber
			&=\pi_0\circ \alpha_0\circ\Theta\circ\gamma^{H}_{0;1}\circ\gamma^{\Hsplit}_{1;0}\circ\nu^{-s}\circ\tau^{-p}\circ\eta_g\circ\beta_g^U\circ\tau^p\circ\nu^{s}\circ\gamma^{\Hsplit}_{0;1}\circ\gamma^{H}_{1;0}\circ\Theta^{-1}\circ\alpha_0^{-1}
		\end{align}
		(where $s=0$ in the case where there is just one translation symmetry).
	\end{lemma}
	\begin{proof}
		To show that this indeed satisfies equation \eqref{eq:lemDefinitionOfWgMapCondition} observe that using equation \eqref{eq:DefinitionOfGroupMorphismVAutEquation} we obtain:
		\begin{align}
			\nonumber
			&\Ad{\phi_2(w^{-s}v^{-p}W_g v^{p}w^{s})}\circ\pi_0\\
			&=\pi_0\circ \alpha_0\circ\Theta\circ\Ad{a}\circ\nu^{-s}\circ\tau^{-p}\circ\eta_g\circ\beta_g^U\circ\tau^p\circ\nu^{s}\circ\Ad{a^{\dagger}}\circ\Theta^{-1}\circ\alpha_0^{-1}\\
			\nonumber
			&=\pi_0\circ \alpha_0\circ\Theta\circ\gamma^H_{0;1}\circ\gamma^{H_L}_{1;0}\otimes\gamma^{H_L}_{1;0}\circ\stkout{\Phi_{0;1}^{-1}}\circ\nu^{-s}\circ\tau^{-p}\circ\eta_g\circ\tau^p\circ\nu^{s}\circ\beta_g^{\tau^{-p}(U)}\circ\stkout{\Phi_{0;1}}\circ\gamma^{\Hsplit}_{0;1}\circ\gamma^H_{1;0}\circ\Theta^{-1}\circ\alpha_0^{-1}.
		\end{align}
		Since $\Phi$ commutes with both the $\nu^{-s}\circ\tau^{-p}\circ\eta_g\circ\tau^p\circ\nu^s$ and with the $\beta_g^{\tau^{-p}(U)}$ the result follows.
	\end{proof}
	\begin{lemma}\label{lem:phi1phi2matchingCondition}
		Take $\phi_1$ and $\phi_2$ as defined previously. Take $s\in\{-1,0,1\}$ and take $p\in\{-1,0,1\}$ arbitrary in the case where $\omega$ and $H$ have two translation symmetries while we take $p=0$ in the case where we only have one translation symmetry. For any $x\in\IAC_{L\times R}(\alpha_0,\theta,\Theta)$, we have that
		\begin{align}
			\phi_1(\Ad{w^{-p} W_g w^{p}}(x))&=\Ad{\phi_2(w^{-p} W_g w^{p})}(\phiTilde_1(x))\\
			\phi_2(\Ad{v^s}(w^{-p} W_gw^{p}))&=\Ad{v^s}(\phi_2(w^{-p} W_g w^{p}))
		\end{align}
		If additionally $x\in\IAC_{L\times R}^W(\alpha_0,\theta,\Theta)$ we get that
		\begin{align}
			\phi_1(\Ad{v^{s}}(x))&=\Ad{v^{s}}(\phi_1(x)).
		\end{align}
	\end{lemma}
	\begin{proof}
		Take $x\in \IAC_R(\alpha_0,\theta,\Theta)$ arbitrary (the proof for $x\in \IAC_L(\alpha_0,\theta,\Theta)$ is analogous). Take $A\in\UU(\AA_R)$ and $y\in\cone_R(\alpha_0,\theta)$ such that $x=\pi_0\circ\alpha_0\circ\Theta(A)y$ and $\xi\in\Aut{\AA_{C_\theta}\cap \AA_R}$ the automorphism such that
		\begin{equation}
			\Ad{y}\circ\pi_0\circ\alpha_0\circ\Theta=\pi_0\circ\alpha_0\circ\Theta\circ\xi.
		\end{equation}
		Take $a$ and $\Phi$ as given in equation \eqref{eq:DefinitionOfGroupMorphismVAutEquation} and take $\tilde{a}=\pi_0\circ\alpha_0\circ\Theta(a)$. We now get
		\begin{align}
			\Ad{w^{-p}W_gw^{p}} (x) &=\Ad{w^{-p}W_gw^{p}}\circ\pi_0\circ\alpha_0\circ\Theta(A) \Ad{w^{-p}W_gw^{p}}(y)\\
			\nonumber
			&=\pi_0\circ\alpha_0\circ\Theta\circ\nu^{-p}\circ\eta_g\circ\nu^{p}\circ\beta_g^U(A) \Ad{w^{-p}W_gw^{p}}(y).
		\end{align}
		We still have that $\Ad{w^{-p}W_gw^{p}}(y)\in\cone_R(\alpha_0,\theta)$. This is because there exists a $\tilde{\xi}\in\Aut{\AA_{C_\theta}\cap \AA_R}$ such that
		\begin{equation}
			\Ad{\Ad{w^{-p}W_gw^{p}}(y)}\circ \pi_0\circ\alpha_0\circ\Theta=\pi_0\circ\alpha_0\circ\Theta\circ\tilde\xi
		\end{equation}
		(one can just take $\tilde\xi=\nu^{-p}\circ\eta_g\circ\nu^{p}\circ\beta_g^U\circ\xi\circ\beta_{g^{-1}}^U\circ\nu^{-p}\circ\eta_g^{-1}\circ\nu^{p}$). This implies that
		\begin{equation}\label{eq:lem:phi1phi2matchingCondition:Proof_W_g_PartAppendix}
			\phi_1(\Ad{w^{-p}W_gw^{p}}(x))=\pi_0\circ\alpha_0\circ\Theta\circ\gamma^{H}_{0;1}\circ\gamma^{\Hsplit}_{1;0}\circ\nu^{-p}\circ\eta_g\circ\nu^{p}\circ\beta_g^U(A)\tilde{a}\Ad{w^{-p}W_gw^{p}}(y)\tilde{a}^\dagger.
		\end{equation}
		Filling this in gives (where we've used $\tilde{W}_g\defeq w^{-p}W_g w^{p}$ and $\tilde{\eta}_g\defeq \nu^{-p}\circ\eta_g\circ\nu^{p}$ for notational simplicity)
		\begin{align}
			\label{eq:phi1phi2matchingConditionEquationAppendix}
			&\phi_1(\tilde{W}_gx\tilde{W}_g^\dagger)\phi_2(\tilde{W}_g)\phi_1(x)^\dagger\phi_2(\tilde{W}_g)^\dagger\\
			\nonumber
			&=\pi_0\circ\alpha_0\circ\Theta\circ\gamma^{H}_{0;1}\circ\gamma^{\Hsplit}_{1;0}\circ\tilde{\eta}_g\circ\beta_g^U(A)\tilde{a}\tilde{W}_g\stkout{y\tilde{W}_g^\dagger\tilde{a}^\dagger}\stkout{\tilde{a}\tilde{W}_g\tilde{a}^\dagger\tilde{a}y^\dagger}\tilde{a}^\dagger\pi_0\circ\alpha_0\circ\Theta\circ\gamma^{H}_{0;1}\circ\gamma^{\Hsplit}_{1;0}(A)^\dagger\tilde{a}\tilde{W}_g^\dagger\tilde{a}^\dagger.
		\end{align}
		To prove that this is the identity operator observe that
		\begin{align}
			\nonumber
			&\Ad{\tilde{a}\tilde{W}_g\tilde{a}^\dagger}\circ\pi_0\circ\alpha_0\circ\Theta\circ \gamma^{H}_{0;1}\circ\gamma^{\Hsplit}_{1;0}(A)=\Ad{\tilde{a}\tilde{W}_g}\circ\pi_0\circ\alpha_0\circ\Theta\circ\Ad{a^\dagger}\circ \gamma^{H}_{0;1}\circ\gamma^{\Hsplit}_{1;0}(A)\\
			&=\Ad{\tilde{a}\tilde{W}_g}\circ\pi_0\circ\alpha_0\circ\Theta\circ\Phi(A)=\Ad{\tilde{a}}\circ\pi_0\circ\alpha_0\circ\Theta\circ\tilde{\eta}_g\circ\beta_g^U\circ\Phi(A)\\
			\nonumber
			&=\Ad{\tilde{a}}\circ\pi_0\circ\alpha_0\circ\Theta\circ\Phi\circ\tilde{\eta}_g\circ\beta_g^U(A)=\pi_0\circ\alpha_0\circ\Theta\circ\Ad{a}\circ\Phi\circ\tilde{\eta}_g\circ\beta_g^U(A)\\
			\nonumber
			&=\pi_0\circ\alpha_0\circ\Theta\circ \gamma^{H}_{0;1}\circ\gamma^{\Hsplit}_{1;0} \circ\tilde{\eta}_g\circ\beta_g^U(A).
		\end{align}
		Filling this in equation \eqref{eq:phi1phi2matchingConditionEquationAppendix} concludes the proof of the first item. The proof of the last item looks very similar. Now take $x\in \IAC_R^W(\alpha_0,\theta,\Theta)$ arbitrary. Take $A\in\UU(\AA_R)$ and $y\in\text{Cone}_R^W(\alpha_0,\theta)$ such that $x=\pi_0\circ\alpha_0\circ\Theta(A)y$. We now get that (the part with $v$ replaced with $v^\dagger$ and $\tau$ by $\tau^{-1}$ is equivalent)
		\begin{equation}
			vxv^\dagger=\pi\circ\alpha_0\circ\Theta\circ\tau(A)vyv^\dagger.
		\end{equation}
		When we now take $\tilde\xi=\tau\circ\xi\circ\tau^{-1}$ we still obtain an equation of the form
		\begin{equation}
			\Ad{vyv^\dagger}\circ \pi_0\circ\alpha_0\circ\Theta=\pi_0\circ\alpha_0\circ\Theta\circ\tilde\xi.
		\end{equation}
		This again (in analogy to equation \eqref{eq:lem:phi1phi2matchingCondition:Proof_W_g_PartAppendix}) implies that we can take
		\begin{equation}
			\phi_1(vxv^\dagger)=\pi_0\circ\alpha_0\circ\Theta\circ\gamma^{H}_{0;1}\circ\gamma^{\Hsplit}_{1;0}\circ\tau(A)\tilde{a}vyv^\dagger\tilde{a}^\dagger.
		\end{equation}
		Filling this in now gives
		\begin{align}
			&\phi_1(vxv^\dagger)v\phi_1(x)^\dagger v^\dagger\\
			\label{eq:lem:phi1phi2matchingCondition:Proof_v_Part1Appendix}
			&=\pi_0\circ\alpha_0\circ\Theta\circ\gamma^{H}_{0;1}\circ\gamma^{\Hsplit}_{1;0}\circ\tau(A)\tilde{a}vyv^\dagger\tilde{a}^\dagger v \tilde{a}y^\dagger\tilde{a}^\dagger \pi_0\circ\alpha_0\circ\Theta\circ\gamma^{H}_{0;1}\circ\gamma^{\Hsplit}_{1;0}(A)^\dagger v^\dagger.
		\end{align}
		Before we proceed we will first show the equation
		\begin{equation}\label{eq:lem:phi1phi2matchingCondition:Proof_v_Part2Appendix}
			\Ad{\tilde{a}vyv^\dagger\tilde{a}^\dagger v\tilde{a}y^\dagger\tilde{a}^\dagger}\circ\pi_0\circ\alpha_0\circ\Theta=\pi_0\circ\alpha_0\circ\Theta\circ\tau.
		\end{equation}
		To prove this, observe that
		\begin{align}
			&\Ad{\tilde{a}vyv^\dagger\tilde{a}^\dagger v\tilde{a}y^\dagger\tilde{a}^\dagger}\circ\pi_0\circ\alpha_0\circ\Theta=\Ad{\tilde{a}vyv^\dagger\tilde{a}^\dagger v}\circ\pi_0\circ\alpha_0\circ\Theta\circ\Ad{a^\dagger}\circ\xi^{-1}\circ\Ad{a}.
		\end{align}
		Now inserting equation \eqref{eq:DefinitionOfGroupMorphismVAutEquation} in this gives
		\begin{align}
			\nonumber
			&=\Ad{\tilde{a}vyv^\dagger\tilde{a}^\dagger v}\circ\pi_0\circ\alpha_0\circ\Theta\circ\gamma^{H}_{0;1}\circ\gamma^{\Hsplit}_{1;0}\circ\stkout{(\Phi)^{-1}}\circ\xi^{-1}\circ\stkout{\Phi}\circ\gamma^{\Hsplit}_{0;1}\circ\gamma^H_{1;0}\\
			&=\Ad{\tilde{a}vyv^\dagger\tilde{a}^\dagger }\circ\pi_0\circ\alpha_0\circ\Theta\circ\gamma^{H}_{0;1}\circ\gamma^{\Hsplit}_{1;0}\circ\tau\circ\xi^{-1}\circ\underline{\tau^{-1}\circ\tau}\circ\gamma^{\Hsplit}_{0;1}\circ\gamma^H_{1;0}\\
			\nonumber
			&=\pi_0\circ\alpha_0\circ\Theta\circ\Ad{a}\circ\tau\circ\xi\circ\tau^{-1}\circ\Ad{a^\dagger}\circ\gamma^{H}_{0;1}\circ\gamma^{\Hsplit}_{1;0}\circ\tau\circ\xi^{-1}\circ\tau^{-1}\circ\gamma^{\Hsplit}_{0;1}\circ\gamma^H_{1;0}\circ\tau.
		\end{align}
		Using again equation \eqref{eq:DefinitionOfGroupMorphismVAutEquation} now yields:
		\begin{align}
			&=\pi_0\circ\alpha_0\circ\Theta\circ\gamma^{H}_{0;1}\circ\gamma^{\Hsplit}_{1;0}\circ\stkout{\Phi^{-1}}\circ\tau\circ\xi\circ\tau^{-1}\circ\stkout{\Phi}\circ\tau\circ\xi^{-1}\circ\tau^{-1}\circ\gamma^{\Hsplit}_{0;1}\circ\gamma^H_{1;0}\circ\tau\\
			\nonumber
			&=\pi_0\circ\alpha_0\circ\Theta\circ\stkout{\gamma^{H}_{0;1}\circ\gamma^{\Hsplit}_{1;0}\circ\tau\circ\xi\circ\tau^{-1}\circ\tau\circ\xi^{-1}\circ\tau^{-1}\circ\gamma^{\Hsplit}_{0;1}\circ\gamma^H_{1;0}}\circ\tau
		\end{align}
		concluding the proof of equation \eqref{eq:lem:phi1phi2matchingCondition:Proof_v_Part2Appendix}. We will now show that the expression in equation \eqref{eq:lem:phi1phi2matchingCondition:Proof_v_Part1Appendix} is the identity. First, we will show that the $A$'s in equation \eqref{eq:lem:phi1phi2matchingCondition:Proof_v_Part1Appendix} cancel. By equation \eqref{eq:lem:phi1phi2matchingCondition:Proof_v_Part2Appendix} we get that
		\begin{equation}
			\Ad{\tilde{a}vyv^\dagger\tilde{a}^\dagger v\tilde{a}y^\dagger\tilde{a}^\dagger}\circ\pi_0\circ\alpha_0\circ\Theta\circ\gamma^{H}_{0;1}\circ\gamma^{\Hsplit}_{1;0}(A^\dagger)=\pi_0\circ\alpha_0\circ\Theta\circ\gamma^{H}_{0;1}\circ\gamma^{\Hsplit}_{1;0}\circ\tau(A^\dagger).
		\end{equation}
		Filling this in shows that the $A$'s cancel. Now, what is left to show is that $\tilde{a}vyv^\dagger\tilde{a}^\dagger v \tilde{a}y^\dagger\tilde{a}^\dagger  v^\dagger=1$ which is equivalent to showing that $\Ad{y}(v^\dagger \tilde{a}^\dagger v \tilde{a})=v^\dagger \tilde{a}^\dagger v \tilde{a}$. To show this, first, we use the definitions of these objects to see that this is equivalent to showing that
		\begin{equation}\label{eq:proofOf:lem:phi1phi2matchingCondition:LastThingToProveAppendix}
			\xi(\tau^{-1}(a^\dagger)a)=\tau^{-1}(a^\dagger)a.
		\end{equation}
		However, since
		\begin{equation}
			\Ad{\tau^{-1}(a^\dagger)a}=\tau^{-1}\circ\Phi\circ\gamma^{\Hsplit}_{0;1}\circ\gamma^H_{1;0}\circ\tau\circ\gamma^H_{0;1}\circ\gamma^{\Hsplit}_{1;0}\circ\Phi^{-1}=\tau^{-1}\circ\Phi\circ\tau\circ\Phi^{-1},
		\end{equation}
		we get that $\tau^{-1}(a^\dagger)a\in\UU(\AA_{W(C_\theta)})'=\UU(\AA_{W(C_\theta)^c})$, where the last equality is again because of section 3 in \cite{NaScWe_2013}. This concludes the proof of \ref{eq:proofOf:lem:phi1phi2matchingCondition:LastThingToProveAppendix}. The proof of the second item is analogous concluding the proof of this lemma.
	\end{proof}
	\subsection{Bringing \texorpdfstring{$w$}{} inside \texorpdfstring{$\phi_1$}{} and \texorpdfstring{$\phi_2$}{}}
	In the last section, we constructed our morphisms $\phi_1$ and $\phi_2$ in such a way that we could take the translation operator $v$ outside of the morphism (see lemma \ref{lem:phi1phi2matchingCondition}). To prove the invariance of the $H^2(G,\TT)$-valued index, this is sufficient and the rest of this section can be skipped. However, if we want to show the invariance of the $H^1(G,\TT)$-valued index under LGAs generated by a one-parameter family of interactions $H\in\BB_{F_\phi}([0,1])$ that is invariant under both translation symmetries, we need some additional structure. We need to have a mechanism to bring the additional translation operator, $w$ outside of the morphisms $\phi_1$ and $\phi_2$.
	\begin{lemma}\label{lem:TranslatingSplittedTimeEvolution}
		There exist maps
		\begin{align}
			A_L&:[0,1]\rightarrow \UU(\AA_L):\lambda\mapsto A_{L}(\lambda)&A_R&:[0,1]\rightarrow \UU(\AA_R):\lambda\mapsto A_{R}(\lambda)
		\end{align}
		both continuous in norm topology and
		\begin{align}
			\Phi^{\mu\nu}:[0,1]\rightarrow\Aut{\AA_{W(C_\theta)^c\cap\mu\cap\nu}}:\lambda\mapsto \Phi^{\mu\nu}(\lambda)
		\end{align}
		where $\mu\in\{U,D\}$ and $\nu\in\{L,\nu(R)\}$, all four continuous in strong\footnote{Meaning that $\Phi^{\mu\nu}(\lambda)(A)$ is continuous for all $A\in\AA$.} topology, satisfying that
		\begin{align}
			\nu\circ\gamma^{\Hsplit}_{0;\lambda}\circ\nu^{-1}\circ\gamma^{\Hsplit}_{\lambda;0}&=\gamma^{H_L}_{0;\lambda}\otimes\gamma^{H_{\nu\circ\nu(R)}}_{0;\lambda}\circ\gamma^{\Hsplit}_{\lambda;0}\\
			\label{eq:TranslatingSplittedTimeEvolution}
			&=\Ad{A_L(\lambda)}\otimes\Ad{A_R(\lambda)}\circ\bigotimes_{\substack{\mu\in\{U,D\},\\\nu\in\{L,\nu(R)\}}}\Phi^{\mu\nu}(\lambda)
		\end{align}
		and that $\Phi^{\mu\nu}(\lambda)\circ\beta_g=\beta_g\circ\Phi^{\mu\nu}(\lambda)$.
	\end{lemma}
	\begin{proof}
		This lemma is proven in \versionDifference{\ref{lem:SplittedAutomorphismAfterTranslatedIsVertical}}{C.8. of \cite{jappens2023spt}} which was based on similar statements made in \cite{ogata2021h3gmathbb}.
	\end{proof}
	In what follows, let $A_L\defeq A_L(1)$, $A_R\defeq A_R(1)$ and $\Phi^{\mu\nu}=\Phi^{\mu\nu}(1)$. We will now show some properties for these operators. The first one is that they can indeed be used to exchange $w$ and $\phi_1/\phi_2$:
	\begin{lemma}\label{lem:TranslationOutOfPhi_KW}
		Define $a_L\defeq \pi_0\circ\alpha_0\circ\Theta(A_L)$ and $a_R\defeq \pi_0\circ\alpha_0\circ\Theta(A_R)$, then for any $K\in\text{Cone}_{L\times\nu(R)}(\alpha_0,\theta)$ we get that
		\begin{equation}\label{eq:TranslationOutOfPhi_K}
			w^\dagger \phi_1(K)w=\phi_1(w^\dagger \: a_L\otimes a_R \: K \: a_L^\dagger\otimes a_R^\dagger \: w).
		\end{equation}
		Similarly, we have that
		\begin{equation}\label{eq:TranslationOutOfPhi_W}
			w^\dagger \phi_2(W_g)w=\phi_1(w^\dagger \: a_L\otimes a_R \: w)\phi_2(w^\dagger W_g w)\phi_1(w^\dagger \: a_L^\dagger\otimes a_R^\dagger \: w).
		\end{equation}
	\end{lemma}
	\begin{proof}
		We will only show the first equality as the second one is completely analogous. We need to show that
		\begin{align}
			\phi_1(K)&=\Ad{w\phi_1(w^\dagger a_L\otimes a_R w)}(\phi_1(w^\dagger K w))=\Ad{w\phi_1(w^\dagger a_L\otimes a_R w)}(\tilde{a}w^\dagger K w\tilde{a}^\dagger)\\
			\nonumber
			&=\Ad{w\phi_1(w^\dagger a_L\otimes a_R w)}(\tilde{a}w^\dagger \tilde{a}^\dagger \phi_1(K) \tilde{a}^\dagger w\tilde{a}^\dagger)=\Ad{w\phi_1(w^\dagger a_L\otimes a_R w)w^\dagger (w \tilde{a}w^\dagger \tilde{a}^\dagger)}(\phi_1(K)).
		\end{align}
		To do this, we first find an $A\in\UU(\AA)$ such that
		\begin{equation}
			\pi_0\circ\alpha_0\circ\Theta\circ\gamma^H_{0;1}\circ\gamma^{\Hsplit}_{1;0}(A)=w\phi_1(w^\dagger a_L\otimes a_R w)w^\dagger (w \tilde{a}w^\dagger \tilde{a}^\dagger)
		\end{equation}
		and then we will prove that $A\in\UU(\AA_{W(C_\theta)^c})$. It should be clear that the result follows from this\footnote{This is most easily seen from\begin{equation}\Ad{\phi_1(K)}\circ\pi_0\circ\alpha_0\circ\Theta\circ\gamma^H_{0;1}\circ\gamma^{\Hsplit}_{1;0}(A)=\phi_1(\Ad{K}\circ\pi_0\circ\alpha_0\circ\Theta(A))=\phi_1(\pi_0\circ\alpha_0\circ\Theta(A))=\pi_0\circ\alpha_0\circ\Theta\circ\gamma^H_{0;1}\circ\gamma^{\Hsplit}_{1;0}(A). \end{equation}}. Finding this $A$ can be done by working out
		\begin{equation}
			w\phi_1(w^\dagger a_L\otimes a_R w)w^\dagger (w \tilde{a}w^\dagger \tilde{a}^\dagger)=\pi_0\circ\alpha_0\circ\Theta\left(\nu\circ\gamma^H_{0;1}\circ\gamma^{\Hsplit}_{1;0}\circ\nu^{-1}(A_L\otimes A_R)\nu(a)a^\dagger\right).
		\end{equation}
		First we insert $\gamma^H_{0;1}\circ\gamma^{\Hsplit}_{1;0}\circ\gamma^{\Hsplit}_{0;1}\circ\gamma^H_{1;0}=\text{Id}$ to obtain
		\begin{equation}
			=\pi_0\circ\alpha_0\circ\Theta\circ\underline{\gamma^H_{0;1}\circ\gamma^{\Hsplit}_{1;0}\circ\gamma^{\Hsplit}_{0;1}\circ\gamma^H_{1;0}}\left(\nu\circ\gamma^H_{0;1}\circ\gamma^{\Hsplit}_{1;0}\circ\nu^{-1}(A_L\otimes A_R)\nu(a)a^\dagger\right).
		\end{equation}
		Using the fact that $H$ is translation invariant allows us to cancel it giving
		\begin{align}
			&=\pi_0\circ\alpha_0\circ\Theta\circ\gamma^H_{0;1}\circ\gamma^{\Hsplit}_{1;0}(\gamma^{\Hsplit}_{0;1}\circ\stkout{\gamma^H_{1;0}}\circ\nu\circ\stkout{\gamma^H_{0;1}}\circ\gamma^{\Hsplit}_{1;0}\circ\nu^{-1}(A_L\otimes A_R)\gamma^{\Hsplit}_{0;1}\circ\gamma^H_{1;0}(\nu(a)a^\dagger))\\
			\nonumber
			&=\pi_0\circ\alpha_0\circ\Theta\circ\gamma^H_{0;1}\circ\gamma^{\Hsplit}_{1;0}(\gamma^{\Hsplit}_{0;1}\circ\nu\circ\gamma^{\Hsplit}_{1;0}\circ\nu^{-1}(A_L\otimes A_R)\gamma^{\Hsplit}_{0;1}\circ\gamma^H_{1;0}(\nu(a)a^\dagger)).
		\end{align}
		This shows that we can take
		\begin{equation}
			A\defeq \gamma^{\Hsplit}_{0;1}\circ\nu\circ\gamma^{\Hsplit}_{1;0}\circ\nu^{-1}(A_L\otimes A_R)\gamma^{\Hsplit}_{0;1}\circ\gamma^H_{1;0}(\nu(a)a^\dagger).
		\end{equation}
		Now we only have to show that indeed $A\in\UU(\AA_{W(C_\theta)^c})$. To do this, take $X\in\AA_{W(C_\theta)}$ arbitrary. If we can show that $\Ad{A}(X)=X$ for any such $X$ then we get that $A\in\UU(\AA_{W(C_\theta)}')$. Since by corollary 3.1 of \cite{NaScWe_2013} we get that $\UU(\AA_{W(C_\theta)}')=\UU(\AA_{W(C_\theta)^c})$ this is sufficient to conclude the proof. To show this, observe that
		\begin{align}
			\Ad{A}(X)&=\gamma^{\Hsplit}_{0;1}\circ\nu\circ\gamma^{\Hsplit}_{1;0}\circ\nu^{-1}\circ\Ad{A_L\otimes A_R}\circ\nu\circ\gamma^{\Hsplit}_{0;1}\circ\stkout{\nu^{-1}}\circ\stkout{\gamma^{\Hsplit}_{1;0}\circ\gamma^{\Hsplit}_{0;1}}\\
			&\nonumber
			\qquad\circ\gamma^H_{1;0}\circ\stkout{\nu}\circ\Ad{a}\circ\nu^{-1}\circ\Ad{a^\dagger}\circ\gamma^H_{0;1}\circ\gamma^{\Hsplit}_{1;0}(X)
		\end{align}
		where the cancellation is done using the fact that $H$ is translation invariant. Using equation \eqref{eq:TranslatingSplittedTimeEvolution} we obtain
		\begin{equation}
			\Ad{A}(X)=(\bigotimes_{\mu\nu}\Phi^{\mu\nu})^{-1}\circ\nu\circ\gamma^{\Hsplit}_{0;1}\circ\gamma^H_{1;0}\circ\Ad{a}\circ\nu^{-1}\circ\Ad{a^\dagger}\circ\gamma^H_{0;1}\circ\gamma^{\Hsplit}_{1;0}(X).
		\end{equation}
		Now using equation \eqref{eq:DefinitionOfGroupMorphismVAutEquation} we get
		\begin{equation}
			\Ad{A}(X)=(\bigotimes_{\mu\nu}\Phi^{\mu\nu})^{-1}\circ\nu\circ(\Phi^U\otimes\Phi^D)^{-1}\circ\nu^{-1}\circ\Phi^U\otimes\Phi^D(X)=X
		\end{equation}
		concluding the proof.
	\end{proof}
	We will now prove a small lemma that we will need to prove lemma \ref{lem:DefinitionOfEpsilons}.
	\begin{lemma}\label{lem:EqualityTwoTranslationsUsingConnectedPath}
		For all $\sigma\in\{L,R\}$, the following equality holds:
		\begin{equation}
			\Ad{W_g}(a_\sigma^\dagger v a_\sigma v^\dagger)=a_\sigma^\dagger v a_\sigma v^\dagger.
		\end{equation}
	\end{lemma}
	\begin{proof}
		We will prove this equality in two steps. First, we will find some $B_\sigma\in\UU(\AA)$ such that $\pi_0\circ\alpha_0\circ\Theta(B_\sigma)=a_\sigma^\dagger v a_\sigma v^\dagger$, then we will show that $B_\sigma\in\AA_{(W(C_\theta)\cap\sigma)^c}$ (his already proves invariance under $\eta_g$) and that $\beta_g^{U}(B_\sigma)=B_\sigma$ (concluding the proof of being invariant under $\eta_g\circ\beta_g^U$). To find $B_\sigma$, we write out                                                                                                                                                                                                              
		\begin{equation}
			a_\sigma^\dagger v a_\sigma v^\dagger=\pi_0\circ\alpha_0\circ\Theta(A_\sigma^\dagger\tau(A_\sigma))
		\end{equation}
		and observe that we can take $B_\sigma \defeq A_\sigma^\dagger\tau(A_\sigma)$. Now we will prove that $B_\sigma\in\UU(\AA_{C_\theta^c\cap\sigma})=\UU(\AA_{(C_\theta^c\cap\sigma)^c}')$. Take $X\in\AA_{(C_\theta^c\cap\sigma)^c}$ arbitrary. Using equation \eqref{eq:TranslatingSplittedTimeEvolution} we have that
		\begin{align}
			\Ad{B_L\otimes B_R}(X)&=\Ad{A_L^\dagger\otimes A_R^\dagger}\circ\tau\circ\Ad{A_L\otimes A_R}\circ\tau^{-1}(X)\\
			&=\bigotimes_{\mu\nu}\Phi^{\mu\nu}\circ\stkout{\gamma^{\Hsplit}_{0;1}\circ\nu\circ \gamma^{\Hsplit}_{1;0}\circ\nu^{-1}}\\
			\nonumber
			&\qquad\circ\tau\circ\stkout{\nu\circ\gamma^{\Hsplit}_{0;1}\circ\nu^{-1}\circ\gamma^{\Hsplit}_{1;0}}\circ(\bigotimes_{\mu\nu}\Phi^{\mu\nu})^{-1}\circ\tau^{-1}(X)=X
		\end{align}
		concluding the proof. The fact that $\Ad{\beta_g^{U}(B_\sigma)}=\Ad{B_\sigma}$ follows from this same equation as well. This already shows that that there exists a $G$-dependent phase $\alpha(g)$ such that $\beta_g^{U}(B_\sigma)B_\sigma^\dagger=\alpha(g)\id_\AA$. To show that $\alpha(g)=1$ we have to use two things. First, we use the fact that $\alpha$ has to be a $U(1)$-representation. Secondly, because we can make this construction for every $\lambda$ (I mean the $\lambda$ from lemma \ref{lem:TranslatingSplittedTimeEvolution}) we need that $\alpha$ must be connected to the trivial $U(1)$-representation through a continuous path. However, because our group is finite and finite groups have discrete $U(1)$-representations (see e.g. lemma \ref{lem:FiniteGroupsHaveDiscreteU(1)Representations}) we have that $\alpha$ is trivial, and hence the result follows.
	\end{proof}
	We conclude this subsection by showing that the following lemma holds:
	\begin{lemma}\label{lem:DefinitionOfEpsilons}
		Define $\epsilon^L_g\in\IAC_L(\alpha_0,\theta)$ and $\epsilon^R_g\in\IAC_R(\alpha_0,\theta)$ through
		\begin{align}
			\epsilon_g^\sigma\defeq \pi_0\circ\alpha_0\circ\Theta\circ\nu^{-1}(A_\sigma\:\eta_g^\sigma\circ\beta_{g}^{U\sigma}(A_\sigma^\dagger))
		\end{align}
		(where $A_\sigma$ was defined in lemma \ref{lem:TranslatingSplittedTimeEvolution}) then
		\begin{align}
			\label{eq:TransformationOfKUnderEpsilon}
			w^\dagger \phi_2(W_g)w &= \phi_1(\epsilon_g^L\otimes\epsilon_g^R)\phi_2(w^\dagger W_g w)&
			w^\dagger \phi_1(K_g^\sigma)w&=v^\dagger\phi_1(\epsilon_g^\sigma)v\phi_1(w^\dagger K_g^\sigma w)\phi_1(\epsilon_g^\sigma)^\dagger.
		\end{align}
	\end{lemma}
	\begin{proof}
		By equation \eqref{eq:TranslationOutOfPhi_W} we get that
		\begin{align}
			w^\dagger\phi_2(W_g)w&=\phi_1(w^\dagger \: a_L\otimes a_R \: w)\phi_2(w^\dagger W_g w)\phi_1(w^\dagger \: a_L^\dagger\otimes a_R^\dagger \: w)\\
			&=\phi_1(w^\dagger \: a_L\otimes a_R \: w)\Ad{\phi_2(w^\dagger W_g w)}(\phi_1(w^\dagger \: a_L^\dagger\otimes a_R^\dagger \: w))\phi_2(w^\dagger W_g w).
		\end{align}
		Using result 2 and 3 of lemma \ref{lem:phi1phi2matchingCondition} on this now gives
		\begin{align}
			&=\phi_1(w^\dagger \: a_L\otimes a_R \:\Ad{W_g}(a_L^\dagger\otimes a_R^\dagger)w)\phi_2(w^\dagger W_g w)\\
			&=\phi_1(w^\dagger a_L\Ad{W_g}(a_L^\dagger)\otimes a_R \Ad{W_g}( a_R^\dagger)w)\phi_2(w^\dagger W_g w).
		\end{align}
		Since
		\begin{align}
			a_\sigma\Ad{W_g}(a_\sigma^\dagger)&=\pi_0\circ\alpha_0\circ\Theta(A_\sigma\eta_g^\sigma\circ\beta_g^{\sigma U}(A_\sigma^\dagger))=w\epsilon_g^\sigma w^\dagger
		\end{align}
		the first result follows. For the second result, we start by using equation \eqref{eq:TranslationOutOfPhi_K} which gives us
		\begin{align}
			w^\dagger \phi_1(K_g^\sigma)w&=\phi_1(w^\dagger a_L\otimes a_R K_g^\sigma a_L^\dagger\otimes a_R^\dagger w)=\phi_1(w^\dagger a_\sigma K_g^\sigma a_\sigma^\dagger w).
		\end{align}
		Inserting $v^\dagger W_g va_\sigma^\dagger a_\sigma v^\dagger W_g^\dagger v=\id$ in this gives us
		\begin{equation}
			=\phi_1(w^\dagger a_\sigma \underline{\Ad{v^\dagger W_g v}(a_\sigma^\dagger a_\sigma)} K_g^\sigma a_\sigma^\dagger w)=\phi_1(w^\dagger a_\sigma \Ad{v^\dagger W_g v}(a_\sigma^\dagger)K_g^\sigma  \Ad{(K_g^\stkout{\sigma})^\dagger v^\dagger W_g v}(a_\sigma) a_\sigma^\dagger w).
		\end{equation}
		We now use the fact that $K_g W_g=v^\dagger W_g v$ to get
		\begin{align}
			&=\phi_1(w^\dagger a_\sigma \Ad{v^\dagger W_g v}(a_\sigma^\dagger)K_g^\sigma  \Ad{W_g}(a_\sigma) a_\sigma^\dagger w)\\
			&=\phi_1(w^\dagger a_\sigma \Ad{v^\dagger W_g v}(a_\sigma^\dagger)K_g^\sigma  (\epsilon_g^\sigma)^\dagger w).
		\end{align}
		Inserting $w w^\dagger$ in this and using lemma \ref{lem:phi1phi2matchingCondition} gives
		\begin{align}
			&=\phi_1(w^\dagger a_\sigma \Ad{v^\dagger W_g v}(a_\sigma^\dagger)\underline{w w^\dagger}K_g^\sigma \underline{w w^\dagger}  (\epsilon_g^\sigma)^\dagger w)\\
			&=\phi_1(w^\dagger a_\sigma \Ad{v^\dagger W_g v}(a_\sigma^\dagger)w)\phi_1( w^\dagger K_g^\sigma w ) \phi_1( w^\dagger  (\epsilon_g^\sigma)^\dagger w)\\
			&=\underline{v^\dagger}\phi_1(\underline{v} w^\dagger a_\sigma \Ad{v^\dagger W_g v}(a_\sigma^\dagger)w \underline{v^\dagger})\underline{v}\phi_1( w^\dagger K_g^\sigma w ) \phi_1( w^\dagger  (\epsilon_g^\sigma)^\dagger w).
		\end{align}
		The only thing we now have to prove is that
		\begin{equation}\label{eq:5.126Appendix}
			a_\sigma \Ad{v^\dagger W_g v}(a_\sigma^\dagger)=v^\dagger a_\sigma \Ad{W_g}(a_\sigma^\dagger)v=v^\dagger w \epsilon_g^\sigma w^\dagger v
		\end{equation}
		since inserting this, concludes the proof. To do this notice that we can rewrite equation \eqref{eq:5.126Appendix} as
		\begin{equation}
			\Ad{v^\dagger W_g v}(v^\dagger a_\sigma^\dagger v a_\sigma)=v^\dagger a_\sigma^\dagger v a_\sigma.
		\end{equation}
		This equality is proven in lemma \ref{lem:EqualityTwoTranslationsUsingConnectedPath}.
	\end{proof}
	\subsection{Operators belonging to \texorpdfstring{$\omega\circ\gamma^H_{0;1}$}{}}
	The goal of this subsection is to take automorphisms and operators belonging to $\omega$ and relate them to operators belonging to $\omega\circ\gamma^H_{0;1}$. To this end, take $\alpha\in\QAut{\AA}$ such that $\omega=\omega_0\circ\alpha$, for some product state, $\omega_0$. Take $0<\theta_1<\theta_2<\theta_3<\pi/2$ arbitrary. Take $\alpha_L\in\Aut{\AA_L},\alpha_R\in\Aut{\AA_R},V_1\in\UU(\AA)$ and $\Theta\in\Aut{\AA_{W(C_{\theta_3})^c}}$ such that
	\begin{equation}
		\alpha=\Ad{V_1}\circ\alpha_L\otimes\alpha_R\circ\Theta.
	\end{equation}
	Take $\tilde{\beta}_{g,1}$ such that $\omega\circ\tilde{\beta}_{g,1}=\omega$ and such that there exists some $V_{2,g}\in\UU(\AA)$ and some $\eta_g^{\sigma}\in\Aut{\AA_{\nu^\sigma(C_{\theta_1}\cap \sigma)}}$ satisfying
	\begin{equation}
		\tilde{\beta}_{g,1}=\Ad{V_{2,g}}\circ\eta_{g}^L\otimes\eta_{g}^R\circ\beta_g^U.
	\end{equation}
	Take $W_g,u_{L/R}(g,h),v$ and $K_g^{L/R}$ operators belonging to these automorphisms (see lemmas \ref{lem:Definition_W_And_u} and \ref{lem:Definition_K}). Clearly $\alpha\circ\gamma^H_{0;1}$ satisfies $\omega\circ\gamma^{H}_{0;1}=\omega_0\circ\alpha\circ\gamma^{H}_{0;1}$. We also have that $\alpha\circ\gamma^{H}_{0;1}\in\QAut{\AA}$. To show this notice that
	\begin{align}
		\alpha\circ\gamma^{H}_{0;1}&=\Ad{V_1}\circ(\alpha_L\circ\gamma^{H_L}_{0;1}\otimes\alpha_R\circ\gamma^{H_R}_{0;1})\circ\gamma^{\Hsplit}_{1;0}\circ\Theta\circ\gamma^H_{0;1}\\
		&=\Ad{V_1}\circ(\alpha_L\circ\gamma^{H_L}_{0;1}\otimes\alpha_R\circ\gamma^{H_R}_{0;1})\circ\gamma^{\Hsplit}_{1;0}\circ\gamma^{H}_{0;1}\circ\gamma^{H}_{1;0}\circ\Theta\circ\gamma^H_{0;1}.
	\end{align}
	Because of lemmas \versionDifference{\ref{lem:TwoAngleLemmaPart1} and \ref{lem:PropertiesLocallyGeneratedAutomorphisms} part 2}{C.5. and C.4. part 2 of \cite{jappens2023spt}} (which where based on similar lemmas in \cite{ogata2021h3gmathbb}) there exists a $\tilde{\Theta}\in\Aut{\AA_{W(C_{\theta_2})^c}}$ and some $\tilde V_1,A_1\in\UU(\AA)$ such that
	\begin{align}
		\alpha\circ\gamma^H_{0;1}&=\Ad{\tilde V_1}\circ(\alpha_L\circ\gamma^{H_L}_{0;1}\otimes\alpha_R\circ\gamma^{H_R}_{0;1})\circ\tilde\Theta\\
		\label{eq:ProofThatTheH2IndexIsInvariantUnderLGA's_Definition_A1}
		(\alpha_L\circ\gamma^{H_L}_{0;1}\otimes\alpha_R\circ\gamma^{H_R}_{0;1})\circ\tilde\Theta&=\Ad{A_1}\circ \alpha_L\otimes\alpha_R\circ\Theta\circ\gamma^{H}_{0;1}.
	\end{align}
	If we now define the automorphism $\tilde{\beta}_{g,2}=\gamma^{H}_{1;0}\circ\tilde\beta_{g,1}\circ\gamma^{H}_{0;1}$ then this indeed satisfies that $\omega\circ\gamma^{H}_{0;1}\circ\tilde{\beta}_{g,2}=\omega\circ\gamma^{H}_{0;1}$. Define $\tilde\eta_g^\sigma\in\Aut{\AA_{\nu^{\sigma}(C_{\theta_2}\cap\sigma)}}$ through lemma \versionDifference{\ref{lem:TwoAngleLemmaPart3}}{C.7. of \cite{jappens2023spt}} (which was again based on similar lemmas from \cite{ogata2021h3gmathbb}). It satisfies that there exists a $\tilde V_{2,g}\in\UU(\AA),A_{2,g}^{\sigma}\in\UU(\AA_{\nu^\sigma(\sigma)})$ such that
	\begin{align}
		\label{eq:IndexInvariantUnderLGAEtaTildeDefiniton1}
		\gamma^{H}_{1;0}\tilde{\beta}_g\gamma^{H}_{0;1}&=\Ad{\tilde V_{2,g}}\circ\tilde{\eta}_g^L\otimes\tilde{\eta}_g^R\circ\beta_g^U\\
		\label{eq:IndexInvariantUnderLGAEtaTildeDefiniton2}
		\tilde{\eta}_g^\sigma&=\Ad{A_{2,g}^\sigma}\circ\gamma^{H_{\nu^{\sigma}(\sigma)}}_{1;0}\circ\eta_g^\sigma\circ\beta_g^{\sigma U}\circ\gamma^{H_{\nu^{\sigma}(\sigma)}}_{0;1}\circ(\beta_g^{\sigma U})^{-1}.
	\end{align}
	Take some $a$ and $\Phi$ like in equation \eqref{eq:DefinitionOfGroupMorphismVAutEquation} and use this to define $\phi_1$, the group homomorphism defined from lemma \ref{lem:DefinitionOfGroupMorphism} and $\phi_2$, the map defined in lemma \ref{lem:DefinitionOfWgMap}. Now take $A_1$ as in equation \eqref{eq:ProofThatTheH2IndexIsInvariantUnderLGA's_Definition_A1} and define $\delta_g^\sigma\defeq \pi_0\circ\alpha_0\circ\Theta\circ\gamma^{H_\sigma}_{0;1}(A^\sigma_{2,g})$ where $A^\sigma_{2,g}$ is as defined in equation \eqref{eq:IndexInvariantUnderLGAEtaTildeDefiniton2}. The following lemma now holds:
	\begin{lemma}\label{lem:OperatorsBelongingToOmegaAfterH}
		The following operators belong to $\omega\circ\gamma^{H}_{0;1}:$
		\begin{align}
			\tilde u_\sigma(g,h)&=\pi_0(A_1) \phi_1\left(\delta^\sigma_g W_g\delta^\sigma_h W_g^\dagger u_\sigma(g,h)(\delta^\sigma_{gh})^\dagger\right)\pi_0(A_1)^\dagger&\tilde{v}&=\pi_0(A_1)v\pi_0(A_1)^\dagger\\
			\tilde{W}_g&=\pi_0(A_1)\phi_1(\delta^L_g\otimes\delta^R_g)\phi_2( W_g)\pi_0(A_1^\dagger)&\tilde{K}_g^\sigma&=\pi_0(A_1)\phi_1(v^\dagger \delta_g^\sigma v K_g^\sigma (\delta_g^\sigma)^\dagger) \pi_0(A_1)^\dagger.
		\end{align}
	\end{lemma}
	\begin{proof}
		To do this, we first state that a small calculation shows that $\forall x\in\UU(\HH_0)$, satisfying that $\exists\xi\in\Aut{\AA}$, such that
		\begin{equation}
			\Ad{x}\circ\pi_0\circ\alpha_0\circ\Theta=\pi_0\circ\alpha_0\circ\Theta\circ\xi,
		\end{equation}
		we get that
		\begin{equation}\label{eq:IndexInvariantUnderLGA_A1_Transformation}
			\Ad{\pi_0(A_1)x\pi_0(A_1^\dagger)}\circ\pi_0\circ\alpha_0\circ\gamma^{\Hsplit}_{0;1}\circ\tilde{\Theta}=\pi_0\circ\alpha_0\circ\gamma^{\Hsplit}_{0;1}\circ\tilde{\Theta}\circ\gamma^H_{1;0}\circ\xi\circ\gamma^H_{0;1}.
		\end{equation}
		This already proves that
		\begin{align}
			\Ad{\pi_0(A_1)v\pi_0(A_1^\dagger)}\circ\pi_0\circ\alpha_0\circ\gamma^{\Hsplit}_{0;1}\circ\tilde{\Theta}&=\pi_0\circ\alpha_0\circ\gamma^{\Hsplit}_{0;1}\circ\tilde{\Theta}\circ\gamma^H_{1;0}\circ\tau\circ\gamma^H_{0;1}\\
			&=\pi_0\circ\alpha_0\circ\gamma^{\Hsplit}_{0;1}\circ\tilde{\Theta}\circ\tau.
		\end{align}
		To show the second one we get that
		\begin{align}
			&\Ad{\phi_1(\delta_g)\phi_2(W_g)}\circ\pi_0\circ\alpha_0\circ\Theta\\
			&=\Ad{\phi_1(\delta_g)}\circ\pi_0\circ\alpha_0\circ\Theta\circ\gamma^{H}_{0;1}\circ\gamma^{\Hsplit}_{1;0}\circ\eta_g\circ\beta_g^U\circ\gamma^{\Hsplit}_{0;1}\circ\gamma^H_{1;0}\\
			&=\pi_0\circ\alpha_0\circ\Theta\circ\gamma^{H}_{0;1}\circ\gamma^{\Hsplit}_{1;0}\circ\Ad{\gamma^{H_L}_{0;1}(A_{2,g}^L)\otimes\gamma^{H_R}_{0;1}(A_{2,g}^R)}\circ\eta_g\circ\beta_g^U\circ\gamma^{\Hsplit}_{0;1}\circ\gamma^H_{1;0}\\
			&=\pi_0\circ\alpha_0\circ\Theta\circ\gamma^{H}_{0;1}\circ\Ad{A_{2,g}^L\otimes A_{2,g}^R}\circ\gamma^{\Hsplit}_{1;0}\circ\eta_g\circ\beta_g^U\circ\gamma^{\Hsplit}_{0;1}\circ\gamma^H_{1;0}.
		\end{align}
		Using equation \eqref{eq:IndexInvariantUnderLGAEtaTildeDefiniton2} on this we obtain
		\begin{equation}
			=\pi_0\circ\alpha_0\circ\Theta\circ\gamma^H_{0;1}\circ\tilde{\eta}_g\circ\beta_g^U\circ\gamma^H_{1;0}.
		\end{equation}
		Inserting this in equation \eqref{eq:IndexInvariantUnderLGA_A1_Transformation} with $x=\phi_1(\delta_g)\phi_2(W_g)$ and $\xi=\gamma^H_{0;1}\circ\tilde{\eta}_g\circ\beta_g^U\circ\gamma^H_{1;0}$ now proves the fact that $\tilde{W}_g$ is an operator belonging to $\omega\circ\gamma^H_{0;1}$. Proving that $\tilde u_\sigma(g,h)$ and $\tilde{K}_g^\sigma$ are operators belonging to $\omega\circ\gamma_{0;1}^H$ is completely analogous. This concludes the proof.
	\end{proof}
	Additionally, in the case where both $\omega$ and $H$ are translation invariant in both directions, take $w$, $b_g^{L/R}$, and $h_g$, the additional operators belonging to this state. Take $\epsilon_g^\sigma$ as defined in lemma \ref{lem:DefinitionOfEpsilons}. The following lemma now holds:
	\begin{lemma}\label{lem:OperatorsBelongingToOmegaAfterH_TwoTranslations}
		If both $\omega$ and $H$ are translation invariant in both directions then, additionally to the operators presented in lemma \ref{lem:OperatorsBelongingToOmegaAfterH}, the following operators belong to $\omega\circ\gamma_{0;1}^H$ as well:
		\begin{equation}
			\tilde{w}=\pi_0(A_1)w\pi_0(A_1)^\dagger\quad\tilde{b}_g^\sigma=\pi_0(A_1)w^\dagger \phi_1(\delta_g^R)w \phi_1(\epsilon_g^R)\phi_1(b_g^R)\phi_1(\delta_g^R)^\dagger\pi_0(A_1)^\dagger\quad\tilde{h}_g=\pi_0(A_1)\phi_1(h_g)\pi_0(A_1).
		\end{equation}
	\end{lemma}
	\begin{proof}
		The proof that the first four operators belong to $\omega\circ\gamma^H_{0;1}$ is completely analogous to what was done in the proof of theorem \ref{lem:OperatorsBelongingToOmegaAfterH}. We will therefore focus on the $\tilde{b}_g^\sigma$ part (the other two are trivial). To show this, observe that we have
		\begin{align}
			&\Ad{\phiTilde_1(\epsilon_g^R b_g^R)}\circ\pi_0\circ\alpha_0\circ\Theta\\
			&=\pi_0\circ\alpha_0\circ\Theta\circ\gamma^H_{0;1}\circ\gamma^{\Hsplit}_{1;0}\circ\nu^{-1}\circ\Ad{A_R\: \eta_g^R\circ\beta_g^{RU}(A_R^\dagger)}\circ\stkout{\nu\circ\nu^{-1}}\circ\eta_g^R\circ\nu\circ(\eta_g^R)^{-1}\circ\gamma^{\Hsplit}_{0;1}\circ\gamma^{H}_{1;0}\\
			&=\pi_0\circ\alpha_0\circ\Theta\circ\gamma^H_{0;1}\circ\gamma^{\Hsplit}_{1;0}\circ\nu^{-1}\circ\Ad{A_R\: \eta_g^R\circ\beta_g^{RU}(A_R^\dagger)}\circ\eta_g^R\circ\underline{\beta_g^{RU}}\circ\nu\\
			\nonumber
			&\qquad\circ\underline{(\beta_g^{\nu^{-1}(RU)})^{-1}}\circ(\eta_g^R)^{-1}\circ\gamma^{\Hsplit}_{0;1}\circ\gamma^{H}_{1;0}\\
			&=\pi_0\circ\alpha_0\circ\Theta\circ\gamma^H_{0;1}\circ\gamma^{\Hsplit}_{1;0}\circ\nu^{-1}\circ\Ad{A_R}\circ\eta_g^R\circ\beta_g^{RU}\circ\Ad{A_R^\dagger}\circ\nu\\
			\nonumber
			&\qquad\circ(\beta_g^{\nu^{-1}(RU)})^{-1}\circ(\eta_g^R)^{-1}\circ\gamma^{\Hsplit}_{0;1}\circ\gamma^{H}_{1;0}\\
			&=\pi_0\circ\alpha_0\circ\Theta\circ\gamma^H_{0;1}\circ\underline{\nu^{-1}\circ\nu}\circ\gamma^{\Hsplit}_{1;0}\circ\nu^{-1}\circ\Ad{A_R}\circ\eta_g^R\circ\beta_g^{RU}\circ\Ad{A_R^\dagger}\circ\nu\circ\underline{\gamma^{\Hsplit}_{0;1}}\\
			\nonumber
			&\qquad \circ\underline{\nu^{-1}\circ\nu\circ\gamma^{\Hsplit}_{1;0}}\circ(\beta_g^{\nu^{-1}(RU)})^{-1}\circ(\eta_g^R)^{-1}\circ\gamma^{\Hsplit}_{0;1}\circ\gamma^{H}_{1;0}.
		\end{align}
		After using the restriction to $R$ of equation \eqref{eq:TranslatingSplittedTimeEvolution}, this leads to
		\begin{align}
			&=\pi_0\circ\alpha_0\circ\Theta\circ\nu^{-1}\circ\gamma^H_{0;1}\circ\underline{\gamma^{\Hsplit}_{1;0}\circ\stkout{(\otimes_\mu\Phi^{\mu\nu(R)})^{-1}}}\circ\eta_g^R\circ\beta_g^{RU}\circ\underline{\stkout{(\otimes_\mu\Phi^{\mu\nu(R)})}\circ\gamma^{\Hsplit}_{0;1}}\circ\nu\\
			\nonumber
			&\qquad \circ\gamma^{\Hsplit}_{1;0}\circ(\beta_g^{\nu^{-1}(RU)})^{-1}\circ(\eta_g^R)^{-1}\circ\gamma^{\Hsplit}_{0;1}\circ\gamma^{H}_{1;0}.
		\end{align}
		On the other hand, we have that
		\begin{equation}
			\Ad{w^\dagger \phiTilde_1(\delta_g^\sigma)w}\circ\pi_0\circ\alpha_0\circ\Theta=\pi_0\circ\alpha_0\circ\Theta\circ\nu^{-1}\circ\gamma^H_{0;1}\circ\Ad{A^\sigma_{2,g}}\circ\gamma^H_{1;0}\circ\nu.
		\end{equation}
		This leads us to
		\begin{align}
			&\Ad{w^\dagger\phiTilde_1(\delta_g^R)w\phiTilde_1(\epsilon_g^R b_g^R)}\circ\pi_0\circ\alpha_0\circ\Theta\\
			&=\pi_0\circ\alpha_0\circ\Theta\circ\nu^{-1}\circ\gamma^H_{0;1}\circ\Ad{A_{2,g}^R}\circ\gamma^{\Hsplit}_{1;0}\circ\eta_g^R\circ\beta_g^{RU}\circ\gamma^{\Hsplit}_{0;1}\circ\nu\\
			\nonumber
			&\qquad \circ\gamma^{\Hsplit}_{1;0}\circ(\beta_g^{\nu^{-1}(RU)})^{-1}\circ(\eta_g^R)^{-1}\circ\gamma^{\Hsplit}_{0;1}\circ\gamma^{H}_{1;0}\\
			&=\pi_0\circ\alpha_0\circ\Theta\circ\gamma^H_{0;1}\circ\nu^{-1}\circ\tilde\eta_g^R\circ\beta_g^{RU}\circ\nu\circ\gamma^{\Hsplit}_{1;0}\circ(\beta_g^{\nu^{-1}(RU)})^{-1}\circ(\eta_g^R)^{-1}\circ\gamma^{\Hsplit}_{0;1}\circ\gamma^{H}_{1;0}.
		\end{align}
		Now define $S=\{(0,y)|y\in\ZZ\}$ then
		\begin{equation}
			=\pi_0\circ\alpha_0\circ\Theta\circ\gamma^H_{0;1}\circ\nu^{-1}\circ\tilde\eta_g^R\circ\beta_g^{\nu(RU)}\circ\stkout{\beta_g^S}\circ\nu\circ\gamma^{\Hsplit}_{1;0}\circ\stkout{(\beta_g^{\nu^{-1}(S)})^{-1}}\circ(\beta_g^{RU})^{-1}\circ(\eta_g^R)^{-1}\circ\gamma^{\Hsplit}_{0;1}\circ\gamma^{H}_{1;0}.
		\end{equation}
		Doing the same thing on the right will give us
		\begin{align}
			&\Ad{w^\dagger\phiTilde_1(\delta_g^R)w\phiTilde_1(\epsilon_g^R b_g^R)\phiTilde_1(\delta_g^R)^\dagger}\circ\pi_0\circ\alpha_0\circ\Theta\\
			&=\pi_0\circ\alpha_0\circ\Theta\circ\gamma^H_{0;1}\circ\nu^{-1}\circ\tilde\eta_g^R\circ\beta_g^{\nu(RU)}\circ\nu\circ\gamma^{\Hsplit}_{1;0}\circ(\beta_g^{RU})^{-1}\circ(\eta_g^R)^{-1}\circ\gamma^{\Hsplit}_{0;1}\circ\Ad{(A_{2,g}^R)^\dagger}\circ\gamma^{H}_{1;0}\\
			&=\pi_0\circ\alpha_0\circ\Theta\circ\gamma^H_{0;1}\circ\nu^{-1}\circ\tilde\eta_g^R\circ\stkout{\beta_g^{\nu(RU)}}\circ\nu\circ\stkout{(\beta_g^{RU})^{-1}}\circ(\tilde\eta_g^R)^{-1}\circ\circ\gamma^{H}_{1;0}.
		\end{align}
		The role that the $A_1$ will play is analogous to the proof of theorem \ref{thrm:IndexInvariantUnderLGA} (see equation \eqref{eq:IndexInvariantUnderLGA_A1_Transformation}), concluding the proof.
	\end{proof}
	\subsection{The indices are invariant under LGAs}\label{sec:IndexInvariantUnderLGA}
	\begin{theorem}\label{thrm:IndexInvariantUnderLGA}
		$\textrm{Index}(\omega)=\textrm{Index}(\omega\circ\gamma^H_{0;1})$
	\end{theorem}
	\begin{proof}
		To show this, we merely have to show that the index is invariant under the transformations indicated in lemma \ref{lem:OperatorsBelongingToOmegaAfterH}. The fact that the index is invariant under $\Ad{\pi_0(A_1)}$ and under $\phi(\cdot)$ is just because both transformations are homomorphisms. To show that it is invariant under the transformation with de $\delta^\sigma_g$ we invoke lemma \ref{lem:TransformationUnderDelta}.
	\end{proof}
	\begin{theorem}\label{thrm:IndexInvariantUnderLGA_TwoTrans}
		$\textrm{Index}_{\text{2 trans}}(\omega)=\textrm{Index}_{\text{2 trans}}(\omega\circ\gamma^H_{0;1}).$
	\end{theorem}
	\begin{proof}
		To show this, we merely have to show that the index is invariant under the transformations indicated in lemma \ref{lem:OperatorsBelongingToOmegaAfterH} and \ref{lem:OperatorsBelongingToOmegaAfterH_TwoTranslations}. Let $\alpha\in\hom(G,U(1))$ be the old index and $\tilde{\alpha}\in\hom(G,U(1))$ the new one. We have that
		\begin{align}
			\alpha(g)&=v^\dagger b_g^R v h_g K_g^R (b_g^R)^\dagger w^\dagger (K_g^R)^\dagger w\\
			&=\phi_1(v^\dagger b_g^R v h_g K_g^R (b_g^R)^\dagger w^\dagger (K_g^R)^\dagger w)\\
			&=\phi_1(v^\dagger b_g^R v h_g K_g^R (b_g^R)^\dagger)\phi_1( w^\dagger (K_g^R)^\dagger w).
		\end{align}
		Using equation \eqref{eq:TransformationOfKUnderEpsilon} on this gives us
		\begin{align}
			&=\phi_1(v^\dagger b_g^R v h_g K_g^R (b_g^R)^\dagger)\phi_1((\epsilon_g^R)^\dagger)w^\dagger\phi_1(  (K_g^R)^\dagger )wv^\dagger \phi_1(\epsilon_g^R)v\\
			&=\phi_1(v^\dagger \epsilon_g^R b_g^R v h_g K_g^R (b_g^R)^\dagger (\epsilon_g^R)^\dagger)w^\dagger\phi_1(  (K_g^R)^\dagger )w.
		\end{align}
		This shows that the index is invariant under the $\phi_1$ and $\epsilon^R_g$ part (simultaneously). The fact that the index is invariant under the $\phi_1(\delta_g)$ part is now completely equivalent to the proof of lemma \ref{lem:TransformationUnderDeltaTwoTranslations}. The fact that the index remains unchanged by the $\Ad{\pi_0(A_1)}$ transformation is trivial. This concludes the proof.
	\end{proof}
	\section{The \texorpdfstring{$H^1(G,\TT)$}{H1}-valued index is invariant under rotations by \texorpdfstring{$90^\circ$}{90}}\label{sec:H1ValuedIndexInvariantUnderRotations}
	Take
	\begin{equation}
		\omega\in\left\{\omega\in\PP(\AA)\left| \begin{matrix}\omega\text{ is an SRE state,}\\\omega\circ\beta_g=\omega,\:\omega\circ\tau=\omega\text{ and }\omega\circ\nu=\omega\end{matrix} \right.\right\}
	\end{equation}
	arbitrary and let $\mu$ be the automorphism that rotates the lattice by $90^\circ$. This section will be dedicated to proving that
	\begin{equation}
		\textrm{Index}_{\text{2 trans}}^{\AA,U(g)}(\omega)=\textrm{Index}_{\text{2 trans}}^{\AA,U(g)}(\omega\circ\mu).
	\end{equation}
	For the $H^2(G,\TT)$-valued index, the similar statement is not true, that is why in equation \eqref{eq:2TranslationsIntroduction}, we have two $H^2(G,\TT)$-valued indices. This is because if we have two translation symmetries, $\omega\mapsto\textrm{Index}^{\AA,U(g)}(\omega)$ and $\omega\mapsto\textrm{Index}^{\AA,U(g)}(\omega\circ\mu)$ are both well-defined indices.
	\\\\
	In what follows, let $\omega_0$ be the product state and let $\gamma^\Phi_{0;1}$ be the disentangler such that $\omega=\omega_0\circ\gamma^\Phi_{0;1}$. Sometimes we will need a specific decomposition of this locally generated Automorphism:

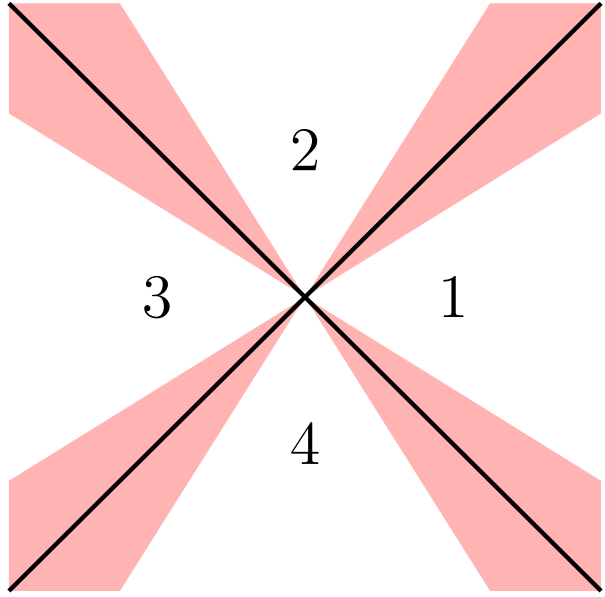
\begin{figure}
	\centering
	\def\s{0.5}
	\resizebox{0.35\textwidth}{!}{%
		\begin{tikzpicture}
			\fill[fill=red!30!white] (0,0) -- (2.5*\s,4*\s) -- (4*\s,4*\s) -- (4*\s,2.5*\s);
			\fill[fill=red!30!white] (0,0) -- (2.5*\s,-4*\s) -- (4*\s,-4*\s) -- (4*\s,-2.5*\s);
			\fill[fill=red!30!white] (0,0) -- (-2.5*\s,4*\s) -- (-4*\s,4*\s) -- (-4*\s,2.5*\s);
			\fill[fill=red!30!white] (0,0) -- (-2.5*\s,-4*\s) -- (-4*\s,-4*\s) -- (-4*\s,-2.5*\s);
			
			\draw[draw=black,line width=0.3mm] (-4*\s,4*\s) -- (4*\s,-4*\s);
			\draw[draw=black,line width=0.3mm] (-4*\s,-4*\s) -- (4*\s,4*\s);
			
			\node at (2*\s,0) {$1$};
			\node at (0,2*\s) {$2$};
			\node at (-2*\s,0) {$3$};
			\node at (0,-2*\s) {$4$};
		\end{tikzpicture}}
	\caption{This figure indicates four different areas of $\ZZ^2$ and we will take $\pi_0=\bigotimes_{i=1}^4\pi_{0,i}$ where each of the $\pi_{0,i}$ is an irreducible representation of the restriction of $\AA$ to each of the areas. The red areas are chosen such that they do not overlap with the support of $\tilde{\eta}_g$ and $\eta_g$.}
	\label{fig:IndexInvariantUnderRotationDecomposition}
\end{figure}

	\begin{lemma}\label{lem:DecompositionOfLGAInCross}
		Let $\{1,2,3,4\}$ be regions indicated in figure \ref{fig:IndexInvariantUnderRotationDecomposition} (including the red areas) and let $\AA_{\text{red}}$ be the part of the $C^*$ algebra that has support in the red regions of figure \ref{fig:IndexInvariantUnderRotationDecomposition}. There exists a $B\in\UU(\AA)$ and a $\Theta\in\Aut{\AA_{\text{red}}}$ such that $\beta^{\mu\nu}_g\circ\Theta=\Theta\circ\beta^{\mu\nu}_g$ (for all $\mu\in\{U,D\}$ and $\nu\in\{L,R\}$) satisfying
		\begin{equation}
			\gamma^\Phi_{0;1}=\Ad{B}\circ\gamma^{\sum_{i=1}^4\Phi_i}_{0;1}\circ\Theta
		\end{equation}
		where $\Phi_i$ is the restriction of $\Phi$ to region $i\in\{1,2,3,4\}$.
	\end{lemma}
	\begin{proof}
		Completely analogous to the proof of (for example) lemma \versionDifference{\ref{lem:PropertiesLocallyGeneratedAutomorphisms}}{C.4. of \cite{jappens2023spt}} which was in turn based on similar lemmas in \cite{ogata2021h3gmathbb}.
	\end{proof}
	Let $A_{g,1},A_{g,2}\in\UU(\AA),\eta^\sigma_g\in\Aut{\AA_{\nu^{\sigma}(C_\theta\cap\sigma)}}$ and $\eta^\rho_{2,g}\in\Aut{\AA_{\tau^{\rho}(\mu(C_\theta)\cap\rho)}}$ be such that the automorphisms
	\begin{align}
		\tilde{\beta}_{g,1}&=\Ad{A_{g,1}}\circ\eta_g^L\otimes\eta_g^R\circ\beta_g^U&\tilde{\beta}_{g,2}&=\Ad{A_{g,2}}\circ\eta_{2,g}^U\otimes\eta_{2,g}^D\circ\beta_g^L
	\end{align}
	satisfy $\omega\circ\tilde{\beta}_{g,1}=\omega\circ\tilde{\beta}_{g,2}=\omega$. Let $(\HH_0=\bigotimes_i\HH_{0,i},\pi_0=\bigotimes_{i=1}^4\pi_{0,i},\bigotimes_i\Omega_{0,i})$ be a GNS triple for $\omega_0$ (see the caption of figure \ref{fig:IndexInvariantUnderRotationDecomposition}). The following now holds:
	\begin{lemma}\label{lem:DefinitionX_gLemma}
		There exists an $X_g\in\UU(\HH_0)$ satisfying that
		\begin{equation}\label{eq:DefinitionX_gLemma}
			\Ad{X_g}\circ\pi_0\circ\gamma^\Phi_{0;1}=\pi_0\circ\gamma^\Phi_{0;1}\circ\tilde{\eta}^U_g\circ\eta^R_g\circ\beta_g^{RU}
		\end{equation}
		where $\tilde{\eta}_g=\beta_g^U\circ\eta_{2,g^{-1}}\circ(\beta_g^U)^{-1}$.
	\end{lemma}
	\begin{proof}
		Let $W_g$ and $\tilde W_g$ be such that
		\begin{align}
			\Ad{W_g}\circ\pi_0\circ\gamma^\Phi_{0;1}& =\pi_0\circ\gamma^\Phi_{0;1}\circ\eta_g\circ\beta_g^U
			&\Ad{\tilde W_g} \circ\pi_0\circ\gamma^\Phi_{0;1}& =\pi_0\circ\gamma^\Phi_{0;1}\circ\eta_{2,g}\circ\beta_g^L
		\end{align}
		then
		\begin{align}
			\Ad{W_g\tilde{W}_{g^{-1}}}\circ\pi_0\circ\gamma^\Phi_{0;1}&=\pi_0\circ\gamma^\Phi_{0;1}\circ\eta_g\circ\beta_g^U\circ\eta_{2,g^{-1}}\circ(\beta_g^L)^{-1}\\
			&=\pi_0\circ\gamma^\Phi_{0;1}\circ\eta_g\circ\tilde{\eta}_g\circ\beta_g^{RU}\circ(\beta_g^{LD})^{-1}.
		\end{align}
		Now we use lemma \ref{lem:DecompositionOfLGAInCross} and obtain
		\begin{align}
			\Ad{W_g\tilde{W}_{g^{-1}}}\circ\pi_0\circ\Ad{B}\circ\gamma^{\sum_i\Phi_i}_{0;1}\circ\stkout{\Theta}&=\pi_0\circ\Ad{B}\circ\gamma^{\sum_i\Phi_i}_{0;1}\circ\stkout{\Theta}\circ\eta_g\circ\tilde{\eta}_g\circ\beta_g^{RU}\circ(\beta_g^{LD})^{-1}\\
			\Ad{\pi_0(B^\dagger)W_g\tilde{W}_{g^{-1}}\pi_0(B)}\circ\pi_0\circ\gamma^{\sum_i\Phi_i}_{0;1}&=\pi_0\circ\gamma^{\sum_i\Phi_i}_{0;1}\circ\eta_g\circ\tilde{\eta}_g\circ\beta_g^{RU}\circ(\beta_g^{LD})^{-1}\\
			\Ad{\pi_0(B^\dagger)W_g\tilde{W}_{g^{-1}}\pi_0(B)}\circ\pi_0&=\left(\pi_{0,1}\otimes\pi_{0,2}\circ\gamma^{\Phi_1+\Phi_2}_{0;1}\circ\eta_g^R\circ\tilde{\eta}_g^U\circ\beta_g^{RU}\circ\gamma^{\Phi_1+\Phi_2}_{1;0}\right)\\
			\nonumber
			&\qquad\otimes \left(\pi_{0,3}\otimes\pi_{0,4}\circ\gamma^{\Phi_3+\Phi_4}_{0;1}\circ\eta_g^L\circ\tilde{\eta}_g^D\circ(\beta_g^{LD})^{-1}\circ\gamma^{\Phi_3+\Phi_4}_{1;0}\right).
		\end{align}
		By lemma \ref{lem:SplittingOfUnitary}, there exist $\tilde{X}_g\in\UU(\HH_{0,1}\otimes\HH_{0,2})$ and $\tilde{Y}_g\in\UU(\HH_{0,3}\otimes\HH_{0,4})$ such that $\tilde{X}_g\otimes\tilde{Y}_g=\pi_0(B^\dagger)W_g\tilde{W}_{g^{-1}}\pi_0(B)$. Define $X_g=\pi_0(B)(\tilde{X}_g\otimes\id)\pi_0(B^\dagger)$. This indeed satisfies equation \eqref{eq:DefinitionX_gLemma} concluding the proof.
	\end{proof}
	\begin{lemma}
		Let $v,w,K_g^R$ and $b_g^R$ be such that
		\begin{align}
			\Ad{v}\circ\pi&=\pi\circ\tau&\Ad{K_g^R}\circ\pi&=\pi\circ\tau^{-1}\circ\eta_g^R\circ\beta_g^{RU}\circ\tau\circ(\beta_g^{RU})^{-1}\circ(\eta_g^R)^{-1}\\
			\Ad{w}\circ\pi&=\pi\circ\nu&\Ad{b_g^R}\circ\pi&=\pi\circ\nu^{-1}\circ\eta_g^R\circ\nu\circ(\eta_g^R)^{-1}
		\end{align}
		(where $\pi=\pi_0\circ\gamma^\Phi_{0;1}$) then
		\begin{equation}\label{eq:H^1IndexRotationInvariantHorizontalIndex}
			v^\dagger b_g^R vh_g K_g^R (b_g^R)^\dagger=\alpha(g)w^\dagger K_g^R w
		\end{equation}
		holds. Moreover $\alpha=\textrm{Index}_{\text{2 trans}}(\omega)$. If one now defines $\tilde{b}_g^R$ and $\tilde{K}_g^U$ such that
		\begin{align*}
			\tilde{b}_g^UK_g^R X_g&=v^\dagger X_g v\Label{eq:H^1IndexRotationInvariantFirstDefinition}&b_g^R\tilde{K}_g^UX_g&=w^\dagger X_g w\Label{eq:H^1IndexRotationInvariantSecondDefinition}
		\end{align*}
		then
		\begin{equation}\label{eq:H^1IndexRotationInvariantVerticalIndex}
			w^\dagger \tilde b_g^U wh_g \tilde K_g^U (\tilde b_g^U)^\dagger=\tilde\alpha(g)v^\dagger \tilde K_g^U v
		\end{equation}
		holds. Moreover $\tilde\alpha=\textrm{Index}_{\text{2 trans}}(\omega\circ\mu)$.
	\end{lemma}
	\begin{proof}
		The existence of the $v,w,K_g^R$ and $b_g^R$ is due to the fact that $\gamma^\Phi_{0;1}\in \QAut{\AA}$. This follows from lemma \versionDifference{\ref{lem:PropertiesLocallyGeneratedAutomorphisms}}{C.4. of \cite{jappens2023spt}} (which was based on similar lemmas in \cite{ogata2021h3gmathbb}) and from the fact that $\textrm{SQAut}_1(\AA)\subset \QAut{\AA}$ (see definition \ref{def:SQAut2}). By construction this also implies that indeed $\alpha=\textrm{Index}_{\text{2 trans}}(\omega)$. We now only have to show the second part of the proof. We have that
		\begin{align}
			\Ad{\tilde{K}_g^U}\circ\pi_0\circ\gamma^\Phi_{0;1}&=\Ad{(b_g^R)^\dagger w^\dagger X_g w X_g^\dagger}\circ\pi_0\circ\gamma^\Phi_{0;1}\\
			&=\Ad{(b_g^R)^\dagger w^\dagger X_g w}\circ\pi_0\circ\gamma^\Phi_{0;1}\circ(\beta_g^{RU})^{-1}\circ(\eta_g^R)^{-1}\circ(\tilde{\eta}_g^U)^{-1}\\
			&=\Ad{(b_g^R)^\dagger}\circ\pi_0\circ\gamma^\Phi_{0;1}\circ\nu^{-1}\circ\tilde{\eta}_g^U\circ\eta_g^R\circ\beta_g^{RU}\circ\nu\circ(\beta_g^{RU})^{-1}\circ(\eta_g^R)^{-1}\circ(\tilde{\eta}_g^U)^{-1}\\
			&=\pi_0\circ\gamma^\Phi_{0;1}\circ\nu^{-1}\circ\tilde{\eta}_g^U\circ\beta_g^{RU}\circ\nu\circ(\beta_g^{RU})^{-1}\circ(\tilde{\eta}_g^U)^{-1}.
		\end{align}
		By similar arguments, we get that
		\begin{equation}
			\Ad{\tilde{b}_g^U}\circ\pi_0\circ\gamma^\Phi_{0;1}=\pi_0\circ\gamma^\Phi_{0;1}\circ\tau^{-1}\circ\tilde{\eta}_g^U\circ\tau\circ(\tilde{\eta}_g^U)^{-1}.
		\end{equation}
		This also shows that
		\begin{equation}
			\Ad{\tilde{K}_g^U}\circ\pi_0\circ\gamma^\Phi_{0;1}\circ\mu=\pi_0\circ\gamma^\Phi_{0;1}\circ\mu\circ\mu^{-1}\circ\nu^{-1}\circ\tilde{\eta}_g^U\circ\beta_g^{RU}\circ\nu\circ(\beta_g^{RU})^{-1}\circ(\tilde{\eta}_g^U)^{-1}\circ\mu
		\end{equation}
		and similarly for $\tilde{b}_g^U$ and therefore these operators also belong to $\omega\circ\mu$. This together with the independence of the choice of $\tilde{\eta}_g$ proves the second result.
	\end{proof}
	\begin{lemma}
		The following equalities hold
		\begin{align*}
			[h_g ,w^\dagger \tilde{b}_g^U w]&=0 \Label{eq:H^1IndexRotationInvariantCommutator1}& [w^\dagger \tilde{b}_g^U w,v^\dagger b_g^R v]&=0 \Label{eq:H^1IndexRotationInvariantCommutator4}\\
			[h_g, v^\dagger b_g^R v]&=0 \Label{eq:H^1IndexRotationInvariantCommutator2}& [K_g^R,\tilde{b}_g^U] &= 0 \Label{eq:H^1IndexRotationInvariantCommutator5}\\
			[h_g ,K_g^R]&=0 \Label{eq:H^1IndexRotationInvariantCommutator3}& [K_g^R,v^\dagger \tilde{K}^U_g v] &= 0 \Label{eq:H^1IndexRotationInvariantCommutator6}
		\end{align*}
	\end{lemma}
	\begin{proof}
		All of the above commute as automorphisms on the $C^*$ algebra, we now have to show that the operators in the GNS space commute as well. First we will start by proving equations \eqref{eq:H^1IndexRotationInvariantCommutator1}, \eqref{eq:H^1IndexRotationInvariantCommutator2} and \eqref{eq:H^1IndexRotationInvariantCommutator3}. For any operator $x\in\UU(\HH_0)$ satisfying that there exists a $\xi\in\Aut{\AA}$ such that
		\begin{equation}
			\Ad{x}\circ\pi_0\circ\gamma^{\Phi}_{0;1}=\pi_0\circ\gamma^{\Phi}_{0;1}\circ\xi
		\end{equation}
		we get that
		\begin{equation}
			\Ad{x}(h_g)=\pi_0\circ\gamma^\Phi_{0;1}\circ\xi(U_{(-1,-1)}(g)).
		\end{equation}
		Applying this to these equations gives the desired result. To show the remaining equations decompose $\pi_0$ and $\gamma^\Phi_{0;1}$ as $\gamma^\Phi_{0;1}=\Ad{B}\circ\bigotimes_{i=1}^4\gamma^{\Phi_i}_{0;1}\circ\Theta$ and $\pi_0=\bigotimes_{i=1}^4\pi_{0,i}$ where the $\pi_{0,i}$ are irreducible representations in $\AA_i$, the areas $i$ are indicated in figure \ref{fig:IndexInvariantUnderRotationDecomposition} and the $\Theta$ have support in the areas shaded red. This shows that up to some inner automorphisms, the operators with an $R$ on top act on different parts of the decomposition of $\pi_0$ than the operators with a $U$ on top do. Combining this with lemma \ref{lem:SplittingOfUnitary} concludes the proof.
	\end{proof}
	\begin{lemma}
		$\tilde{\alpha}(g)=\alpha(g)$.
	\end{lemma}
	\begin{proof}
		We will work out the expression
		\begin{equation}\label{eq:H^1IndexRotationInvariantExpression}
			h_g v^\dagger b_g^R v w^\dagger\tilde{b}_g^U w K_g^R\tilde{K}_g^U X_g
		\end{equation}
		in two different ways. On the one hand we will use \eqref{eq:H^1IndexRotationInvariantCommutator4} together with equation \eqref{eq:H^1IndexRotationInvariantCommutator1} to get
		\begin{equation}
			\eqref{eq:H^1IndexRotationInvariantExpression}= \underline{w^\dagger\tilde{b}_g^U w h_g v^\dagger b_g^R v} K_g^R\tilde{K}_g^U X_g.
		\end{equation}
		Now using equation \eqref{eq:H^1IndexRotationInvariantHorizontalIndex} we get
		\begin{align}
			\eqref{eq:H^1IndexRotationInvariantExpression}&=\alpha(g)w^\dagger\tilde{b}_g^U w w^\dagger K_g^R w b_g^R\tilde{K}_g^U X_g=\alpha(g)w^\dagger\tilde{b}_g^U  K_g^R X_g w=\alpha(g)w^\dagger v^\dagger X_g v w
		\end{align}
		where we've used equation \eqref{eq:H^1IndexRotationInvariantSecondDefinition} in the first equality and \eqref{eq:H^1IndexRotationInvariantFirstDefinition} in the second equality. On the other hand, starting with equation \eqref{eq:H^1IndexRotationInvariantCommutator2}, \eqref{eq:H^1IndexRotationInvariantCommutator5} and \eqref{eq:H^1IndexRotationInvariantCommutator3} gives
		\begin{align}
			\eqref{eq:H^1IndexRotationInvariantExpression}&= \underline{v^\dagger b_g^R v h_g} w^\dagger\tilde{b}_g^U w K_g^R\tilde{K}_g^U X_g =v^\dagger b_g^R v \underline{K_g^R h_g w^\dagger\tilde{b}_g^U w} \tilde{K}_g^U X_g.
		\end{align}
		Now using equation \eqref{eq:H^1IndexRotationInvariantVerticalIndex} we get
		\begin{align}
			\eqref{eq:H^1IndexRotationInvariantExpression}&=\tilde{\alpha}(g)v^\dagger b_g^R v K_g^R v^\dagger \tilde{K}_g^U v \tilde{b}_g^U X_g=\tilde{\alpha}(g)v^\dagger b_g^R v \underline{v^\dagger \tilde{K}_g^U v \tilde{b}_g^U K_g^R} X_g\\
			&=\tilde{\alpha}(g)v^\dagger b_g^R \stkout{v v^\dagger} \tilde{K}_g^U \stkout{v v^\dagger} X_g v=\tilde{\alpha}(g) v^\dagger w^\dagger X_g w v
		\end{align}
		where we've used equation \eqref{eq:H^1IndexRotationInvariantCommutator6} and \eqref{eq:H^1IndexRotationInvariantCommutator5} in the first equality, equation \eqref{eq:H^1IndexRotationInvariantFirstDefinition} in the second equality and equation \eqref{eq:H^1IndexRotationInvariantSecondDefinition} in the last equality. These two results can only be consistent if $\alpha(g)=\tilde{\alpha}(g)$ for all $g\in G$.
	\end{proof}
	\section{The role of stacking}\label{sec:the-role-of-stacking}
	Take $\AA$ a quasi-local $C^*$ algebra on a two-dimensional lattice and take $U\in\hom(G,U(\CC^d))$ an on-site group action arbitrary. The goal of this section is to prove the following theorem:
	\begin{theorem}\label{th:IndexMorphismUnderStacking}
		Let $\omega_1,\omega_2\in\PP(\AA)$ satisfy assumption \ref{assumption} then
		\begin{equation}\label{eq:IndexMorphismUnderStacking}
			\textrm{Index}^{\AA^2,U\otimes U}(\omega_1\otimes_{\text{stack}}\omega_2)=\textrm{Index}^{\AA,U}(\omega_1)+\textrm{Index}^{\AA,U}(\omega_2).
		\end{equation}
		If additionally, $\omega_1$ and $\omega_2$ satisfy assumption \ref{assumption:2Translations} then
		\begin{equation}\label{eq:IndexMorphismUnderStackingTwoTranslations}
			\textrm{Index}^{\AA^2,U\otimes U}_{\text{2 trans}}(\omega_1\otimes_{\text{stack}}\omega_2)=\textrm{Index}^{\AA,U}(\omega_1)+\textrm{Index}^{\AA,U}_{\text{2 trans}}(\omega_2).
		\end{equation}
	\end{theorem}
	To prove this theorem, let $\omega_{0i}$ and $\gamma_{0;1}^{\Phi_i}$ (for all $i\in\{1,2\}$) be product states and disentanglers satisfying that $\omega_i=\omega_{0i}\circ\gamma_{0;1}^{\Phi_i}$. Let $(\HH_{0i},\pi_{0i},\Omega_{0i})$ be GNS triples for the $\omega_{0i}$. Clearly, $(\HH_{01}\otimes\HH_{02},\pi_{01}\otimes\pi_{02},\Omega_{01}\otimes\Omega_{02})$ is a GNS triple for $\omega_{01}\otimes_{\text{stack}}\omega_{02}$. The proof of the theorem therefore follows by using the following lemma:
	\begin{lemma}
		Let $W_{g,i},u^\sigma_i(g,h),K_{g,i}^\sigma$ and $v_{i}$ be operators belonging to $\omega_i$ then $W_{g,1}\otimes W_{g,2},u^\sigma_1(g,h)\otimes u^\sigma_2(g,h),K_{g,1}^\sigma\otimes K_{g,2}^\sigma$ and $v_{1}\otimes v_{2}$ are operators belonging to $\omega_1\otimes_{\text{stack}}\omega_2$. When $\omega_1$ and $\omega_2$ are translation invariant in both directions, the similar statement holds for $w,b_g^\sigma$ and $h_g$.
	\end{lemma}
	\begin{proof}
		To show this we simply have to apply the adjoint action of these operators on $\pi_{01}\otimes\pi_{02}$ and see that these are indeed implementations of the right automorphisms. This is so by construction.
	\end{proof}
	To conclude the proof theorem \ref{th:IndexMorphismUnderStacking}, observe that
	\begin{align}
		&C(g,h)\\
		\nonumber
		&=(K_{g,1}^R\otimes K_{g,2}^R)\Ad{W_{g,1}\otimes W_{g,2}}\left((K_{h,1}^R\otimes K_{h,2}^R)\Ad{(W_{h,1}\otimes W_{h,2})(W_{gh,1}\otimes W_{gh,2})^\dagger}\left((K_{gh,1}^R\otimes K_{gh,2}^R)^\dagger\right)\right)\\
		&(u^R_1(g,h)\otimes u^R_2(g,h))v^\dagger (u^R_1(g,h)\otimes u^R_2(g,h))^\dagger v=C_1(g,h)C_2(g,h)
	\end{align}
	concluding the proof of the first part. The second part is analogous.
	\begin{remark}\label{rem:StackingH3ValuedIndex}
		The same argument can be used to argue that the $H^3$-valued index is invariant under stacking.
	\end{remark}
	\section{The indices are independent on the choice of on-site Hilbert space}\label{sec:the-indices-are-independent-on-the-choice-of-on-site-hilbert-space}
	Let $V \in U(\CC^d)$ be a unitary operator on the on-site Hilbert space and take $i_V$ as defined in subsection \ref{sec:QuasiLocalC*Algebra}. In this section we will prove the following lemma:
	\begin{lemma}\label{lem:IndependenceOnChoiceOnSitHilbertSpace}
		Let $\omega$ satisfy assumption \ref{assumption} then for any choice of automorphisms $(\theta,\tilde{\beta}_g,\eta_g,\alpha_0,\Theta,\omega,\omega_0)$ we get that
		\begin{align}
			&\textrm{Index}^{\AA,\Ad{V^\dagger}(U)}(\theta,i_V^{-1}\circ\tilde{\beta}_g\circ i_V,i_V^{-1}\circ\eta_g\circ i_V,i_V^{-1}\circ\alpha_0\circ i_V,i_V^{-1}\circ\Theta\circ i_V,\omega\circ i_V,\omega_0\circ i_V)\\
			&=\textrm{Index}^{\AA,U}(\theta,\tilde{\beta}_g,\eta_g,\alpha_0,\Theta,\omega,\omega_0).
		\end{align}
		Additionally, when $\omega$ satisfies assumption \ref{assumption:2Translations}, we get that
		\begin{align}
			&\textrm{Index}^{\AA,\Ad{V^\dagger}(U)}_{\textrm{2 trans}}(\theta,i_V^{-1}\circ\tilde{\beta}_g\circ i_V,i_V^{-1}\circ\eta_g\circ i_V,i_V^{-1}\circ\alpha_0\circ i_V,i_V^{-1}\circ\Theta\circ i_V,\omega\circ i_V,\omega_0\circ i_V)\\
			&=\textrm{Index}^{\AA,U}_{\textrm{2 trans}}(\theta,\tilde{\beta}_g,\eta_g,\alpha_0,\Theta,\omega,\omega_0).
		\end{align}
	\end{lemma}
	We do this by proving the following lemma:
	\begin{lemma}\label{lem:IndependenceOnChoiceOnSitHilbertSpaceOperatorsBelongingTo}
		Let $W_g$, $u_R(g,h)$, $v$ and $K^R_g$ be operators belonging to
		\begin{equation}
			\textrm{Index}^{\AA,U}(\theta,\tilde{\beta}_g,\eta_g,\alpha_0,\Theta,\omega,\omega_0)
		\end{equation}
		for a GNS triple for $\omega_0$, $(\HH_0,\pi_0,\Omega_0)$, then $W_g$, $u_R(g,h)$, $v$ and $K^R_g$ are also operators belonging to
		\begin{equation}
			\textrm{Index}^{\AA,\Ad{V^\dagger}(U)}(\theta,i_V^{-1}\circ\tilde{\beta}_g\circ i_V,i_V^{-1}\circ\eta_g\circ i_V,i_V^{-1}\circ\alpha_0\circ i_V,i_V^{-1}\circ\Theta\circ i_V,\omega\circ i_V,\omega_0\circ i_V)
		\end{equation}
		for the GNS triple for $\omega_0\circ i_V$ given by $(\HH_0,\pi_0\circ i_V,\Omega_0)$. In the case where $\omega$ is translation invariant in both directions, the same holds for the additional operators $w$, $b_g^R$, and $h_g$.
	\end{lemma}
	\begin{proof}
		To show this, we merely have to observe that for any $x\in U(\HH_0)$ satisfying that
		\begin{equation}
			\Ad{x}\circ\pi_0\circ\alpha_0\circ\Theta =\pi_0\circ\alpha_0\circ\Theta\circ\xi
		\end{equation}
		for some $\xi\in\Aut{\AA}$ we get that
		\begin{equation}
			\Ad{x}\circ\pi_0\circ i_V\circ i_V^{-1}\circ\alpha_0\circ i_V\circ i_V^{-1}\circ\Theta\circ i_V =\pi_0\circ i_V\circ i_V^{-1}\circ\alpha_0\circ i_V\circ i_V^{-1}\circ\Theta\circ i_V\circ i_V^{-1}\circ\xi\circ i_V
		\end{equation}
		concluding the proof.
	\end{proof}
	The proof that lemma \ref{lem:IndependenceOnChoiceOnSitHilbertSpaceOperatorsBelongingTo} implies lemma \ref{lem:IndependenceOnChoiceOnSitHilbertSpace} is trivial.
	\begin{remark}\label{rem:OnSiteUnitaryTransformationH3ValuedIndex}
		The same argument can be used to argue that the $H^3$-valued index is invariant under the same transformation.
	\end{remark}
	\section{Proof of the main theorems}\label{sec:ProofOfMainTheorems}
	We first summarise the proof of theorem \ref{thrm:ExistenceFirstIndex}:
	\begin{proof}
		The 2-cochain was defined in subsection \ref{sec:DefinitionH2Index}. The proof that it wasn't dependent on any of the choices made in assumption \ref{assumption} was done in section \ref{sec:IndexIsInvariantUnderChoices}. The first item in the list concerning the invariance under locally generated automorphisms was shown in subsection \ref{sec:IndexInvariantUnderLGA}. The proof of the second item concerning the relation between the 1d SPT index and the 2d translation SPT index is done in subsection \ref{sec:ExampleOneTranslation}. The proof of the third item concerning the stacking is done in section \ref{sec:the-role-of-stacking}. The proof of the last item concerning a change of basis of the on-site Hilbert space is done in section \ref{sec:the-indices-are-independent-on-the-choice-of-on-site-hilbert-space}. This concludes the proof.
	\end{proof}
	We now summarise the proof of theorem \ref{thrm:ExistenceSecondIndex}:
	\begin{proof}
		The 2-cochain was defined in subsection \ref{sec:DefinitionH1Index}. The proof that it wasn't dependent on any of the choices made in assumption \ref{assumption:2Translations} was done in subsection \ref{sec:IndexIsInvariantUnderChoices}. The first item in the list concerning the invariance under locally generated automorphisms was shown in subsection \ref{sec:IndexInvariantUnderLGA}. The proof of the second item concerning the relation between the 1d SPT index and the 2d translation SPT index is done in subsection \ref{sec:ExampleTwoTranslations}. The proof of the third item concerning the stacking is done in section \ref{sec:the-role-of-stacking}. The proof of the last item concerning a change of basis of the on-site Hilbert space is done in section \ref{sec:the-indices-are-independent-on-the-choice-of-on-site-hilbert-space}. This concludes the proof.
	\end{proof}
	\section{Proof of theorem \ref{lem:IndicesConstantOnStableEquivalenceClasses}}\label{sec:ProofOfMainResult1}\label{sec:ProofOf:lem:IndicesConstantOnStableEquivalenceClasses}
	\begin{figure}
		\includegraphics[width=0.5\textwidth]{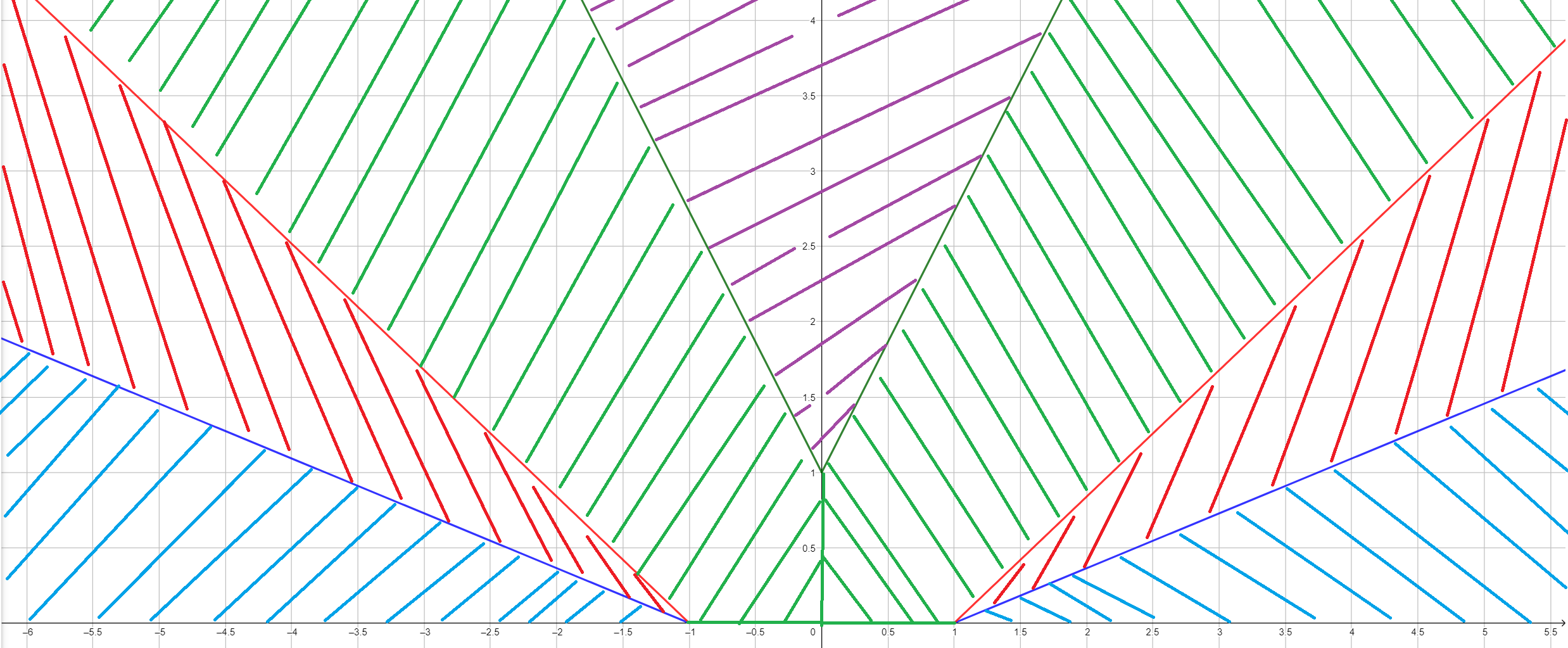}
		\includegraphics[width=0.5\textwidth]{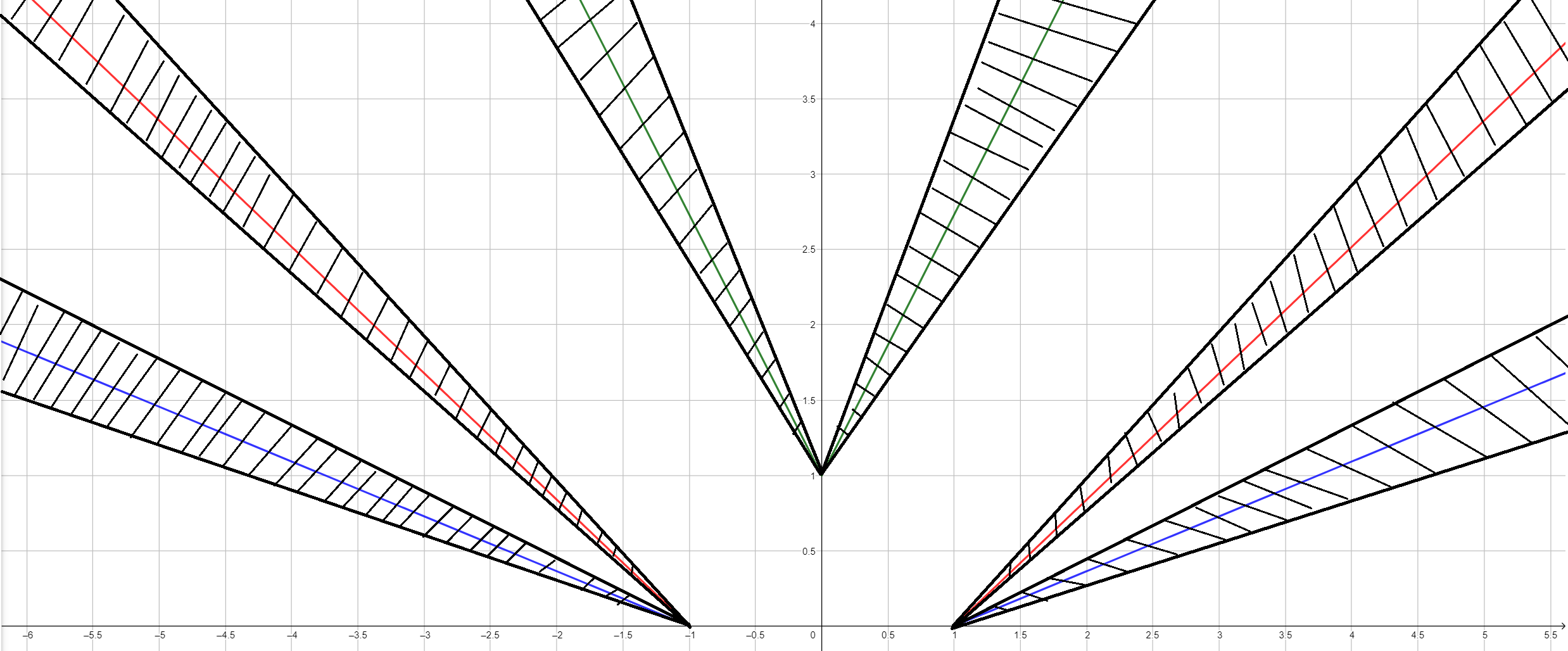}
		\caption{This figure shows the support of the different automorphisms present in the decomposition of an SQ$_1$-automorphism acting on the upper half-plane.}
		\label{fig:SetupSQAut2}
	\end{figure}
	We will use the notation
	\begin{align}
		C_{[0,\theta]}&\defeq C_{\theta}&C_{]\theta_1,\theta_2]}&\defeq C_{\theta_2}/C_{\theta_1}&C_{]\theta,\pi/2]}&\defeq \ZZ^2/C_{\theta}
	\end{align}
	where $0<\theta<\pi/2$ and $0<\theta_1<\theta_2<\pi_2$. In analogy with section 2.1 from \cite{ogata2021h3gmathbb} we will define some classes of automorphisms (see figure \ref{fig:SetupSQAut2}):
	\begin{definition}\label{def:SQAut2}
		Take $\alpha\in\Aut{\AA}$. We say that $\alpha\in\textrm{SQAut}_1(\AA)$ if and only if for any $0<\theta_{0.8}<\theta_{1}<\theta_{1.2}<\theta_{1.8}<\theta_{2}<\theta_{2.2}<\theta_{2.8}<\theta_3<\theta_{3.2}<\pi/2$ there exist $a\in\AA$, $\alpha_{[0,\theta_1], \sigma}\in\Aut{\AA_{\nu^{\sigma}(C_{[0,\theta_1]}\cap\sigma)}},\:\alpha_{]\theta_1,\theta_2],\sigma,\rho}\in\Aut{\AA_{\nu^{\sigma}(C_{]\theta_1,\theta_2]}\cap\rho\cap\sigma)}},\:\alpha_{]\theta_2,\theta_3],\sigma,\rho}\in\Aut{\AA_{(W(C_{[0,\theta_3]})/\nu^\sigma(C_{[0,\theta_2]}\cap\sigma))\cap\rho\cap\sigma)}},\\
		\alpha_{]\theta_3,\pi/2],\rho}\in\Aut{\AA_{\tau^\rho(C_{[\theta_3,\pi/2]}\cap\rho)}},\:\alpha_{]\theta_{0.8},\theta_{1.2}],\sigma,\rho}\in\Aut{\AA_{\nu^{\sigma}(C_{]\theta_{0.8},\theta_{1.2}]}\cap\rho\cap\sigma)}},\\
		\alpha_{]\theta_{1.8},\theta_{2.2}],\sigma,\rho}\in\Aut{\AA_{\nu^{\sigma}(C_{]\theta_{1.8},\theta_{2.2}]}\cap\rho\cap\sigma)}}$ and  $\alpha_{]\theta_{2.8},\theta_{3.2}],\sigma,\rho}\in\Aut{\AA_{\tau^{\rho}(C_{]\theta_{2.8},\theta_{3.2}]}\cap\rho\cap\sigma)}}$ for any $\rho\in\{U,D\}$ and $\sigma\in\{L,R\}$ (here $\tau^U=\tau$, $\tau^D=\tau^{-1}$, $\nu^R=\nu$ and $\nu^L=\nu^{-1}$) such that
		\begin{equation}
			\alpha=\Ad{a}\circ \bigotimes_{\sigma}\alpha_{[0,\theta_1],\sigma}\circ\bigotimes_{\sigma,\rho}\alpha_{]\theta_1,\theta_2],\sigma,\rho}\circ\bigotimes_{\rho}\alpha_{[\theta_3,\pi/2]}\circ\bigotimes_{\rho,\sigma}\left(\alpha_{]\theta_{0.8},\theta_{1.2}],\rho,\sigma}\otimes\alpha_{]\theta_{1.8},\theta_{2.2}],\rho,\sigma}\otimes\alpha_{]\theta_{2.8},\theta_{3.2}],\rho,\sigma} \right).
		\end{equation}
		If additionally everything (except $a$) commutes with $\beta_g^U$ we say that $\alpha\in \textrm{GSQAut}_1(\AA)$.
	\end{definition}
	As a remark, notice that this definition is not identical to the de definition given in \cite{ogata2021h3gmathbb} due to the translation operators. Before we prove the theorem, we will first list some required lemmas along with their proof.
	\begin{lemma}\label{lem:SRE_Implies_QDisentanglable_OneTranslation}
		Take $\omega\in\PP(\AA)$ a short-range entangled state that is $G$-invariant and translation invariant then it satisfies assumption \ref{assumption}.
	\end{lemma}
	\begin{proof}
		Let $\gamma^{\Phi'}_{0;1}$ be the disentangler with $\Phi'\in\BB_{F_\phi}([0,1])$ (for some $0<\phi<1$) a one-parameter family of interactions. By lemma \versionDifference{\ref{lem:PropertiesLocallyGeneratedAutomorphisms}}{C.4. of \cite{jappens2023spt}} part 5 (which is just a slight modification of section 5 from \cite{ogata2021h3gmathbb}) we have that $\gamma^{\Phi'}_{0;1}\in\textrm{SQAut}_1(\AA)$ since $\textrm{SQAut}_1(\AA)\subset\QAut{\AA}$ this implies that $\gamma^{\Phi'}_{0;1}\in\QAut{\AA}$. The existence of the $\tilde{\beta}_g$ was done in \cite{ogata2021h3gmathbb} section 3, where one has to substitute the support of the cones used there by the shifted cones seen in the definition of the $\textrm{SQAut}_2$. This concludes the proof that assumption \ref{assumption} is satisfied.
	\end{proof}
	\begin{lemma}\label{lem:SRE_Implies_QDisentanglable_TwoTranslations}
		Let $\omega$ be a short-range entangled state that is $G$-invariant and translation invariant in two directions then it satisfies assumption \ref{assumption:2Translations}.
	\end{lemma}
	\begin{proof}
		Completely analogous to the proof of lemma \ref{lem:SRE_Implies_QDisentanglable_OneTranslation}.
	\end{proof}
	\begin{lemma}\label{lem:RotationDoesn'tChangeH1Index}
		Let $\mu$ be the automorphism that rotates every element of the $C^*$ algebra by $90^\circ$ degrees, then $\textrm{Index}_{\text{2 trans}}^{\AA,U}(\omega)=\textrm{Index}_{\text{2 trans}}^{\AA,U}(\omega\circ\mu)$.
	\end{lemma}
	\begin{proof}
		This is proven in subsection \ref{sec:H1ValuedIndexInvariantUnderRotations}.
	\end{proof}
	Finally, we prove Theorem \ref{lem:IndicesConstantOnStableEquivalenceClasses}:
	\begin{proof}
		The last few lemmas show that the indices are all well-defined. Now we only have to show that they are independent on the choice of representative.	We will only prove the first item as all the other proofs are analogous. First notice that the equivalence implies that there exist product states $\phi_1\in\PP(\tilde\AA_1)$ and $\phi_2\in\PP(\tilde\AA_2)$ and group actions $\tilde U_1$ and $\tilde U_2$ such that the combinations $(\phi_1,\tilde U_1)$ and $(\phi_2,\tilde U_2)$ are trivial (see \ref{sec:States} for what this means). Moreover, these will satisfy that there exists a unitary transformation of the on-site Hilbert space, $V$ such that
		\begin{align}
			\omega_1\otimes_{\text{stack}}\phi_1\circ i_V&=\omega_2\otimes_{\text{stack}}\phi_2\circ\gamma^\Phi_{0;1}&V^\dagger U_{1}(g)\otimes \tilde{U}_{1}(g)V&=U_{2}(g)\otimes \tilde{U}_{2}(g)
		\end{align}
		for some $G$-invariant locally generated automorphism. Using item 1 and item 3 of \ref{thrm:ExistenceOriginalIndex} this implies that
		\begin{equation}
			\textrm{Index}^{\AA_1\otimes_{\text{stack}}\tilde \AA_1,U_1\otimes \tilde U_1}_{\text{no trans}}(\omega_1\otimes_{\text{stack}}\phi_1) =\textrm{Index}^{\AA_2\otimes_{\text{stack}}\tilde \AA_2,U_2\otimes\tilde U_2}_{\text{no trans}}(\omega_2\otimes_{\text{stack}}\phi_2).
		\end{equation}
		Using item 2 of \ref{thrm:ExistenceOriginalIndex} this is satisfied if $\textrm{Index}^{\tilde \AA_1,\tilde U_1}_{\text{no trans}}(\phi_1)=\textrm{Index}^{\tilde \AA_2,\tilde U_2}_{\text{no trans}}(\phi_2)=1$. This is trivially true. Only in the last part with the $H^1$-valued index does one need that the $(\phi,\tilde U)$ pairs are trivial (see remark \ref{rem:NontrivialProductState}). For the other three statements, it is enough that we have $G$-invariant product states.
	\end{proof}
	
	\versionDifference{
	\appendix
	\section{Properties of locally generated automorphisms: 1d}\label{sec:properties-of-locally-generated-automorphisms-1d}
	In this section, let $\AA$ be a quasi-local $C^*$ algebra over $\ZZ$.
	\begin{lemma}\label{lem:PropertiesLocallyGeneratedAutomorphisms1d}
		Take $\Phi\in\BB_{F_\phi}([0,1])$ (with $F_\phi$ as presented in \eqref{eq:OurFFunction}) then there exists an $A\in\UU(\AA)$ satisfying
		\begin{equation}
			\Ad{A}=\gamma^{\Phi_L+\Phi_R}_{0;1}\circ\gamma^\Phi_{1;0}.
		\end{equation}
		Moreover, this $A$ is summable (this was defined in subsection \ref{sec:QuasiLocalC*Algebra}). More generally there exists a continuous (in norm topology) one-parameter family $\lambda\in[0,1]\mapsto A_\lambda\in\UU(\AA)$ such that $A_\lambda$ is summable for each $\lambda$ and such that
		\begin{equation}
			\Ad{A_\lambda}=\gamma^{\Phi_L+\Phi_R}_{0;\lambda}\circ\gamma^\Phi_{\lambda;0}.
		\end{equation}
	\end{lemma}
	\begin{proof}
		We have for any $B\in\AA$ that
		\begin{align}
			\gamma^{\Phi_L+\Phi_R}_{0;t}\circ\gamma^\Phi_{t;0}(B)&=B+\int_{0}^{t}\d s\frac{\d}{\d s}\left(\gamma^{\Phi_L+\Phi_R}_{0;s}\circ\gamma^\Phi_{s;0}(B)\right)\\
			&=B-i\int_{0}^{t}\d s\sum_{I\in\mathfrak{G}_\ZZ}\gamma^{\Phi_L+\Phi_R}_{0;s}([\Phi_L(s,I)+\Phi_R(s,I)-\Phi(s,I),\gamma^{\Phi}_{s;0}(B)])\\
			\label{eq:EquationShowingBounded_H}
			&=B-i\int_{0}^{t}\d s\gamma^{\Phi_L+\Phi_R}_{0;s}\circ\gamma^{\Phi}_{s;0}\left(\left[\sum_{I\in\mathfrak{G}_\ZZ}\gamma^{\Phi}_{0;s}(\Phi_L(s,I)+\Phi_R(s,I)-\Phi(s,I)),B\right]\right).
		\end{align}
		To simplify notation we will define an interaction
		\begin{equation}
			\tilde\Phi(s,I)=\Phi_L(s,I)+\Phi_R(s,I)-\Phi(s,I).
		\end{equation}
		We will in essence have to find a 1D analogue of the proof of Theorem 5.2 of \cite{ogata2021h3gmathbb} with a few exceptions. Most notably we need the summability condition. First, let us still define the extensions
		\begin{equation}
			X(m)\defeq \{x\in\ZZ|\exists y\in X:\textrm{dist}(x,y)\leq m\}
		\end{equation}
		and
		\begin{equation}
			\Delta_{X(m)}\defeq \Pi_{X(m)}-\Pi_{X(m-1)}
		\end{equation}
		the differences between the conditional expectation values. Define an interaction
		\begin{equation}\label{eq:RecastingTheInteraction}
			\Xi(Z,t)\defeq\sum_{m\geq 0}\sum_{\substack{X\subset Z:\\X(m)=Z}}\Delta_{X(m)}(\gamma^{\Phi}_{0;t}(\tilde{\Phi}(X,t)))
		\end{equation}
		then we get by construction that
		\begin{equation}
			\gamma^{\Phi_L+\Phi_R}_{0;t}\circ\gamma^{\Phi}_{t;0}=\gamma^{\Xi}_{0;t}.
		\end{equation}
		We will now define
		\begin{equation}
			A_\lambda\defeq \mathcal{T}\exp(-i\int_0^\lambda \d s\: \sum_{Z\in\mathfrak{B}_{\ZZ}}\Xi(Z,s)).
		\end{equation}
		By construction, if this exists it is continuous in $\lambda$. To show that it exists and is summable, define some sets
		\begin{align}
			B_n&\defeq [-n,n]\cap\ZZ&B_{n,L}&\defeq ]-\infty,-n]\cap\ZZ&B_{n,R}&\defeq[n,\infty[\cap\ZZ
		\end{align}
		and let
		\begin{equation}
			A_n\defeq \mathcal{T}\exp(-i\int_0^1 \d s\: \sum_{Z\in\mathfrak{B}_{\ZZ}}\Xi_{B_n}(Z,s)).
		\end{equation}
		In what follows we will find a bound on
		\begin{equation}
			c_n\defeq \sup_{t\in[0,1]}\norm{\sum_{Z\in\mathfrak{B}_\ZZ}(\Xi(Z,t)-\Xi_{B_n}(Z,t))}
		\end{equation}
		that is summable. Finding this would conclude the proof because
		\begin{equation}
			\sum_n\norm{A-A_n}=\sum_n\norm{\int_0^1\d s\: \frac{\d}{\d s}A_{n}(s)^\dagger A(s)}\leq \sum_n \sup_{t\in[0,1]}\norm{\sum_{Z\in\mathfrak{B}_\ZZ}(\Xi(Z,t)-\Xi_{B_n}(Z,t))}=\sum_n c_n.
		\end{equation}
		In analogy with equation (5.22) of \cite{ogata2021h3gmathbb} we get
		\begin{align}
			\label{eq:c_n_Bound1D}
			c_n&\leq \sum_{\substack{Z\in\mathfrak{B}_{\ZZ}:\\Z\cap B_{n,L}\neq\emptyset\\\text{or}\\Z\cap B_{n,R}\neq\emptyset}}\sup_{t\in[0,1]}\norm{\Xi(Z,t)}\\
			&\leq \frac{8}{c_F}(e^{2I_F(\Phi)}-1) \sum_{m\geq 0}\sum_{\substack{X\in\mathfrak{B}_{\ZZ}:\\X(m)\cap B_{n,L}\neq\emptyset\\\text{or}\\X(m)\cap B_{n,R}\neq\emptyset}}\sup_{t\in[0,1]}\left(\norm{\tilde{\Phi}(X,t)}\right)\abs{X}G_F(m).
		\end{align}
		Now using a trick analogous to what was done in equation (5.27) of \cite{ogata2021h3gmathbb} we get
		\begin{equation}
			c_n\leq \norm{\Phi_1}_F\left(\sum_{m\geq 0}G_F(m)\right)(\sum_{x\in B_{n,L}}\sum_{y=B_{0,R}}F(d(x,y))+\sum_{x\in B_{0,L}}\sum_{y=B_{n,R}}F(d(x,y))).
		\end{equation}
		To show that the last bound is summable we only have to observe that
		\begin{equation}
			\sum_{n=0}^\infty\sum_{x\in B_{n,L}}\sum_{y=B_{0,R}}F(d(x,y))\leq \sum_{x\in C_1}\sum_{y=C_2}F(d(x,y)).
		\end{equation}
		where $C_1$ is the cone one obtains by putting the $B_{n,L}$ on top of each other whereas $C_2$ is the right half plane of $\ZZ^2$. The proof that the sum over the cones is bounded is done explicitly in \cite{ogata2021h3gmathbb}.
	\end{proof}
	\section{Properties of locally generated automorphisms: 2d}\label{sec:properties-of-locally-generated-automorphisms-2d}
	In this section, let $\AA$ be a quasi-local $C^*$ algebra over $\ZZ^2$. Take $H\in\BB_{F_\phi}([0,1])$ (for some $0<\phi<1$) and $\gamma^H_{s;t}$ its locally generated automorphism (see section \ref{sec:Interactions}). We will now show some properties of this locally generated automorphism $\gamma^H_{s;t}$. In this section we will heavily rely on the framework and theorems of Yoshiko Ogata \cite{ogata2021h3gmathbb} who in her turn relies heavily on the framework developed in \cite{doi:10.1063/1.5095769}. We now need two additional classes of automorphisms:
	\begin{definition}
		Take $\alpha\in\Aut{\AA}$. We say that $\alpha\in\textrm{HAut}_1(\AA)$ if and only if for any $0<\theta<\pi/2$ there exists an $a\in\UU(\AA)$ and some $\alpha_\sigma\in\Aut{\AA_{\nu^{\sigma}(C_\theta\cap\sigma)}}$ for each $\sigma\in\{L,R\}$ such that
		\begin{equation}
			\alpha=\Ad{a}\circ\alpha_L\otimes\alpha_R.
		\end{equation}
	\end{definition}
	\begin{definition}
		Take $\alpha\in\Aut{\AA}$. We say that $\alpha\in\textrm{VAut}_1(\AA)$ if and only if there exists some $a\in\AA$, $\alpha_U\in\Aut{\tau(\AA_{C_\theta^c}\cap\AA_{U})}$ and an $\alpha_D\in\Aut{\tau^{-1}(\AA_{C_\theta^c}\cap\AA_{D})}$ such that
		\begin{equation}
			\alpha=\Ad{a}\circ\alpha_U\otimes\alpha_D.
		\end{equation}
		If furthermore $\alpha_U\circ\beta_g^U=\beta_g^U\circ\alpha_U$ we say that $\alpha\in\textrm{GVAut}_1(\AA)$.
	\end{definition}
	Additionally, remember the definition of the $\textrm{QAut}_2(\AA)$ that was presented in definition \ref{def:SQAut2}. Now we will state some properties of locally generated automorphisms:
	\begin{lemma}\label{lem:PropertiesLocallyGeneratedAutomorphisms}
		Take $H$ an interaction such that there exists a $0<\phi<1$ satisfying that $\norm{H}_{f_\phi}\leq 1$. The following statements now hold (for any $s,t\in\RR$):
		\begin{enumerate}
			\item $\gamma^H_{s;t}\circ\gamma^{H_D}_{t;s}\otimes\gamma^{H_U}_{t;s}\in\textrm{HAut}_1(\AA)$.
			\item $\gamma^H_{s;t}\circ\gamma^{H_L}_{t;s}\otimes\gamma^{H_R}_{t;s}\in\textrm{VAut}_1(\AA)$. If additionally $H$ is a $G$-invariant interaction we even have $\gamma^H_{s;t}\circ\gamma^{H_L}_{t;s}\otimes\gamma^{H_R}_{t;s}\in\textrm{GVQAut}_1(\AA)$. The similar statement also holds if we replace $L$ and $R$ by $\nu^{-1}(L)$ and $\nu(R)$.
			\item $\gamma^{H_U}_{s;t}\otimes\gamma^{H_D}_{s;t}\in\textrm{SQAut}_1(\AA)$. If additionally $H$ is a $G$-invariant interaction we even have $\gamma^{H_U}_{s;t}\otimes\gamma^{H_D}_{s;t}\in\textrm{GSQAut}_1(\AA)$.
			\item If $H$ is $G$-invariant then $\gamma^{H}_{t;s}\circ\beta_g^U\circ\gamma^{H}_{s;t}\circ(\beta_g^U)^{-1}\in\textrm{HAut}_1(\AA)$.
			\item $\gamma^{H}_{t;s}\in\textrm{SQAut}_1(\AA)$.
		\end{enumerate}
	\end{lemma}
	\begin{proof}
		Part 1 is done in Proposition 5.5 in \cite{ogata2021h3gmathbb} and Part 4 follows trivially from Part 1. Except for the translation operators in my definition of the $\textrm{SQAut}(\AA)$, part 3 follows from Theorem 5.2 in \cite{ogata2021h3gmathbb}. To show part 3 we therefore have to show that Theorem 5.2 in \cite{ogata2021h3gmathbb} still holds if we replace $\mathcal{C}_0$ and $\mathcal{C}_1$ in the proof by the sets highlighted in figure \ref{fig:SetupSQAut2}. Take the $\Psi$ from this proof to be our $H$, take $\Psi^{(0)}$ to be $\sum_{C\in\mathcal{C}_0}H_{C}$ (with our new definition of $\mathcal{C}_0$) and take $\Psi^{(1)}\defeq \Psi-\Psi^{(0)}$. Now define $\Xi^{(s)}(Z,t)$ through
		\begin{equation}\label{eq:PropertiesLocallyGeneratedAutomorphismsProofDefinitionXi}
			\Xi^{(s)}(Z,t)\defeq\sum_{m=0}^\infty \sum_{\substack{X\subset Z,\\X(m)=Z}}\Delta_{X(m)}(\gamma^\Psi_{s;t}(\Psi^{(1)}(X;t)))
		\end{equation}
		with these new definitions of $\Psi^{(0)}$ and $\Psi^{(1)}$. We now want to show that for every $t$,
		\begin{equation}
			\sum_{Z\subset\ZZ^2}\left(\Xi^{(s)}(Z,t)-\sum_{C\in\mathcal{C}_1}\id_{Z\subset C}\Xi^{(s)}(Z,t)\right)
		\end{equation}
		is bounded. Following the arguments in equation (5.22) to (5.24) in \cite{ogata2021h3gmathbb} we still get that
		\begin{equation}
			\sum_{\substack{Z\subset\ZZ^2,\\\nexists C\in\mathcal{C}_1:Z\subset C}}\sup_{t\in[0,1]}\norm{\Xi^{(1)}(Z,t)}\leq \frac{8}{C_F}(e^{2I_F(\Psi)}-1)\sum_{\substack{C_1,C_2\in\mathcal{C}_0,\\ C_1\neq C_2}}M(C_1,C_2)
		\end{equation}
		where
		\begin{equation}
			M(C_1,C_2)\defeq \sum_{m\geq 0}\sum_{\substack{X:\\\forall C\in\mathcal{C}_1,X\cap((C^c)(m))\neq\emptyset,\\X\cap C_1\neq\emptyset,X\cap C_2\neq\emptyset}}\left(\sup_{t\in[0,1]}(\norm{\Psi^{(1)}(X;t)})\abs{X}G_F(m)\right)
		\end{equation}
		is now defined using the new $\mathcal{C}_1$. To bound these $M(C_1,C_2)$ we will (just like in \cite{ogata2021h3gmathbb}) differentiate between two cases. That is, the case where $C_1$ and $C_2$ are adjacent\footnote{We say that $C_1$ and $C_2$ are adjacent if and only if $\#(C_1(1)\cap C_2(1))=\infty$.} and the case where they are not. We begin with the latter. In this case, we still have in complete analogy with the proof in \cite{ogata2021h3gmathbb} that
		\begin{align}
			M(C_1,C_2)\leq b_0(C_1,C_2) &\defeq \sum_{m\geq 0}\sum_{\substack{X:X\cap C_1\neq\emptyset,\\X\cap C_2\neq\emptyset}}\left(\sup_{t\in[0,1]}(\norm{\Psi^{(1)}(X;t)})\abs{X}G_F(m)\right)\\
			&\leq \norm{\Psi_1}_F\sum_{\substack{x\in C_1\\y\in C_2}}F(d(x,y))\sum_{m=0}^\infty G_F(m)<\infty.
		\end{align}
		\footnote{Here $\Psi_1$ is defined through $\Psi_1(X)=\abs{X}^1 \Psi(X)$.} What is now left to show is the case where $C_1$ and $C_2$ are adjacent. Take $\tilde{C}\in\mathcal{C}_1$ such that $C_1\cap\tilde{C}\neq\emptyset$ and $C_2\cap\tilde{C}\neq\emptyset$. Take $L_1=\partial\tilde{C}/C_2$\footnote{Here $\partial \tilde{C}$ means the boundary of $\tilde C$.} and $L_2=\partial\tilde{C}/C_1$. By following the same reasoning that led to equation (5.36) in \cite{ogata2021h3gmathbb} we get that
		\begin{align}
			M(C_1,C_2)\leq b_0(C_1,C_2/\tilde{C})+b_0(C_1/\tilde{C},C_2)+b_0(\tilde{C}\cap C_2,(C_1\cup C_2)^c)+b_1(C_1\cap\tilde{C},C_2\cap\tilde{C},L_1,L_2)
		\end{align}
		where
		\begin{align}
			b_1(C_1\cap\tilde{C},C_2\cap\tilde{C},L_1,L_2)&\defeq\sum_{m=0}^\infty \sum_{\substack{X\subset \tilde{C}:\\ X\cap C_1\cap\tilde{C}\neq\emptyset\\X\cap C_2\cap\tilde{C}\neq\emptyset\\ X\cap (\tilde{C}^c(m))\neq\emptyset}}\sup_{t\in[0,1]}\norm{\Psi(X;t)}\abs{X}G_F(m)\\
			&\leq \norm{\Psi_1}_F\sum_{m=0}^{\infty}G_F(m)\left(\sum_{\substack{x\in C_2\cap\tilde{C},\\y\in L_1(m)}}+\sum_{\substack{x\in C_1\cap\tilde{C},\\y\in L_2(m)}}\right)F(d(x,y))<\infty.
		\end{align}
		Since the remainder of the proof can remain unchanged from \cite{ogata2021h3gmathbb}, this concludes the proof of item 3. The proof of item 5 just follows from the fact that for any $\alpha_1\in\textrm{SQAut}(\AA)$ and for any $\alpha_2\in\textrm{HAut}(\AA)$ we have that $\alpha_2\circ\alpha_1\in\textrm{SQAut}(\AA)$. We now only need to comment on item 2. We must show that
		\begin{equation}
			(\gamma^H_{s;t}\circ\gamma^{H_L}_{t;s}\otimes\gamma^{H_R}_{t;s})^{-1}=\gamma^{H_L}_{s;t}\otimes\gamma^{H_R}_{s;t}\circ\gamma^H_{t;s}\in\textrm{GVAut}(\AA).
		\end{equation}
		The proof of this starts analogously to the proof of items 1 and 3. We take $\Psi=H$, $\Psi^{(0)}=H_{\tau(C_\theta^c\cap U)}+H_{\tau^{-1}(C_\theta^c\cap D)}$ and $\Psi^{(1)}=\Psi-\Psi^{(0)}$. Define $\Xi^{(s)}(Z;t)$ again through equation \eqref{eq:PropertiesLocallyGeneratedAutomorphismsProofDefinitionXi}. In analogy to what was done in equation (5.54) in \cite{ogata2021h3gmathbb} we obtain
		\begin{equation}
			\sum_{\substack{Z:Z\nsubseteq \tau(C_\theta^c\cap U)\\\text{and }Z\nsubseteq \tau^{-1}(C_\theta^c\cap D)}}\sup_{t\in[0,1]}\norm{\Xi^{(1)}(Z,t)}\leq \frac{8}{C_F}(e^{2I_F(\Psi)}-1)\sum_{m=0}^\infty\sum_{\substack{X:X(m)\nsubseteq \tau(C_\theta^c\cap U)\\\text{and }X(m)\nsubseteq \tau^{-1}(C_\theta^c\cap D)}} \sup_{t\in[0,1]}\norm{\Psi^{(1)}(X;t)}\abs{X}G_F(m).
		\end{equation}
		If $X$ in the last line has a non-zero contribution, then at least one of the following occurs:
		\begin{enumerate}
			\item $X\cap (W(C_\theta)\cap L)\neq\emptyset$ and $X\cap R\neq\emptyset$.
			\item $X\cap (W(C_\theta)\cap R)\neq\emptyset$ and $X\cap L\neq\emptyset$.
			\item $X\subset W(C_\theta)^c$ and
			\begin{enumerate}
				\item $X\subset U,X\subset D$, or
				\item $X\subset U,X\subset L,X\subset R$ and $X(m)\cap (\tau^{-1}(C_\theta^c\cap D))^c\neq \emptyset$, or
				\item $X\subset D,X\subset L,X\subset R$ and $X(m)\cap (\tau(C_\theta^c\cap U))^c\neq \emptyset$.
			\end{enumerate}
		\end{enumerate}
		This shows that we have a bound
		\begin{align}
			&\sum_{\substack{Z:Z\nsubseteq \tau(C_\theta^c\cap U)\\\text{and }Z\nsubseteq \tau^{-1}(C_\theta^c\cap D)}}\sup_{t\in[0,1]}\norm{\Xi^{(1)}(Z,t)}\\
			\nonumber
			&\leq \frac{8}{C_F}(e^{2I_F(\Psi)}-1)(b_0(W(C_\theta)\cap L,R)+b_0(L,W(C_\theta)\cap R)+b_0(W(C_\theta)^c\cap U,W(C_\theta)^c\cap D)\\
			\nonumber
			&\qquad+b_1(W(C_\theta)^c\cap U\cap L,W(C_\theta)^c\cap U\cap L,L\cap\partial(W(C_\theta)^c\cap U),R\cap\partial(W(C_\theta)^c\cap U))\\
			&\qquad b_1(W(C_\theta)^c\cap D\cap L,W(C_\theta)^c\cap D\cap R,L\cap\partial(W(C_\theta)^c\cap D),R\cap\partial(W(C_\theta)^c\cap D)))<\infty.
		\end{align}
		This concludes the proof.
	\end{proof}
	This implies certain things for our locally generated automorphisms. From these four statements we can prove the following results:
	\begin{lemma}\label{lem:TwoAngleLemmaPart1}
		Take $H$ an interaction such that there exists a $0<\phi<1$ satisfying that $\norm{H}_{f_\phi}\leq 1$. Take $\theta_1$ and $\theta_2$ such that $0<\theta_1<\theta_2<\pi/2$ then for all $\Theta\in\Aut{\AA_{W(C_{\theta_2})^c}}$ and $s,t\in\RR$ there exists an $a_1\in\UU(\AA)$ and a $\tilde{\Theta}\in \Aut{\AA_{W(C_{\theta_1})^c}}$ such that
		\begin{equation}\label{eq:TwoAngleLemmaPart1Equation1}
			\gamma^{H}_{t;s}\circ\Theta\circ\gamma^{H}_{s;t}=\Ad{a_1}\circ\tilde{\Theta}.
		\end{equation}
	\end{lemma}
	\begin{proof}
		We have that
		\begin{equation}
			\gamma^{H}_{t;s}\circ\Theta\circ\gamma^{H}_{s;t}=\gamma^{H_D}_{t;s}\otimes\gamma^{H_U}_{t;s}\circ\gamma^{H_D}_{s;t}\otimes\gamma^{H_U}_{s;t}\circ\gamma^{H}_{t;s}\circ\Theta\circ\gamma^{H}_{s;t}\circ\gamma^{H_D}_{t;s}\otimes\gamma^{H_U}_{t;s}\circ\gamma^{H_D}_{s;t}\otimes\gamma^{H_U}_{s;t}.
		\end{equation}
		Using that $\gamma^{H}_{s;t}\circ\gamma^{H_D}_{t;s}\otimes\gamma^{H_U}_{t;s}\in\textrm{HAut}_1(\AA)$ we get that there exists some $a\in\AA$ and $\eta\in\Aut{\AA_{C_\theta}}$ such that
		\begin{align}
			\gamma^{H}_{t;s}\circ\Theta\circ\gamma^{H}_{s;t}&=\Ad{a}\circ \gamma^{H_D}_{t;s}\otimes\gamma^{H_U}_{t;s}\circ\eta_{s;t}^{-1}\circ\Theta\circ\eta_{s;t}\circ\gamma^{H_D}_{s;t}\otimes\gamma^{H_U}_{s;t}\\
			&=\Ad{a}\circ \gamma^{H_D}_{t;s}\otimes\gamma^{H_U}_{t;s}\circ\Theta\circ\gamma^{H_D}_{s;t}\otimes\gamma^{H_U}_{s;t}.
		\end{align}
		Since by \ref{lem:PropertiesLocallyGeneratedAutomorphisms} part 3 $\gamma^{H_D}_{s;t}\otimes\gamma^{H_U}_{s;t}\in\textrm{GSQAut}_1(\AA)$ the result follows.
	\end{proof}
	\begin{lemma}\label{lem:TwoAngleLemmaPart2}
		Take $H$ an interaction such that there exists a $0<\phi<1$ satisfying that $\norm{H}_{f_\phi}\leq 1$. Take $\theta_1$ and $\theta_2$ such that $0<\theta_1<\theta_2<\pi/2$. Then for all $\eta_{g}^{\sigma}\in\Aut{\AA_{\nu^{\sigma}(C_{\theta_1}\cap\sigma)}}$ (where $\sigma\in\{L,R\}$ and $g\in G$) and $s,t\in\RR$ there exist $a_{2},\in\UU(\AA),a_{3,\sigma}\in\UU(\AA_{\nu^{\sigma}(\sigma)})$ and some $\tilde{\eta}_{\sigma}^g\in \Aut{\nu^{\sigma}(\AA_{C_{\theta_2}\cap\sigma)}}$ such that
		\begin{align}
			\label{eq:TwoAngleLemmaPart2Equation1}
			\gamma^{H}_{t;s}\circ\eta_{g}^L\otimes\eta_{g}^R\circ\gamma^{H}_{s;t}&=\Ad{a_2}\circ(\tilde\eta_{g}^L \otimes\tilde\eta_{g}^R)\\
			\label{eq:TwoAngleLemmaPart2Equation2}
			\gamma^{H_\sigma}_{t;s}\circ\eta_g^\sigma\circ\gamma^{H_\sigma}_{s;t}&=\Ad{a_{3,\sigma}}\circ \tilde\eta_{g}^\sigma.
		\end{align}
	\end{lemma}
	\begin{proof}
		In this proof, take $\Hsplit=H_{\nu^{-1}(L)}+H_{\nu(R)}$. First, we show that equation \eqref{eq:TwoAngleLemmaPart2Equation2} implies equation \eqref{eq:TwoAngleLemmaPart2Equation1}. This is because using equation \eqref{eq:TwoAngleLemmaPart2Equation1} we get that
		\begin{align}
			\gamma^H_{t;s}\circ\eta_g\circ\gamma^{H}_{s;t}&=\gamma^H_{t;s}\circ\gamma^{\Hsplit}_{s;t}\circ\gamma^{\Hsplit}_{t;s}\circ\eta_g\circ\gamma^{\Hsplit}_{s;t}\circ\gamma^{\Hsplit}_{t;s}\circ\gamma^H_{s;t}\\
			&=\gamma^H_{t;s}\circ\gamma^{\Hsplit}_{s;t}\circ\Ad{a_{3,L}\otimes a_{3,R}}\circ\tilde{\eta}_g\circ\gamma^{\Hsplit}_{t;s}\circ\gamma^H_{s;t}.
		\end{align}
		If one now uses the fact that $\gamma^H_{t;s}\circ\gamma^{\Hsplit}_{s;t}\in\textrm{GVAut}_1(\AA)$ (see lemma \ref{lem:PropertiesLocallyGeneratedAutomorphisms} item 2) the implication follows. To finish the proof we now only have to use the fact that the $\gamma^{H_\sigma}_{0;1}$ are in $\textrm{SQAut}_1(\AA_\sigma)$.
	\end{proof}
	Additionally, when we add the group action, we get:
	\begin{lemma}\label{lem:TwoAngleLemmaPart3}
		Take $H$ a $G$-invariant interaction such that there exists a $0<\phi<1$ satisfying that $\norm{H}_{f_\phi}\leq 1$. Take $\theta_1$ and $\theta_2$ such that $0<\theta_1<\theta_2<\pi/2$. Then for all $\eta_{g}^{\sigma}\in\Aut{\AA_{\nu^{\sigma}(C_{\theta_1}\cap\sigma)}}$ (where $\sigma\in\{L,R\}$ and $g\in G$) and $s,t\in\RR$ there exist $a_{2},\in\UU(\AA),a_{3,\sigma}\in\UU(\AA_{\nu^{\sigma}(\sigma)})$ and some $\tilde{\eta}_{\sigma}^g\in \Aut{\nu^{\sigma}(\AA_{C_{\theta_2}\cap\sigma)}}$ such that
		\begin{align}
			\label{eq:TwoAngleLemmaPart3Equation1}
			\gamma^{H}_{t;s}\circ\eta_{g}^L\otimes\eta_{g}^R\circ\beta_g^U\circ\gamma^{H}_{s;t}&=\Ad{a_2}\circ(\tilde\eta_{g}^L \otimes\tilde\eta_{g}^R)\circ\beta_g^U\\
			\label{eq:TwoAngleLemmaPart3Equation2}
			\gamma^{H_\sigma}_{t;s}\circ\eta_g^\sigma\circ\beta_g^{\sigma U}\circ\gamma^{H_\sigma}_{s;t}&=\Ad{a_{3,\sigma}}\circ \tilde\eta_{g}^\sigma\circ\beta_g^{\sigma U}.
		\end{align}
	\end{lemma}
	\begin{proof}
		In this proof, take again $\Hsplit=H_{\nu^{-1}(L)}+H_{\nu(R)}$. First, we show that equation \eqref{eq:TwoAngleLemmaPart3Equation2} implies equation \eqref{eq:TwoAngleLemmaPart3Equation1}. This is because using (the inverse of) equation \eqref{eq:TwoAngleLemmaPart3Equation1} we get that
		\begin{align}
			&\gamma^H_{t;s}\circ\eta_g\circ\beta_g^U\circ\gamma^H_{s;t}\circ\beta_{g^{-1}}^U\circ(\tilde{\eta}_g)^{-1}\\
			&=(\text{Inner})\circ \gamma^H_{t;s}\circ\eta_g\circ\beta_g^U\circ\gamma^H_{s;t}\circ\gamma^\Hsplit_{t;s}\circ\beta_{g^{-1}}^U\circ(\eta_g)^{-1}\circ\gamma^\Hsplit_{s;t}\circ\underline{\gamma^H_{t;s}\circ\gamma^H_{s;t}}.
		\end{align}
		Now using the fact that by lemma \ref{lem:PropertiesLocallyGeneratedAutomorphisms} item 2, $\gamma^H_{s;t}\circ\gamma^\Hsplit_{t;s}\in\textrm{GVAut}_1(\AA)$ this gives us
		\begin{align}
			&=(\text{Inner})\circ\stkout{\gamma^H_{t;s}\circ\eta_g\circ\beta_g^U\circ\beta_{g^{-1}}^U\circ(\eta_g)^{-1}\circ\gamma^H_{s;t}}.
		\end{align}
		By taking $a_2$ such that $\Ad{a_2}$ is this inner automorphism, we can conclude the proof that equation \eqref{eq:TwoAngleLemmaPart3Equation2} implies equation \eqref{eq:TwoAngleLemmaPart3Equation1}. Now we only have to prove equation \eqref{eq:TwoAngleLemmaPart3Equation2}. To finish the proof we now only have to use the fact that the $\gamma^{H_{\nu^{\sigma}(\sigma)}}_{0;1}$ are in $\textrm{GSQAut}_1(\AA_\sigma)$.
	\end{proof}
	\begin{lemma}\label{lem:SplittedAutomorphismAfterTranslatedIsVertical}
		There exist maps
		\begin{align}
			A_L&:[0,1]\rightarrow \UU(\AA_L):\lambda\mapsto A_{L}(\lambda)&A_R&:[0,1]\rightarrow \UU(\AA_R):\lambda\mapsto A_{R}(\lambda)
		\end{align}
		both continuous in norm topology and
		\begin{align}
			\Phi^{\mu\nu}:[0,1]\rightarrow\Aut{\AA_{W(C_\theta)^c\cap\mu\cap\nu}}:\lambda\mapsto \Phi^{\mu\nu}(\lambda)
		\end{align}
		where $\mu\in\{U,D\}$ and $\nu\in\{L,\nu(R)\}$, all four continuous in  strong\footnote{Meaning that $\Phi^{\mu\nu}(\lambda)(A)$ is continuous for all $A\in\AA$.} topology, satisfying that
		\begin{align}
			\nu\circ\gamma^{\HsplitTilde}_{0;\lambda}\circ\nu^{-1}\circ\gamma^{\HsplitTilde}_{\lambda;0}&=\gamma^{H_L}_{0;\lambda}\otimes\gamma^{H_{\nu\circ\nu(R)}}_{0;\lambda}\circ\gamma^{\HsplitTilde}_{\lambda;0}\\
			\label{eq:TranslatingSplittedTimeEvolutionAppendix}
			&=\Ad{A_L(\lambda)}\otimes\Ad{A_R(\lambda)}\circ\bigotimes_{\substack{\mu\in\{U,D\},\\\nu\in\{L,\nu(R)\}}}\Phi^{\mu\nu}(\lambda)
		\end{align}
		and that $\Phi^{\mu\nu}\circ\beta_g=\beta_g\circ\Phi^{\mu\nu}$.
	\end{lemma}
	\begin{proof}
		The proof of this is very similar to all the other proofs in this section. We will go over the highlights. We want to define an $A_L(\lambda)$ and a $\Phi^{U,L}\otimes\Phi^{D,L}$ satisfying
		\begin{equation}
			\Ad{A_L(\lambda)}\circ\Phi^{U,L}\otimes\Phi^{D,L}(\lambda)=\gamma^{H_L}_{0;\lambda}\circ\gamma^{H_{\nu^{-1}(L)}}_{\lambda;0}
		\end{equation}
		and similarly for the right side. We will only work out this side as the other calculation is analogous. We do this in two steps. First, we find an interaction $\Xi$ satisfying
		\begin{equation}
			\gamma^\Xi_{0;\lambda}=\gamma^{H_L}_{0;\lambda}\circ\gamma^{H_{\nu^{-1}(L)}}_{\lambda;0}.
		\end{equation}
		We can do this by taking the usual
		\begin{equation}
			\Xi(Z,t)\defeq \sum_{m=0}^{\infty}\sum_{\substack{X\subset Z\\X(m)=Z}}\Delta_{X(m)}(\gamma^{H_{\nu^{-1}(L)}}_{0;t}(\tilde{H}(X,t)))
		\end{equation}
		where $\tilde{H}\defeq H_L-H_{\nu^{-1}(L)}$. We can now define
		\begin{equation}
			\Phi^{\mu L}(\lambda)\defeq\gamma^{\Xi_{W(C_\theta)^c\cap\mu\cap L}}_{0;\lambda}.
		\end{equation}
		This means that we can satisfy \ref{eq:TranslatingSplittedTimeEvolutionAppendix} if
		\begin{align}
			\Ad{A_L}&=\gamma^{\Xi}_{0;\lambda}\circ\gamma^{\Xi'}_{\lambda;0}&\Xi'&\defeq \Xi_{W(C_\theta)^c\cap U\cap L}+\Xi_{W(C_\theta)^c\cap D\cap L}.
		\end{align}
		Using the same argument as before, this can be realized when we have
		\begin{equation}
			\Ad{A_L(\lambda)}=\gamma^{\tilde\Xi}_{0;\lambda}
		\end{equation}
		where
		\begin{equation}
			\tilde\Xi(Z,t)\defeq \sum_{m=0}^{\infty}\sum_{\substack{X\subset Z\\X(m)=Z}}\Delta_{X(m)}(\gamma^{\Xi'}_{0;t}(\Xi(X,t)-\Xi'(X,t))).
		\end{equation}
		The last part of the proof is to show that $\norm{\sum_{Z\in\mathfrak{B}_{\ZZ^2}}\tilde\Xi(Z,t)}<\infty$ which can be done using similar techniques as before. This means that we can simply define
		\begin{equation}
			A_L(\lambda)\defeq\mathcal{T}\exp(-i\int_0^\lambda \d s\: \sum_{Z\in \mathfrak{B}_{\ZZ^2}}\tilde{\Xi}(Z,s))
		\end{equation}
		and this is by construction bounded and continuous in $\lambda$.
	\end{proof}
	}{}
	
	\bibliography{TSPT}
	\bibliographystyle{plain}
\end{document}